\documentclass[prd,aps,
longbibliography,
nofootinbib,
floatfix,
superscriptaddress]{revtex4}
\usepackage{graphicx}
\usepackage{epsfig}
\usepackage{rotating}
\usepackage{amssymb}
\usepackage{dsfont}
\usepackage{psfrag}
\usepackage{amsmath,euscript,array,mathrsfs}
\usepackage{bbold}
\usepackage{bm}
\usepackage{multirow}
\usepackage{dcolumn}
\RequirePackage{doi}
\usepackage{hyperref}
\usepackage[skip=10pt plus1pt, indent=20pt]{parskip}
\usepackage{hyperref}
\usepackage{float}
\floatstyle{plaintop}
\restylefloat{table}
\usepackage{bbding}
\usepackage{orcidlink}
\usepackage{xcolor}
\colorlet{RED}{red} % can be removed once we no longer use colours in section titles

\allowdisplaybreaks

\usepackage[caption=false]{subfig}
\newcommand{\orcidauthorPIAI}{0000-0002-2251-0111}

\newcommand{\orcidauthorRUCINSKI}{0009-0008-7061-8953}

\newcommand{\beq}{\begin{equation}}
\newcommand{\eeq}{\end{equation}}
\newcommand{\beqs}{\begin{eqnarray}}
\newcommand{\eeqs}{\end{eqnarray}}

\newcommand{\gsim}{\mathrel{\raisebox{-
.6ex}{$\stackrel{\textstyle>}{\sim}$}}}

\renewcommand{\L}{{\cal L}}

\newcommand{\M}{{\cal M}}
\newcommand{\A}{{\cal A}}

\def\hbar{\hspace{0pt}\raisebox{1pt}{$-$} \hspace{-7pt} h}

\def\di{\mbox{d}}
\def\r{\rho}

\newcommand{\be}{\begin{equation}}
\newcommand{\ee}{\end{equation}}

\newcommand{\bea}{\begin{eqnarray}}
\newcommand{\eea}{\end{eqnarray}}
\newcommand{\nn}{\nonumber}

\def\co{{\cal O}}

\def\lbldef#1#2{\expandafter\gdef\csname #1\endcsname {#2}}

\def\href#1#2{#2}

\newcommand{\ber}{\begin{eqnarray}}
\newcommand{\eer}{\end{eqnarray}}
\newcommand{\beqar}{\begin{eqnarray}}

\newcommand{\cO}{{\cal O}}
\newcommand{\cA}{{\cal A}}

\newcommand{\cC}{{\cal C}}
\newcommand{\cR}{{\cal R}}

\newcommand{\eeqar}{\end{eqnarray}}

\newcommand{\dsl}
  {\kern.06em\hbox{\raise.15ex\hbox{$/$}\kern-.56em\hbox{$\partial$}}}

\newcommand{\eeqarr}{\end{eqnarray}}
\newcommand{\ZZ}{{\rm \kern 0.275em Z \kern -0.92em Z}\;}

\def\CC{{\mathchoice
{\rm C\mkern-8mu\vrule height1.45ex depth-.05ex
width.05em\mkern9mu\kern-.05em}
{\rm C\mkern-8mu\vrule height1.45ex depth-.05ex
width.05em\mkern9mu\kern-.05em}
{\rm C\mkern-8mu\vrule height1ex depth-.07ex
width.035em\mkern9mu\kern-.035em}
{\rm C\mkern-8mu\vrule height.65ex depth-.1ex
width.025em\mkern8mu\kern-.025em}}}

\def\RR{{\rm I\kern-1.6pt {\rm R}}}

\def\ZZ{{\rm Z}\kern-3.8pt {\rm Z} \kern2pt}
\def\IB{\relax{\rm I\kern-.18em B}}
\def\ID{\relax{\rm I\kern-.18em D}}
\def\II{\relax{\rm I\kern-.18em I}}
\def\IP{\relax{\rm I\kern-.18em P}}

\newcommand{\bear}{\begin{eqnarray}}
\newcommand{\eear}{\end{eqnarray}}

\newcommand{\F}{{\cal F}}

\def\A6{\mathcal{A}_6}

\def\am6{\mathfrak{a}^{\A6}}
\def\fz{\mathfrak{z}}

\def\a{\alpha}
\def\b{\beta}
\def\c{\gamma}

\def\f{\phi}               %      \varphi

\def\l{\lambda}
\def\m{\mu}

\def\q{\theta}  \def\th{\theta}                  %     \vartheta
\def\r{\rho}
\def\vr{\varrho}   %     \varrho
\def\t{\tau}

\def\6{\partial}

\def\bea{\begin{eqnarray}}
\def\eea{\end{eqnarray}}

\def\beqx{\begin{displaymath}}
\def\eeqx{\end{displaymath}}

\newcommand{\bmat}{\left(\begin{array}}
\newcommand{\emat}{\end{array}\right)}

\def\a{\alpha}
\def\b{\beta}
\def\c{\chi}

\def\f{\phi}

\def\l{\lambda}
\def\m{\mu}

\def\p{\pi}
\def\q{\theta}
    \def\th{\theta}
\def\r{\rho}

\def\t{\tau}

\def\F{\Phi}

\def\L{\Lambda}

\def\P{\Pi}

\def\vr{\varrho}

\def\co{{\cal O}}

\def\fz{\mathfrak{z}}
\def\fa{\mathfrak{a}}

\def\bo{{\raise-.3ex\hbox{\large$\Box$}}}               % D'Alembertian
\def\face{{\raise.2ex\hbox{$\displaystyle \bigodot$}\mskip-2.2mu \llap {$\ddot
        \smile$}}}                                   % happy face
\def\>{\rangle}                                      %right angle
\def\<{\langle}                                      %left angle

\def\leftrightarrowfill{$\mathsurround=0pt \mathord\leftarrow \mkern-6mu
        \cleaders\hbox{$\mkern-2mu \mathord- \mkern-2mu$}\hfill
        \mkern-6mu \mathord\rightarrow$}        % <--> double differential
\def\dvec#1{\vbox{\ialign{##\crcr
        \leftrightarrowfill\crcr\noalign{\kern-1pt\nointerlineskip}
        $\hfil\displaystyle{#1}\hfil$\crcr}}}           % <--> accent
\def\-{\hphantom{-}}
\newcommand{\dd}{\mbox{d}}

\begin{document}

%\title{Stability analysis of holographic confinement with Abelian fluxes in seven dimensions}
\title{Light dilaton from top-down holographic confinement with magnetic fluxes}

\author{Maurizio Piai\,\orcidlink{\orcidauthorPIAI}}
\email{m.piai@swansea.ac.uk}
\affiliation{Department of Physics, Faculty  of Science and Engineering, Swansea University, Singleton Park, SA2 8PP, Swansea, Wales, United Kingdom}
\affiliation{Centre for Quantum Fields and Gravity, Faculty  of Science and Engineering, Swansea University, Singleton Park, SA2 8PP, Swansea, United Kingdom}

\author{James Rucinski\,\orcidlink{\orcidauthorRUCINSKI}}
\email{2315621@swansea.ac.uk}
\affiliation{Department of Physics, Faculty  of Science and Engineering, Swansea University, Singleton Park, SA2 8PP, Swansea, Wales, United Kingdom}
\affiliation{Centre for Quantum Fields and Gravity, Faculty  of Science and Engineering, Swansea University, Singleton Park, SA2 8PP, Swansea, United Kingdom}

\date{\today}

\begin{abstract}

A two-parameter class of higher-dimensional, strongly coupled, confining field theories in the presence of magnetic fluxes for two Abelian gauge groups admits a top-down, holographic dual description. The corresponding two-parameter family of regular background solutions of the classical equations of maximal supergravity in seven dimensions descends from maximal supergravity in eleven dimensions. We study the global and local stability properties of these solutions. We  identify lines of zero-temperature first-order phase transitions, describing a polygon (a square) in the space of parameters, identified with the two fluxes. The transition separates the family of gravity solutions dual to confining theories, inside the polygon, from those outside,  in which the field theory is realised in a conformal phase.  In the spectrum of fluctuations of the supergravity equations, interpreted as bound states of the dual, confining field theories,  we find no evidence of local instabilities (tachyons). Over a significant portion of parameter space, that extends far away from the proximity to the transition, we identify  an approximate dilaton, the mass of which is one order of magnitude smaller than the scale set by confinement. Our findings complement those emerging in other holographic models discussed in the literature, in which either the dilaton mass is only mildly lower than the confinement scale (when approaching a first-order transitions), or parametrically suppressed (when reaching the proximity to a second-order one).

\end{abstract}

\maketitle

\tableofcontents

%%%%%%%%%%%%%%%%%%%%%%%
%%%%%%%%%%%%%%%%%%%%%%%
\section{Introduction}
\label{Sec:introduction}

The advent of gauge-gravity dualities, or holography~\cite{Maldacena:1997re,Gubser:1998bc,Witten:1998qj,Aharony:1999ti}, has provided new tools to gain insight into non-perturbative properties of field theories. This is particularly effective in the strong-coupling regime of special field theories with a large number of degrees of freedom, the dual description of which consists of weakly coupled gravity theories in higher dimensions. Local operators of the field theory and their sources are in one-to-one correspondence with gravity fields and their asymptotic behavior. Non-local, gauge-invariant operators in the field theory correspond to geometric configurations of extended objects in the curved background of the gravity dual. The Wilson loop is a particularly clean example~\cite{Maldacena:1998im,Rey:1998ik}---see also Refs.~\cite{Brandhuber:1998bs,Brandhuber:1998er,Kinar:1998vq,Brandhuber:1999jr,Nunez:2009da}---as it discriminates between conformal and confining field theories. In the holographic treatment, confinement is realised geometrically, by the smooth shrinking of a sub-manifold in the gravity background~\cite{Witten:1998zw}---see also Refs.~\cite{Wen:2004qh,Kuperstein:2004yf,Brower:2000rp,Elander:2013jqa}.\footnote{In the simplest examples, the geometry ends in correspondence to the shrinking to zero size of a circle, which generates a physical scale and  mass gap in the dual theory. Other  examples in which such manifold is larger exist in the context of the conifold~\cite{Candelas:1989js}, to include the celebrated cases of Refs.~\cite{Chamseddine:1997nm,Klebanov:2000hb,Maldacena:2000yy,Butti:2004pk}---see also Refs.~\cite{Klebanov:1998hh,Klebanov:2000nc,Papadopoulos:2000gj,Dymarsky:2005xt,Andrews:2006aw,Hoyos-Badajoz:2008znk,Nunez:2008wi,Elander:2009pk,Cassani:2010na,Bena:2010pr,Bennett:2011va,Dymarsky:2011ve,Maldacena:2009mw,Gaillard:2010qg,Caceres:2011zn,Elander:2011mh,Elander:2012yh,Elander:2017hyr,Elander:2017cle}.}

In  strongly coupled field theories that admit a gravity dual description, the free energy in the presence of external sources, and the associated response functions (condensates), can be extracted from holographic renormalisation~\cite{Bianchi:2001kw,Skenderis:2002wp,Papadimitriou:2004ap}. This is hence the ideal environment for the study of phase transitions, both at finite and zero temperature.  As programmatically outlined in Ref.~\cite{Elander:2020ial}, we can use this technology to investigate the relation between the nature of phase transitions in confining gauge theories and the spectrum of  their bound states,  computed in proximity of the transition. This information is obtained from the study of small fluctuations of the gravity backgrounds, for which we deploy the gauge-invariant formalism developed in Refs.~\cite{Bianchi:2003ug,Berg:2005pd,Berg:2006xy,Elander:2009bm, Elander:2010wd,Elander:2010wn, Elander:2014ola,Elander:2018aub,Elander:2020csd}. Our findings, as we shall see, fill an important, unexpected, gap in the characterisation emerging from this extensive research programme---see also Refs.~\cite{Elander:2020fmv,Elander:2021wkc,Elander:2022ebt,Faedo:2024zib,Fatemiabhari:2024lct,Elander:2025fpk}.\footnote{Related work exists using Monte Carlo numerical techniques in lattice field theory~\cite{Lucini:2013wsa,Bennett:2022yfa} as well as exploiting semi-analytical techniques valid in special  super-renormalisable lower-dimensional field theories~\cite{Cresswell-Hogg:2025kvr}.}

The main motivation of this research programme is to provide the foundations for the physics of the dilaton, the Pseudo-Nambu-Goldstone Boson (PNGB) associated with scale invariance~\cite{Coleman:1985rnk}.\footnote{
A composite dilaton is expected to  exist in special strongly coupled field theories~\cite{Migdal:1982jp,Leung:1985sn,Bardeen:1985sm,Yamawaki:1985zg,Holdom:1984sk}, although there is no clear consensus on such a scenario~\cite{Holdom:1986ub,Holdom:1987yu,Appelquist:2010gy,Grinstein:2011dq}. The emerging low-energy theory, in which the dilaton is coupled to ordinary PNGBs, is called dilaton effective field theory (dEFT),  is discussed in Refs.~\cite{Matsuzaki:2013eva,Golterman:2016lsd,Kasai:2016ifi,Hansen:2016fri,Golterman:2016cdd,Appelquist:2017wcg,Appelquist:2017vyy,Cata:2018wzl,Golterman:2018mfm,Cata:2019edh,Appelquist:2019lgk,Golterman:2020tdq,Golterman:2020utm, Appelquist:2022mjb,Appelquist:2025tol,Cao:2026htw}, and has found additional novel applications, as proposed for example in Refs.~\cite{Appelquist:2020bqj,Appelquist:2022qgl,Cacciapaglia:2023kat,Appelquist:2024koa}.} Its possible emergence in strongly coupled field theories has distinctive implications~\cite{Goldberger:2007zk}, of interest both on phenomenological and theoretical grounds~\cite{Hong:2004td,Dietrich:2005jn,Vecchi:2010gj,Hashimoto:2010nw,DelDebbio:2021xwu,Zwicky:2023fay,Zwicky:2023krx}, especially in view of the Large Hadron Collider (LHC) programme~\cite{Eichten:2012qb,Elander:2012fk,Chacko:2012sy,Bellazzini:2012vz,Abe:2012eu,Bellazzini:2013fga,Hernandez-Leon:2017kea}.\footnote{
The literature on the holographic techni-dilaton is exemplified by Refs.~\cite{Goldberger:1999uk,DeWolfe:1999cp,
 Goldberger:1999un,Csaki:2000zn,Arkani-Hamed:2000ijo,Rattazzi:2000hs,Kofman:2004tk,
 Elander:2011aa,
 Kutasov:2012uq,Evans:2013vca,
 Hoyos:2013gma,Megias:2014iwa,Elander:2015asa,Megias:2015qqh,Athenodorou:2016ndx,
Elander:2018gte,Pomarol:2019aae,CruzRojas:2023jhw,Pomarol:2023xcc,CruzRojas:2025qcj}
(besides the aforementioned Refs.~\cite{Elander:2013jqa,Elander:2018aub,Elander:2009pk,Elander:2012yh,Elander:2017hyr,Elander:2017cle}).}
A particular mechanism leading to the emergence of the dilaton  is linked  to the appearance of local, tachyonic instabilities in some region of  parameter space of the theory. For example, this might happen in association with the existence of complex fixed points in the renormalisation group equations~\cite{Kaplan:2009kr}, characterising theories that violate the Breitenlohner--Freedman (BF) unitarity bound~\cite{Breitenlohner:1982jf}, that underlie the study of non-unitary conformal field theories~\cite{Gorbenko:2018ncu,Gorbenko:2018dtm}.\footnote{Realisations of such scenarios within string theory can be found  in Refs.~\cite{Jensen:2010ga,Faedo:2019nxw}.}

Generalising these ideas, a dilaton might arise in strongly coupled field theories for which the holographic description displays a classical instability that is resolved by the presence of a phase transition. 
A first-order phase transition is characterised by the phenomena of phase coexistence, as stable and metastable solutions appear for the same choices of control parameters, but different response functions. A generic expectation is that additional,  unstable solutions appear as well, for intermediate values of the response functions near the transition. The instability is expected to be signalled by at least one tachyonic state in the dual field theory. Assuming continuity it is therefore expected that the mass of this tachyonic state cross zero somewhere in the approach to the instability, possibly along a metastable branch of solutions. The presence of the phase transition prevents the tachyonic state from being realised physically, yet does allow the possibility of a light state to arise  in its proximity, and it may be associated with the dilaton. This suggestion has been tested both in the context of bottom-up holographic models~\cite{Elander:2022ebt,Fatemiabhari:2024lct,Faedo:2024zib}---see also precursors in Refs.~\cite{Bea:2020ees,Ares:2020lbt,Bea:2021ieq,Bea:2021zsu,Bea:2021zol,Bea:2022mfb,Escriva:2022yaf}---as well as in the rigorous context of top-down constructions derived from complete supergravity theories~\cite{Elander:2020ial,Elander:2020fmv,Elander:2021wkc,Elander:2025fpk}.
 Finding viable realisations of top-down holographic models is challenging, due to the constrained nature of supergravity. 
Yet, fertile ground for the search of relevant supergravity background solutions has been identified in many examples.  The interesting theories include 
Romans's half-maximal supergravity in $D=6$ dimensions~\cite{Romans:1985tw,
DeWitt:1981wm,Giani:1984dw,Romans:1985tz}---see also Refs.~\cite{Ferrara:1998gv,Cvetic:1999un,Brandhuber:1999np,DAuria:2000afl,Nishimura:2000wj,
Andrianopoli:2001rs,Nunez:2001pt,Gursoy:2002tx} and Ref.~\cite{Elander:2013jqa}---the maximal supergravity in
 $D=7$ space-time dimensions~\cite{Pilch:1984xy,Pernici:1984xx,Pernici:1984zw}---see also Refs.~\cite{Nastase:1999cb,Cvetic:1999xp,Lu:1999bc,Cvetic:2000ah,Campos:2000yu,Samtleben:2005bp}---the ${\cal N}=8$ maximal supergravity in $D=5$ dimensions~\cite{Pernici:1985ju,Gunaydin:1984qu,
 Gunaydin:1985cu}---see also Refs.~\cite{Cvetic:2000nc,Pilch:2000ue,Bakas:1999ax} and~\cite{Anabalon:2024che}---and the lower dimensional truncation of maximal  supergravity in $D=11$ dimensions discussed in Refs.~\cite{Faedo:2017fbv,Elander:2020rgv}.
In these explicit examples,  the arising of phase transitions ensures that the tachyonic part of parameter space cannot be physically realised as vacuum of the theory.

These known examples display two possible ways in which the nature of the phase transitions is linked to the dilaton. In the presence of first-order phase transitions, the dilaton has mass just mildly lower than the confinement scale along the stable branches of vacua, while being suppressed along some metastable branches~\cite{Elander:2020ial,Elander:2020fmv,Elander:2021wkc,Elander:2022ebt,Fatemiabhari:2024lct}. Conversely, in the proximity to a second-order phase transition, the dilaton displays a parametrically suppressed mass, approaching zero at the transition~\cite{Faedo:2024zib,Elander:2025fpk}. The common feature to  the two classes is that it appears to be the order of the  phase transition to drive the properties of the dilaton, rather than the precise nature of the instabilities associated with it. As we shall see, our present results  further demonstrate this trend.  While there is evidence of a first-order phase transition in the theories we analyse here,  we do not find direct signals of local instabilities in the spectrum of bound states.

We focus our attention on classes of background solutions in gravity that admit a dual interpretation as confining field theories in the presence of a magnetic flux. The way they are built  is a generalisation of the Melvin flux-tube solutions~\cite{Melvin:1963qx}, that have been studied in Refs.~\cite{Astorino:2012zm,Lim:2018vbq,Kastor:2020wsm}. 
The gravity solutions are related to the literature on holographic treatment of field theories at finite temperature and chemical potential (see Ref.~\cite{Casalderrey-Solana:2011dxg} for a pedagogical introduction).  Soliton solutions with black-brane asymptotics~\cite{Horowitz:1991cd,Gubser:1996de}, in which there is a finite Hawking temperature~\cite{Gibbons:1976ue} for the black brane, are  related by double Wick rotation to the gravity dual of confining field theories~\cite{Witten:1998zw}. In a similar sense, soliton solutions with charged black-brane asymptotics, which have an interpretation as field theories at  finite temperature and finite chemical potential~\cite{Chamblin:1999tk,Chamblin:1999hg,Gubser:1998jb,Cai:1998ji,Cvetic:1999ne,Cvetic:1999rb,Kim:2006gp,Horigome:2006xu,Kobayashi:2006sb,Mateos:2007vc,Nakamura:2006xk, Karch:2007pd}, after double Wick rotation are dual to confining field theories with magnetic fluxes.
 Recent applications of this construction include for example Refs.~\cite{Anabalon:2021tua,Anabalon:2025sok}, that identify a phase transition in the parameter space of the theories, and Refs.~\cite{Nunez:2023xgl, Nunez:2023nnl,Fatemiabhari:2024aua,Chatzis:2024top,Chatzis:2024kdu,Chatzis:2025dnu,Chatzis:2025hek}.
 
We work in the context of the  maximal gauged supergravity in  $D=7$ space-time dimensions~\cite{Pilch:1984xy,Pernici:1984xx,Pernici:1984zw} (see also Refs.~\cite{Nastase:1999cb,Cvetic:1999xp,Lu:1999bc,Cvetic:2000ah,Campos:2000yu,Samtleben:2005bp}), truncated to retain only  the Cartan subgroup of the gauged symmetry, $SO(2)\times SO(2)\subset SO(5)$, together with two real scalars coupled to gravity~\cite{Liu:1999ai,Cvetic:1999ne,Wu:2011gp}. The dual field theory is a special deformation of the ${\cal N}=(2,0)$ superconformal field theory in $D-1=6$ dimensions that lives on a stack of $M5$-branes~\cite{Witten:1995zh,Strominger:1995ac,Witten:1995em}. Furthermore, one of the dimensions is a circle, so that under special conditions its field theory dual confines~\cite{Lambert:2010iw}.  Our analysis is based on the $SO(2)\times SO(2)$ truncation presented in Ref.~\cite{Liu:1999ai}. We find it convenient to construct the soliton solutions of interest by compactifying one dimension on a circle and reducing to $D=6$ dimensions first. We then use this language to compute the free energy and the spectrum of bound states, using the aforementioned holographic techniques. Interesting studies of the finite temperature and finite chemical potential  thermodynamics of related solutions are found also in Refs.~\cite{Cvetic:1999xp,Bobev:2023bxl,Chong:2004dy,Chow:2011fh}.

We identify two different branches of 2-parameter classes of physically inequivalent solutions. We call them, respective, domain-wall and soliton solutions, and holographically interpret the former in terms of conformal field theories, the latter of confining theories with magnetic fluxes. The global analysis based on  the (holographic) calculation of the free energy demonstrates the presence of a closed line  in parameter space, describing a polygon, demarcating the phase transitions between conformal (outer) and confining (inner) portions of space. By analysing the properties of the background solutions near such line, we conclude that 
along the edges of the polygon the transition is of the first order, as in Refs.~\cite{Elander:2020ial,Elander:2020fmv,Elander:2021wkc,Elander:2022ebt,Fatemiabhari:2024lct}.\footnote{At the end point along the branch of solutions the geometry has a singularity at the end of space, while all other solutions along the branch are completely smooth and regular.}
As anticipated, the spectrum of fluctuations of the soliton (confining) solutions does not exhibits pathological behaviour (tachyons) anywhere in parameter space. In the spin-0 spectrum, the lightest state has suppressed mass. This suppression is a tangible effect, that can be as large as an order of magnitude in respect to the mass of the lightest spin-2 state, which we use as  an estimator of the confinement scale. Furthermore, we collect evidence that this state has distinctive properties that allow us to identify it as a dilaton, and that its properties persist along a large portion of parameter space, reaching away from the transition.

The paper is organised as follows.
We present the supergravity formalism in Sect.~\ref{Sec:gravity}. We start from the bosonic part of the action of maximal supergravity in $D=7$ dimensions, recall the properties of its $SO(2)\times SO(2)$ truncation, and dimensionally reduce on a circle, in Sect.~\ref{Sec:dimensional reduction}, to the action in $D=6$ dimensions used in the body of the paper. In Sect.~\ref{Sec:backgrounds} we present the solutions of interest, starting from the classical equations of motion and their properties,  reporting in details the general ultraviolet  (UV) expansions of the solutions, and present the special closed-form functional dependence of the background functions.
We compute the free energy  via holographic renormalisation, in Sect.~\ref{Sec:FreeEnergy}, and the spectrum of bound states in the dual field theory in Sect.~\ref{Sec:MassSpectra}, by applying the gauge-invariant formalism to the system of scalars coupled to gravity in $D=6$ dimensions. We also repeat the calculation of the spectrum in the
probe approximation, that we use as a diagnostic tool to identify the dilaton, in Sect.~\ref{Sec:probe}.
We summarise our findings and outline suggestions for future research in Sect.~\ref{Sec:outlook}.
We complement the main body of the  paper by an Appendix, in which we report additional material that is not essential in order to understand out main result, but that can be useful to reproduce our calculations.

%%%%%%%%%%%%%%%%%%%%%%%
%%%%%%%%%%%%%%%%%%%%%%%
\section{The holographic gravity theory}
\label{Sec:gravity}

In this section we provide a brief review of the field content of maximal supergravity in $D=7$ dimensions, along with its $SO(2)\times SO(2)$-invariant  truncation of interest,\footnote{This is not in general a consistent truncation, but it is for solutions in 
which the product $F^{(1)}\wedge F^{(2)}=0$, so that truncated fields are not excited---see discussion in Sect.~4.1 of Ref.~\cite{Cvetic:1999ne}.}  and the  dimensional reduction on a circle, down to $D=6$ dimensions. We write explicitly the bulk action used later in the paper. In the process, we fix our notation and conventions, and highlight a few technical subtleties.

Supergravity theories in $D=7$ spacetime dimensions can be obtained from the compactification (and reduction) of maximal supergravity in $D=11$ dimensions on an internal 4-dimensional space~\cite{Pilch:1984xy,Pernici:1984xx,Pernici:1984zw} (see also Refs.~\cite{Nastase:1999cb,Cvetic:1999xp,Lu:1999bc,Cvetic:2000ah,Campos:2000yu,Samtleben:2005bp}). While compactification  on  locally flat internal manifolds, (e.g., a 4-torus, $T^4$), leads to ungauged supergravity, replacing the internal space with a 4-sphere, $S^4$, yields the maximal gauged  supergravity in $D=7$ dimensions of interest in this paper. Its gauge group, $SO(5)$, corresponds to the isometries of the internal space.

Upon Kaluza-Klein reduction on $S^4$, the $128$ degrees of freedom of the supergravity in $D=11$  dimensions ($44$ from the graviton and $84$ from the 3-form)  are reorganised into the following field content in $D=7$ dimensions: the metric, providing 14 degrees of freedom, 14 real scalars, 10 massless vector fields, each propagating 5 physical degrees of freedom, and 5 self-dual 3-forms, each providing 10 degrees of freedom. The 128 degrees of freedom match the fermionic field content, but we will only concern ourselves with the bosonic sector. Following the notation of Ref.~\cite{Elander:2021kxk}, but omitting the 3-forms, as they play no part in the following,  the bosonic portion of the action  is the following.
\beqs
{\cal S}_7&=&\int \di^7 x \sqrt{-\hat{g}_7}\left[\frac{{\cal R}_7}{4}
-\frac{1}{4}\hat{g}^{\hat{M}\hat{N}} \delta^{ik}\delta_{j\ell} 
P_{\hat{M}\,i}^{\,\,\,\,\,\,\,\,\,\,j} P_{\hat{N}k}^{\,\,\,\,\,\,\,\,\,\ell}
-\frac{m^2}{8}\left(2\delta^{i}_{\,\,\,k}\delta^{j}_{\,\,\,\ell} -\delta^{ij}\delta_{k\ell} \right)T_{ij}T^{k\ell}
+\frac{1}{8}\left(\Pi_{\alpha}^{\,\,\,i}\Pi_{\beta}^{\,\,\,j}{\cal F}^{\alpha\beta}_{\hat{M}\hat{N}}\right)^2\right]
\,,
\eeqs
where the matrices appearing in the action are defined as
\beqs
P_{\hat{M}\,i}^{\,\,\,\,\,\,\,\,\,\,j}&\equiv&\frac{1}{2}\left((\Pi^{-1})_i^{\,\,\,\a}{\cal D}_{\hat{M}\,\a}^{\,\,\,\,\,\,\,\,\,\,\b}\Pi_{\b}^{\,\,\,\,j}
+(i\leftrightarrow j)\right)\,,\\
{\cal D}_{\hat{M}\,\a}^{\,\,\,\,\,\,\,\,\,\,\b}&\equiv& \delta_{\a}^{\,\,\,\b}\partial_{\hat{M}}
+i g {\cal A}_{\hat{M}\,\a}^{\,\,\,\,\,\,\,\,\,\,\,\b}\,,\\
T_{ij}&\equiv& (\Pi^{-1})_{i}^{\,\,\,\a} (\Pi^{-1})_{j}^{\,\,\,\b}\delta_{\a\b}\,,\\
{\cal F}_{\hat{M}\hat{N}\,\a}^{\,\,\,\,\,\,\,\,\,\,\,\,\,\,\,\,\b}&\equiv&
2\left(\partial_{[\hat{M}}{\cal A}_{\hat{N}]\,\a}^{\,\,\,\,\,\,\,\,\,\,\,\,\,\b}\,+\,i g {\cal A}_{[\hat{M}\,\a}^{\,\,\,\,\,\,\,\,\,\,\,\,\gamma}
{\cal A}_{\hat{N}]\gamma}^{\,\,\,\,\,\,\,\,\,\,\b}\right)\,.
\eeqs
Space-time  indexes in $D=7$ dimensions are denoted by hatted Roman letters, $\hat{M}=0,\, 1,\, 2,\, 3,\, 4,\, 6,\, 7$, while we use  Greek indices, $\alpha=1\,,\cdots\,,5$, for the fundamental representation of the gauge group $SO(5)$, with gauge fields ${\cal A}_{\hat{M}\,\a}^{\,\,\,\,\,\,\,\,\,\,\,\b}$. The scalars are parameterised by the unit-determinant matrix, $\P^i_\a$,  with $i,j=1,\,\cdots\,,5$,  in the $SL(5,\mathbb{R})/SO(5)_c$ right coset, where $SL(5,\mathbb{R})$ is the global symmetry of the system and $SO(5)_c$ the local symmetry that identifies equivalent configurations of the scalar fields.

Following Refs.~\cite{Liu:1999ai, Cvetic:1999ne, Wu:2011gp,Chow:2011fh},  we truncate the theory to preserve the Cartan subgroup, $SO(2)\times SO(2)$ of the rank-2 gauge group $SO(5)$. We choose the two commuting generators, $T_1\,,\, T_2$, to be represented by the following matrixes:
\beqs
T_1 = \begin{pmatrix}
0 & -1 & 0 & 0 & 0 \\
1 & 0 & 0 & 0 & 0 \\
0 & 0 & 0 & 0 & 0 \\
0 & 0 & 0 & 0 & 0 \\
0 & 0 & 0 & 0 & 0
\end{pmatrix}, \quad
T_2 = \begin{pmatrix}
0 & 0 & 0 & 0 & 0 \\
0 & 0 & 0 & 0 & 0 \\
0 & 0 & 0 & -1 & 0 \\
0 & 0 & 1 & 0 & 0 \\
0 & 0 & 0 & 0 & 0
\end{pmatrix}\,,
\eeqs
which generate rotations in the $\{1, 2\}$ and $\{3, 4\}$ planes, respectively. In the truncated theory we retain only gauge fields along the Cartan subgroup.

Out of the $14$  that span the 2-index symmetric, traceless representation of $SO(5)$, we retain the two real scalars, $\phi_1$ and $\phi_2$,  that in the presence of generic vacuum expectation values presence the $SO(2)\times SO(2)$
subgroup defined by $T_1$ and $T_2$.  The specific form of scalar matrix $\P^i_\a$ is hence the exponential of a traceless matrix:
\begin{equation}
    \P^i_\a=\text{diag}\left(e^{\frac{\f_1}{2\sqrt{2}}+\frac{\f_2}{2\sqrt{10}}},e^{\frac{\f_1}{2\sqrt{2}}+\frac{\f_2}{2\sqrt{10}}},e^{-\frac{\f_1}{2\sqrt{2}}+\frac{\f_2}{2\sqrt{10}}},e^{-\frac{\f_1}{2\sqrt{2}}+\frac{\f_2}{2\sqrt{10}}}, e^{-\frac{2\f_2}{\sqrt{10}}}\right)\,.
\end{equation} 
This parametrisation is such that for
 general values of $\phi_1\neq 0$ with $\phi_2=0$, the $SO(5)$ is broken to its maximal $SO(4)$ subgroup, and for general $\phi_1\neq 0$ and $\phi_2 \neq 0$ to its $SO(2)\times SO(2)$
subgroup. In the latter case, the 8 gauge bosons in the coset, $SO(5)/(SO(2)\times SO(2))$, are Higgsed and become massive, by combining with $8$ of the scalars. Of the remaining $14-8=6$ scalars, while we retain the two $SO(2)\times SO(2)$ singlets, we truncate away $4$ degrees of freedom, corresponding to two complex scalars, each of which is charged under one of the Abelian gauge groups.

We identify the two  vector fields we retain  with $\cA_{\hat{M}}^{(1)}\equiv {\cal A}_{\hat{M}\,1}^{\,\,\,\,\,\,\,\,\,\,\,2}$ and $\cA_{\hat{M}}^{(2)}\equiv {\cal A}_{\hat{M}\,3}^{\,\,\,\,\,\,\,\,\,\,\,4}$, respectively. The covariant derivatives reduce to standard derivatives and the action greatly simplifies~\cite{Liu:1999ai, Cvetic:1999ne, Wu:2011gp,Chow:2011fh}:\footnote{This is equivalent to defining $\l_1=-\frac{\f_1}{2\sqrt{2}}-\frac{\f_2}{2\sqrt{10}}$ and $\l_2=\frac{\f_1}{2\sqrt{2}}-\frac{\f_2}{2\sqrt{10}}$, for $\kappa^2=2$ and $m^2=1$, in  Eqs.~(2.5) and~(2.6) of Ref.~\cite{Liu:1999ai}. }
\begin{equation} 
\label{eq:7Daction}
\begin{split}
    \mathcal{S}_7=\frac{1}{2\pi}\int \dd ^7x \sqrt{-\hat{g}_7}\Bigg\{&
    \frac{\mathcal{R}_7}{4}-\frac{1}{8}g^{\hat{M}\hat{N}}(\6_{\hat{M}}\f_1 \6_{\hat{N}}\f_1 +\6_{\hat{M}}\f_2 \6_{\hat{N}}\f_2)- \\
    &\frac{1}{16}g^{\hat{M}\hat{R}}g^{\hat{N}\hat{S}}(e^{\sqrt{2}\f_1+\sqrt{\frac{2}{5}}\f_2} \mathcal{F}_{\hat{M}\hat{N}}^{(1)} \mathcal{F}_{\hat{R}\hat{S}}^{(1)}+e^{-\sqrt{2}\f_1+\sqrt{\frac{2}{5}}\f_2} \mathcal{F}_{\hat{M}\hat{N}}^{(2)} \mathcal{F}_{\hat{R}\hat{S}}^{(2)})-\mathcal{V}_7(\f)\Bigg\},
\end{split}
\end{equation}
where the Abelian field strengths, for $i=1,\,2$, are:
\begin{equation}
    \mathcal{F}_{\hat{M} \hat{N}}^{(i)}=\6_{\hat{M}}\mathcal{A}_{\hat{N}}^{(i)}-\6_{\hat{N}}\mathcal{A}_{\hat{M}}^{(i)}.
\end{equation}
The potential is~\cite{Liu:1999ai}: 
\beqs
\label{Eq:pot7}
    \mathcal{V}_7(\f)
    &=&\frac{1}{8}e^{-\sqrt{\frac{2}{5}}\f_2}\left(e^{\sqrt{10}\f_2}-8e^{\sqrt{\frac{5}{2}}\f_2}\cosh\left(\frac{\f_1}{\sqrt{2}}\right)-8\right)\,,
\eeqs
and it  has a maximum at $\f_1=\f_2=0$ as well as a saddle point at $\f_1=0, \f_2=\sqrt{\frac{2}{5}}\log(2)$.
The potential and the whole action are symmetric for $\phi_1\rightarrow -\phi_1$, provided one also exchanges 
$\mathcal{A}_{\hat{M}}^{(1)}\leftrightarrow \mathcal{A}_{\hat{M}}^{(2)}$. 
Furthermore, $ \mathcal{S}_7$ is also invariant under the separate changes of sign of the two Abelian gauge fields,
$\mathcal{A}_{\hat{M}}^{(1)}\rightarrow -\mathcal{A}_{\hat{M}}^{(1)}$ and $\mathcal{A}_{\hat{M}}^{(2)}\rightarrow -\mathcal{A}_{\hat{M}}^{(2)}$.

%%%
%%%
%%%

%%%%%
\subsection{Dimensional reduction to six dimensions}
\label{Sec:dimensional reduction}

We reduce the theory to six dimensions by assuming one of the space-like dimensions is compactified on a circle, $S^1$, that is parameterised by an angular coordinate, $\eta \in [0, 2 \pi]$,  that all background functions are independent of this angular coordinate, $\eta$, and that any $\eta$-dependent properties can be set to zero---this includes the fact that fluctuations with momentum along the circle are also ignored. Solutions for which the compact dimension smoothly shrinks to zero, at some radial position,  provide a geometric realisation of the notion of confinement in the dual theory. The metric ansatz takes the soliton form:
\begin{equation}\label{eq:metric}
    \dd s^2_7 =e^{\frac{-2}{\sqrt{10}}\c}\dd s_6^2 +e^{\frac{8}{\sqrt{10}}\c}\left(\dd \eta +V_M \dd x^M\right)^2\,,
\end{equation}
where un-hatted Latin indexes run over six dimensions, $M=0,\, 1,\, 2,\, 3,\, 4,\, 6$, with $M=6$ the holographic direction, while the  metric in $D=6$ dimensions can be cast  in domain-wall form, as
\begin{equation}
\begin{split}
     \dd s_6^2&=e^{2A}\dd x_{1, 4}^2+\dd r^2 \\
     \label{Eq:metric6}
    &=e^{2A}\dd x_{1,4}^2+e^{\frac{2}{\sqrt{10}}\c}\dd \r^2\,.
\end{split}
\end{equation}
 The conventions in these equations are chosen so that the AdS$_7$ backgrounds have ${\cal  A}=A-\chi/\sqrt{10}=\rho/2=4\chi/\sqrt{10}$.

Under these assumption, the action,  ${\cal S}_7$, can be written as a lower-dimensional action, ${\cal S}_6$,
 along with an additive total derivative, by performing the integration over the circle:
\beqs
{\cal S}_7&=&{\cal S}_6 + \frac{1}{2}\int \di^6 x \partial_M \left(\sqrt{-g_6}g^{MN}\partial_N\left(\frac{\chi}{\sqrt{10}}\right)\right)\,.
\eeqs
The action of the six dimensional theory is given by\footnote{Notice the different choice of normalisation
between Eqs.~(\ref{eq:7Daction}) and~(\ref{eq:6Daction}), that arises because of the integration over the circle, $\frac{1}{2\pi}\int \di\eta =1$.}
\begin{equation}\label{eq:6Daction}
    \mathcal{S}_6=\int \dd^6x \sqrt{-g_6}\left\{ \frac{\mathcal{R}_6}{4}-\frac{1}{2}g^{MN}G_{ab}\6_M\F^a \6_N \F^b -\frac{1}{4}H_{ab}g^{MR}g^{NS}F_{MN}^aF_{RS}^b -\mathcal{V}(\Phi^a)\right\},
\end{equation}
where the potential is $\mathcal{V}(\Phi^a)=e^{\frac{-2}{\sqrt{10}}\c}\mathcal{V}_7(\f)$.  
The sigma-model metric contains five scalars: besides $\phi_1$ and $\phi_2$, it also includes $\chi$ and the 
two components of the Abelian fields along the circle, $\cA_7^{(1)}$ and $\cA_7^{(2)}$.
In the basis
 $\F^a = \{\f_1,\f_2, \c, \cA_7^{(1)},\cA_7^{(2)}\}$, the sigma-model metric is
\begin{equation}
    G_{ab}={\rm diag}\Bigg(\frac{1}{4}, \frac{1}{4}, 1, \frac{1}{4}e^{\sqrt{2}\f_1+\sqrt{\frac{2}{5}}\f_2-\frac{8\c}{\sqrt{10}}},\frac{1}{4}e^{-\sqrt{2}\f_1+\sqrt{\frac{2}{5}}\f_2-\frac{8\c}{\sqrt{10}}}\Bigg).
\end{equation}
The  vectors, $V^a_M=\{V_M, A_M^{(1)},A_M^{(2)}\}$,  enter the action with the following field-strength tensors, 
\beqs
F^{V}_{MN}&=&\6_M V_N-\6_N V_M \, , \\
F^{\mathcal{A}^{(i)}}_{MN}&=&\6_M\cA^{(i)}_N-\6_N\cA^{(i)}_M + V_M\6_N\cA_7^{(i)} -V_N\6_M\cA_7^{(i)} \,, 
\eeqs
along with their associated sigma-model metric:
\begin{equation}
    H_{ab}=\frac{1}{4} {\rm  diag} \Bigg(e^{\sqrt{10}\c}, e^{\sqrt{2}\f_1+\sqrt{\frac{2}{5}}\f_2+\frac{2}{\sqrt{10}}\c}, e^{-\sqrt{2}\f_1+\sqrt{\frac{2}{5}}\f_2+\frac{2}{\sqrt{10}}\c} \Bigg).
\end{equation}

%%%%%%%%%%%%%%%%%%%%%%%
%%%%%%%%%%%%%%%%%%%%%%%
\subsection{Background solutions} 
\label{Sec:backgrounds}

We allow the background scalars appearing in the six dimensional action, Eq.~(\ref{eq:6Daction}), to develop a non-vanishing bulk profile, as a function of the radial coordinate, $\r$, and back-react their effect on the geometry, so that the metric, which takes the form in Eq.~(\ref{Eq:metric6}),  has non-trivial $A(\rho)$. The vector fields vanish in the background, hence we can use the formalism described in Appendix~\ref{Sec:sigma-model}. We seek background solutions that, after uplifting to the seven dimensional space, have finite curvature everywhere, and are free of conical singularities. We also require that the uplifted 
solution have asymptotically AdS$_7$ geometry,  so that all background functions 
approach the expansion of the backgrounds close to the solution with $\phi_1=0=\phi_2$, when $\rho\rightarrow +\infty$.

\begin{figure*}[t]
    %\centering
    \includegraphics[width=0.4\linewidth]{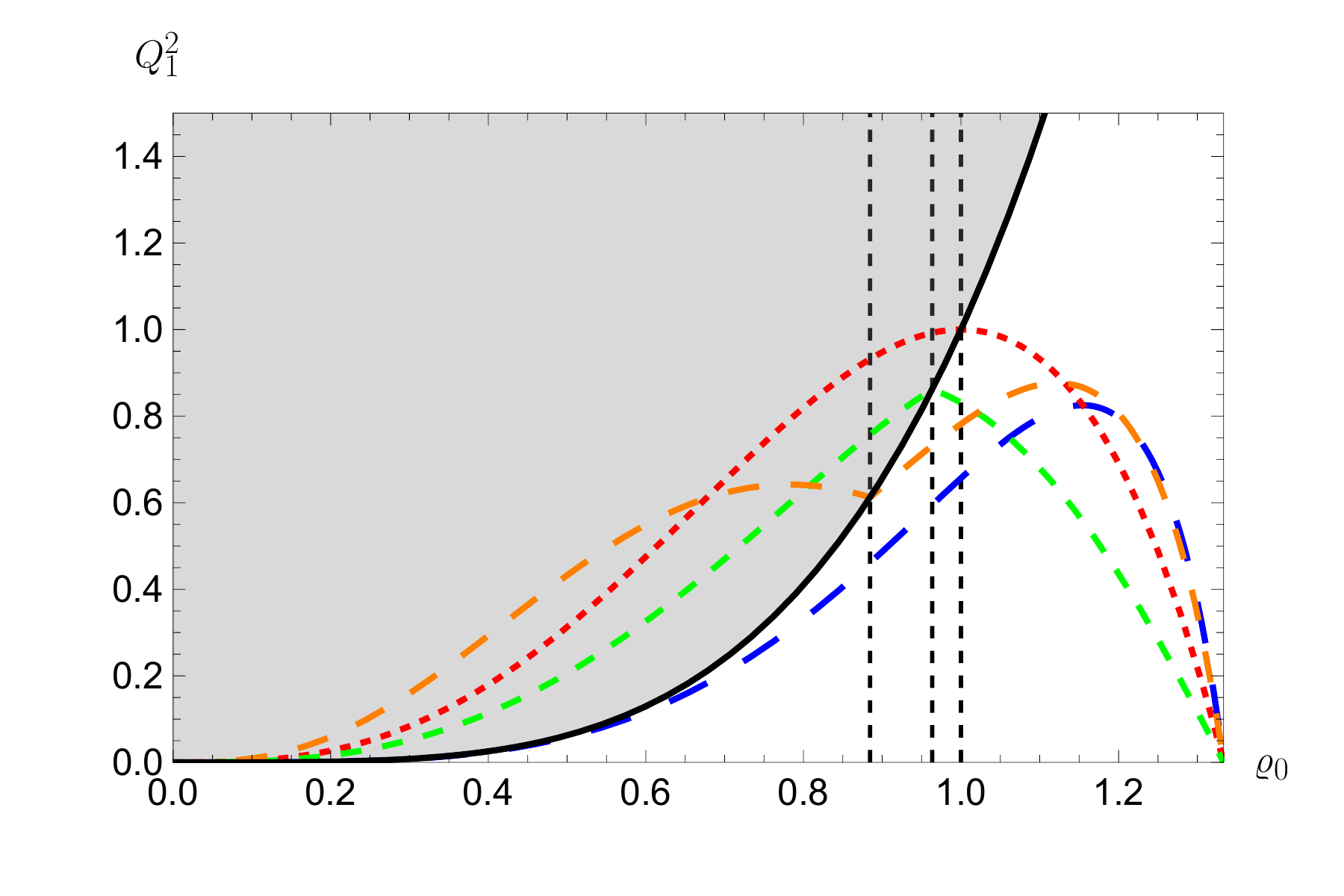}
    \includegraphics[width=0.4\linewidth]{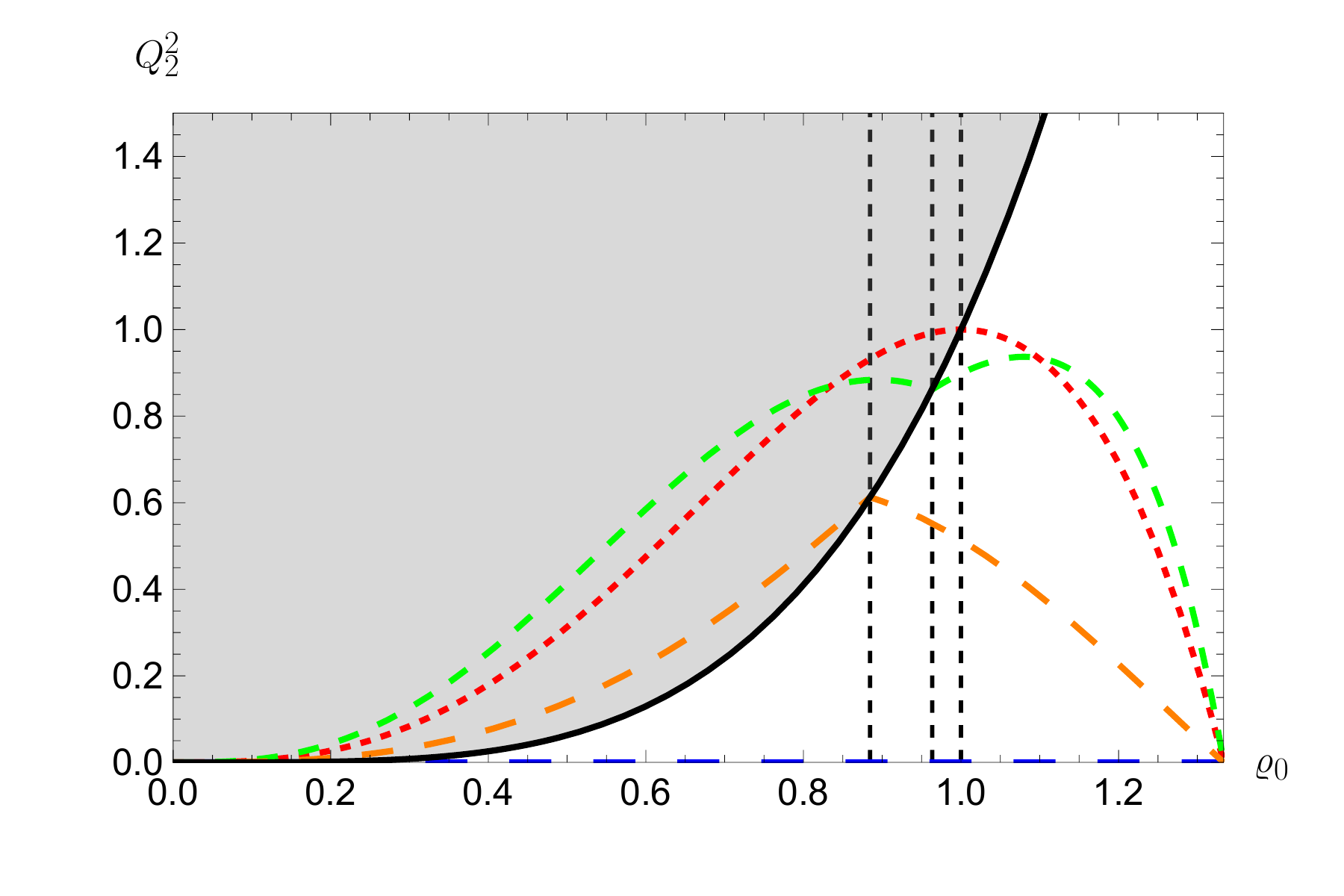}\\
    \vspace{-2.9ex}
    \includegraphics[width=0.4\linewidth]{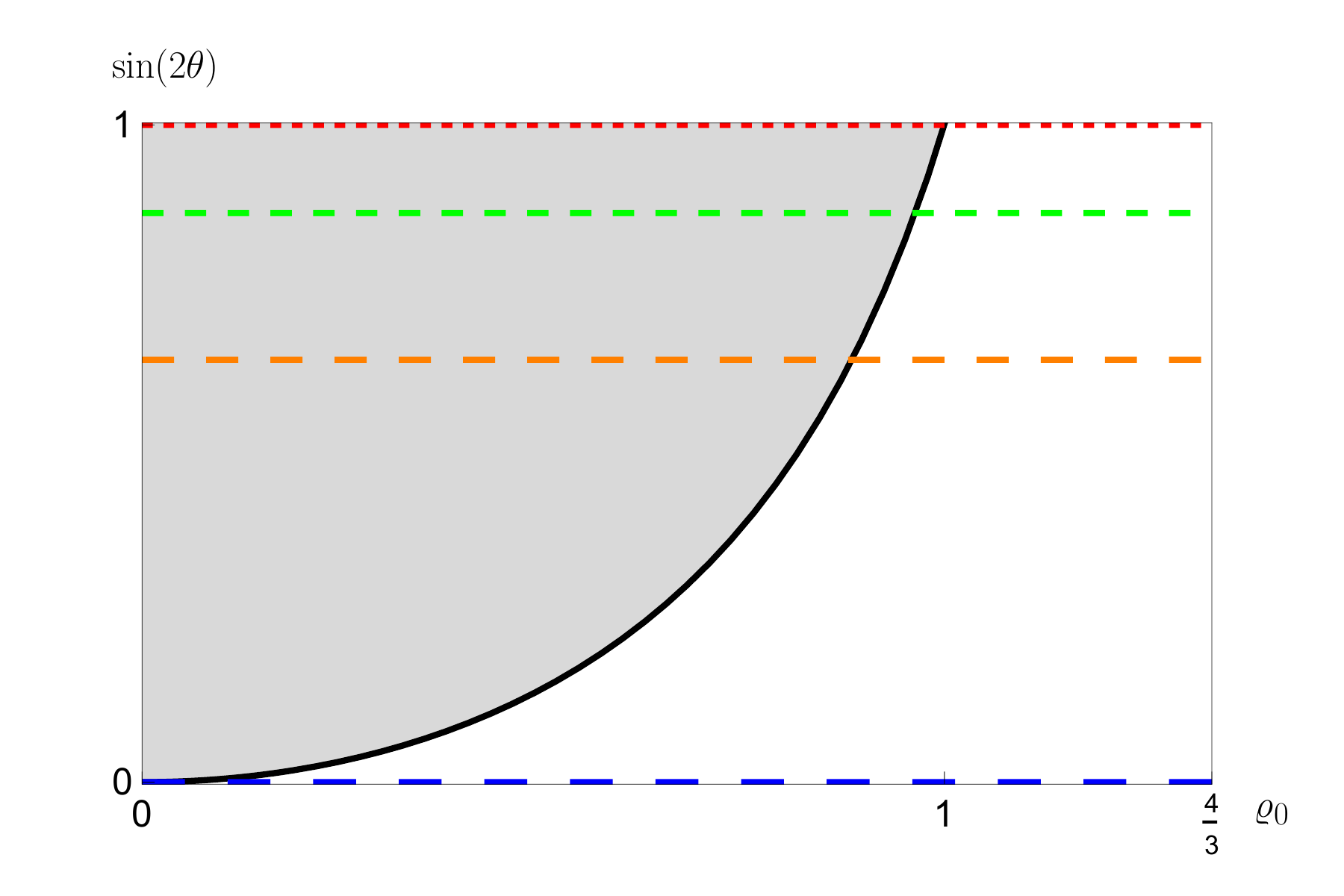}
    \includegraphics[width=0.4\linewidth]{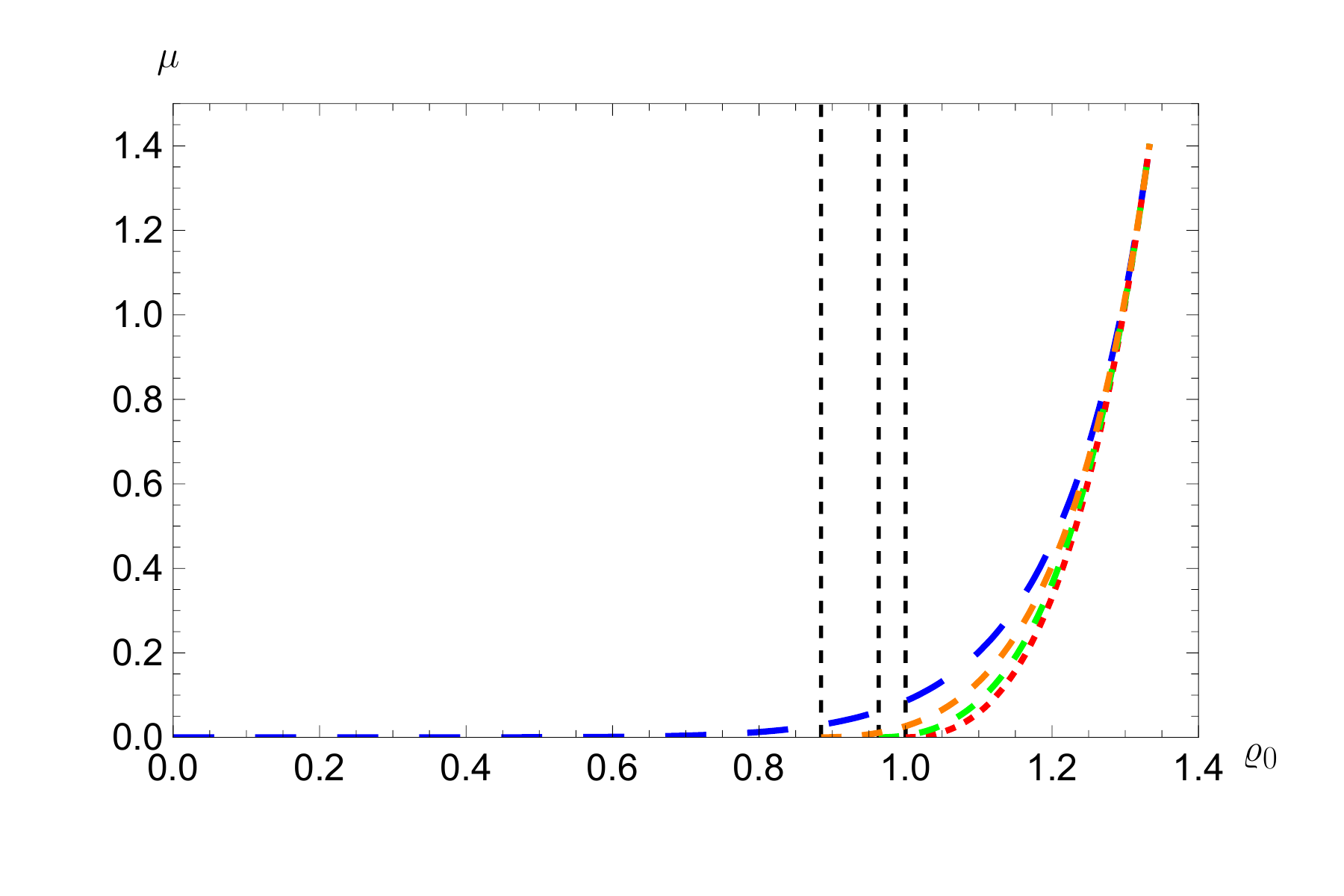}\\
    \vspace{-2.9ex}
    \includegraphics[width=0.4\linewidth]{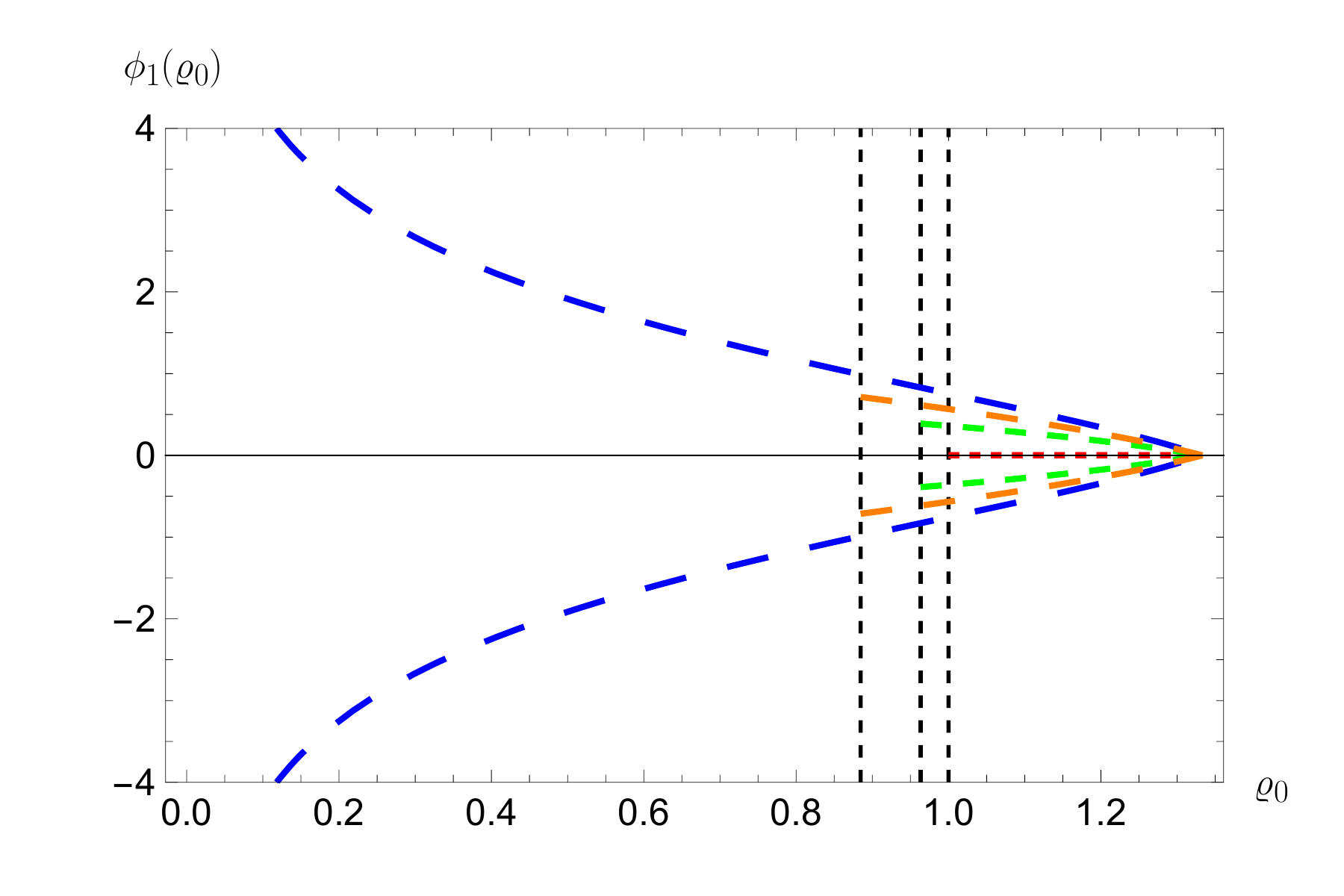}
    \includegraphics[width=0.4\linewidth]{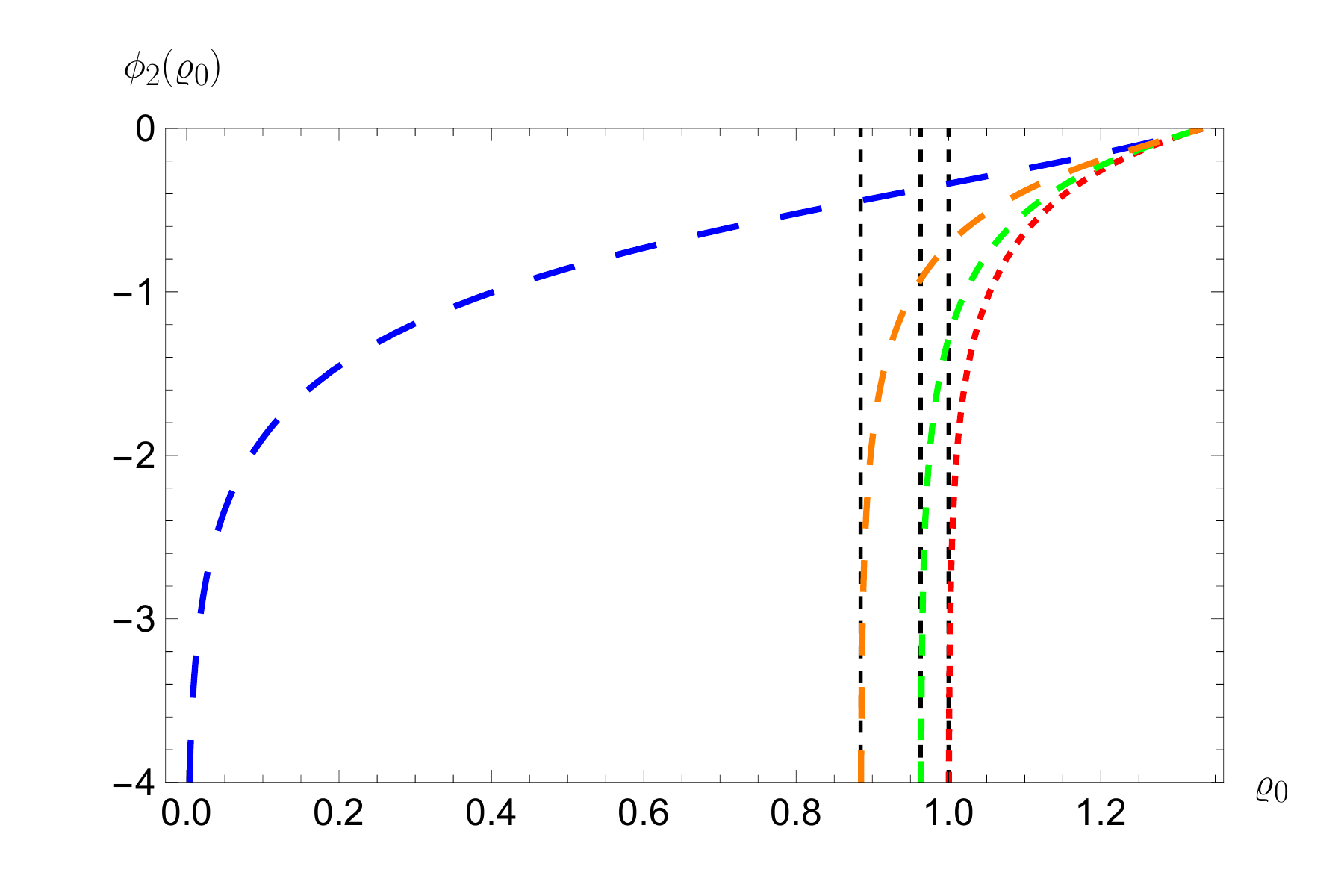}\\
    \vspace{-2.9ex}
    \includegraphics[width=0.4\linewidth]{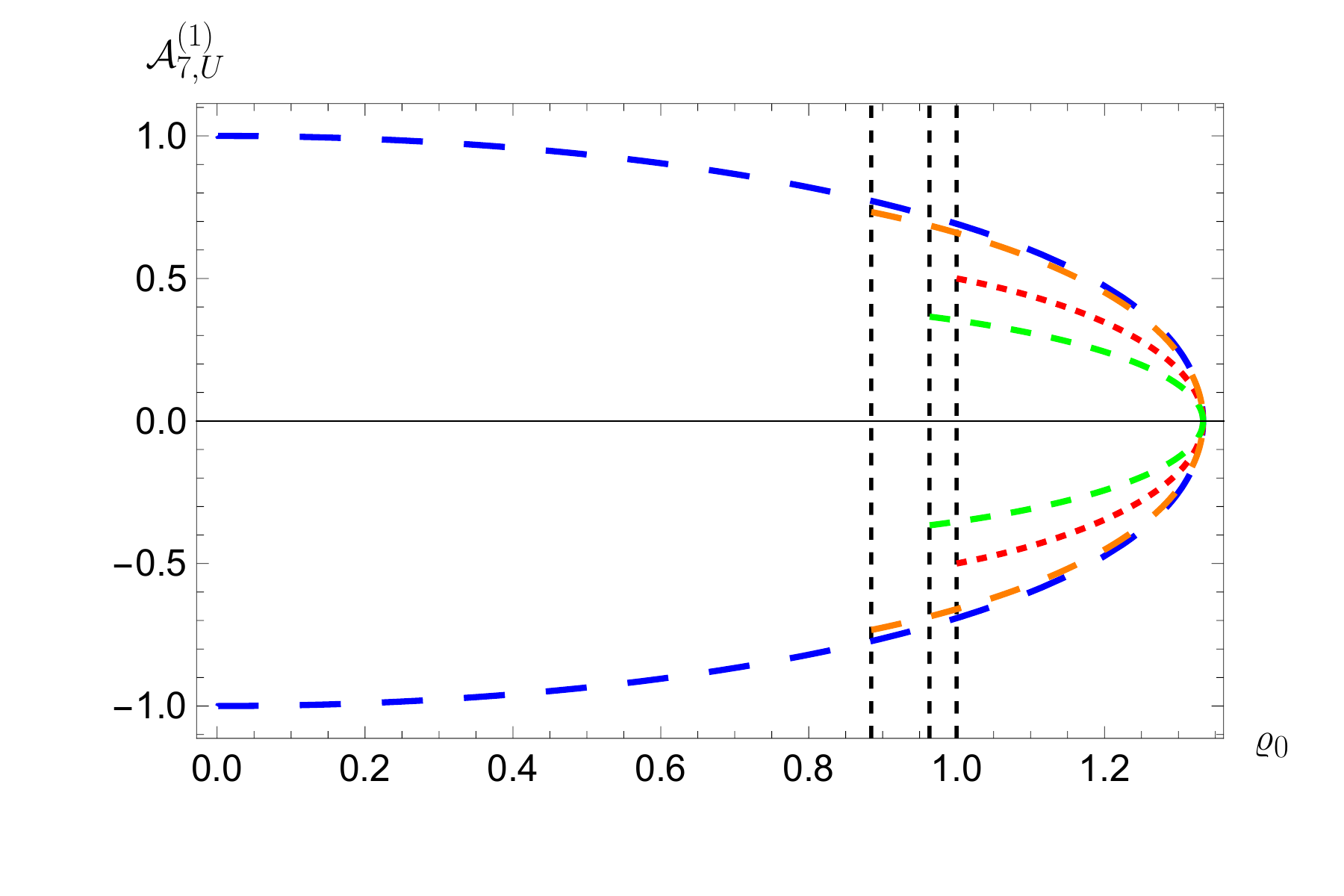}
    \includegraphics[width=0.4\linewidth]{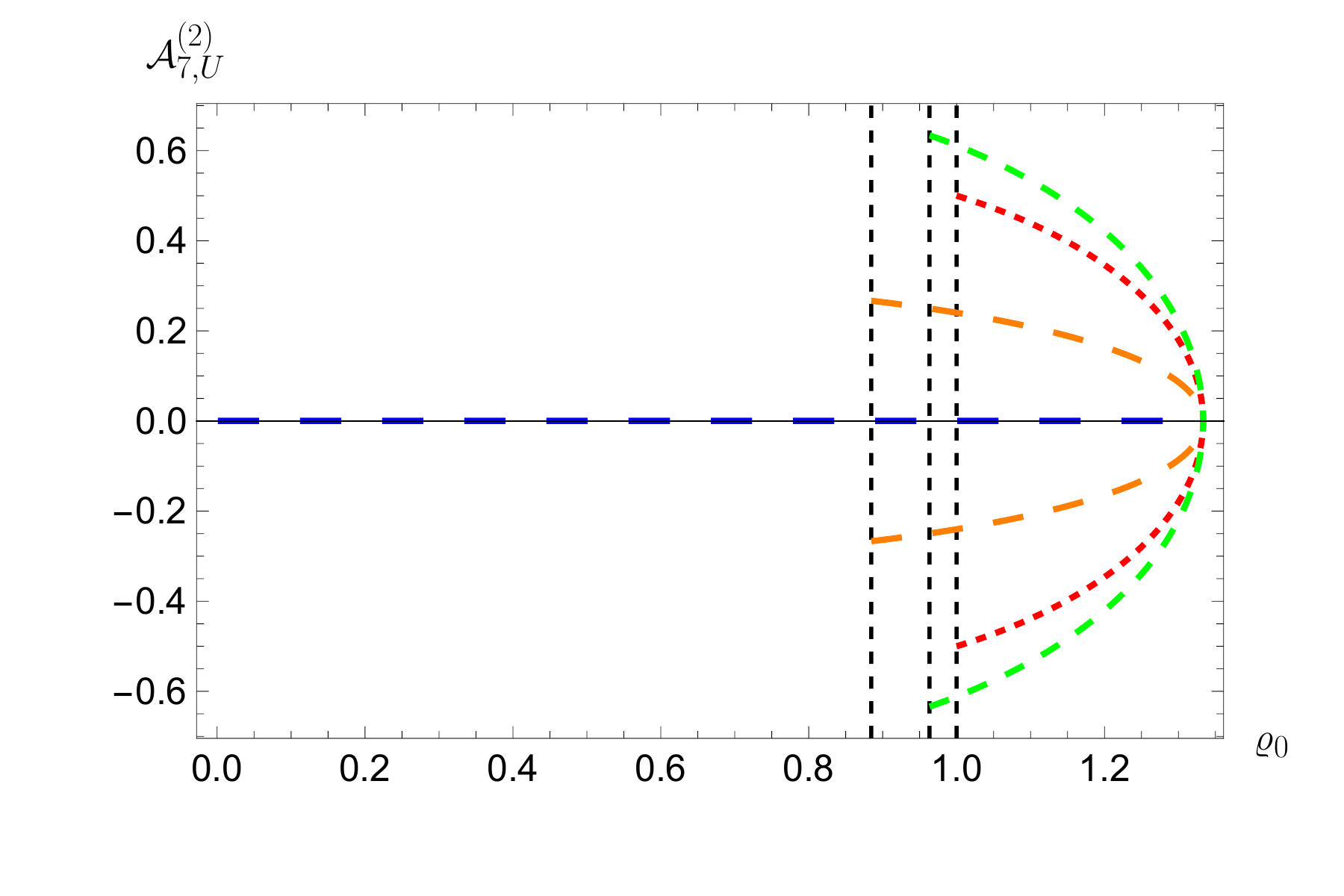}\\
    \vspace{-2.9ex}
    \includegraphics[width=0.4\linewidth]{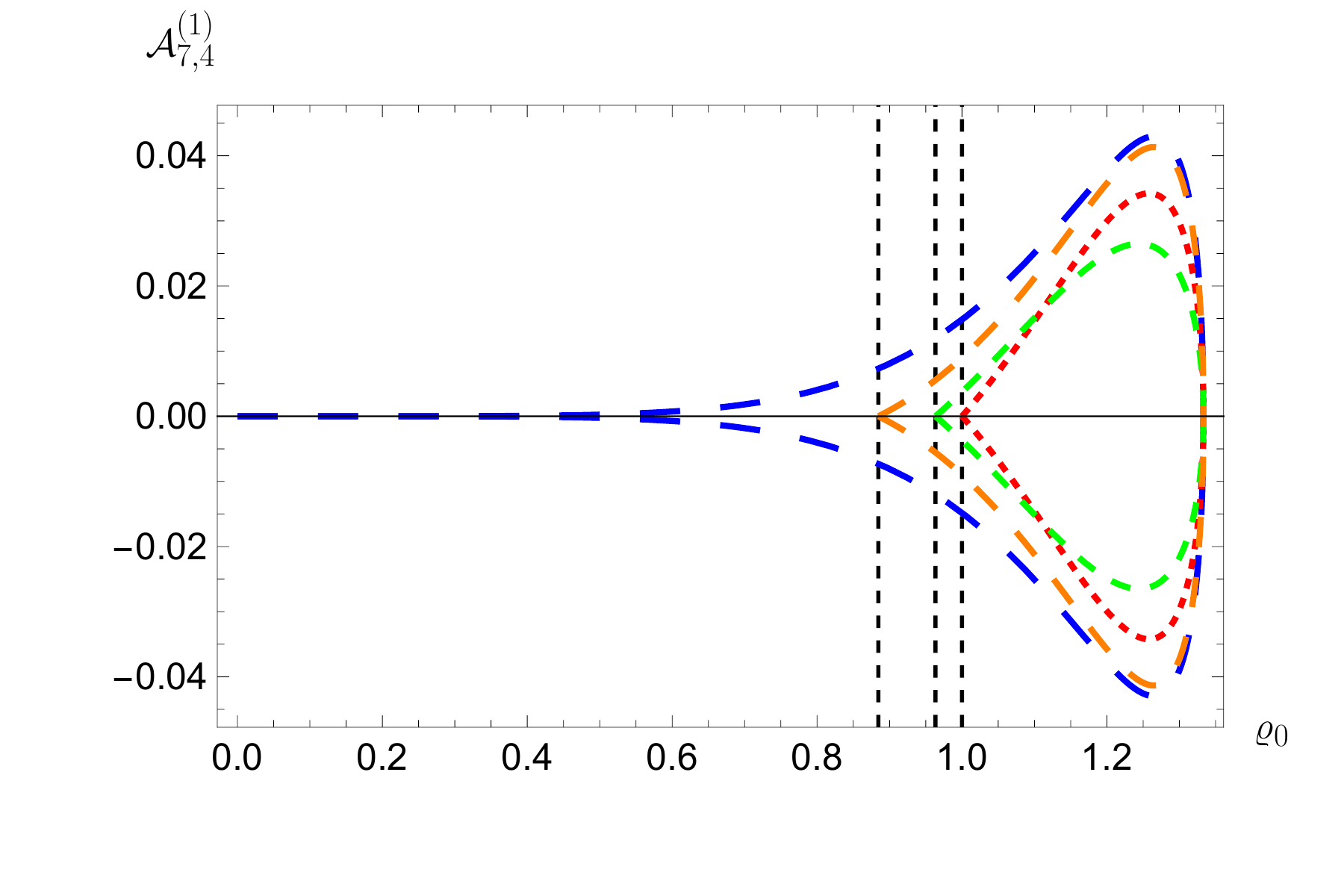}
    \includegraphics[width=0.4\linewidth]{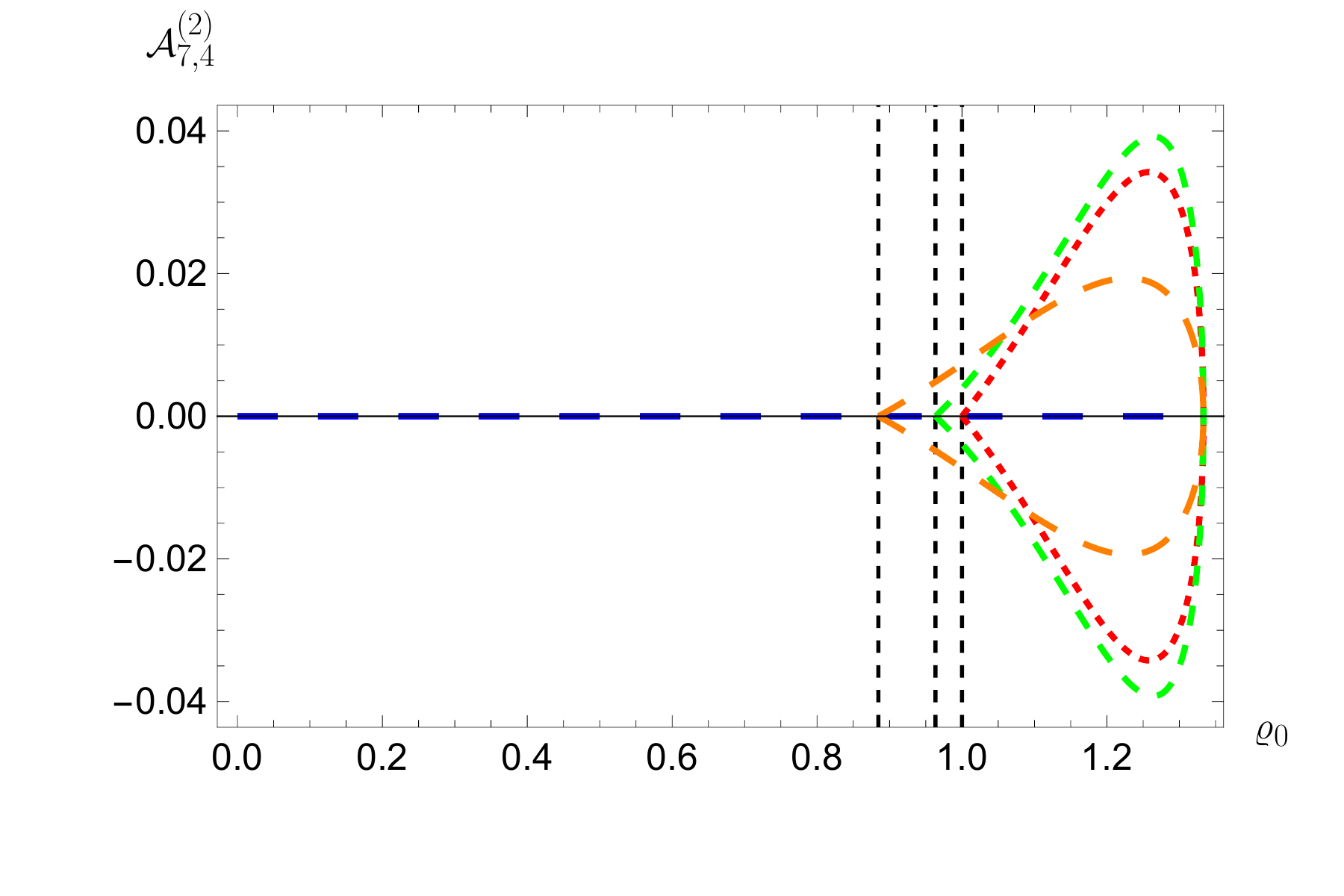}
    \caption{Parameters and functions appearing in the soliton (confining) solutions,  as a function of $\vr_0$, that sets the end of space in the geometry. In each plot the four colours (blue, orange, green, red), with long to short dashing, correspond to branches of solutions with $\cA_{7, U}^{(2)}/\cA_{7, U}^{(1)}=\left(0,\tan\left(\frac{\pi}{9}\right),\tan\left(\frac{\pi}{6}\right),\tan\left(\frac{\pi}{4}\right)\right)$, respectively.
    We shall use the same color-coding also in  Fig.\ref{Fig:PhaseDiagram}. In the first three plots we mark a region of instability in the solutions as a  dark shaded area, this is given when $\vr_0^2\leq Q_i$, or equivalently $\sin(2\theta)\geq \vr_0^2/(2-\vr_0^2)$. In these regions, at least one of the $H_i$ functions changes sign in the region of the geometry corresponding to the deep IR, at which point the metric would become singular and would result in the background values of $\phi_1$ or $\phi_2$ to become imaginary. These cases are hence unphysical, and excluded from the rest of the analysis in the paper. In the fifth and sixth panels we plot the functions $\f_1(\vr), \f_2(\vr)$, evaluated at the end of space, $\vr=\vr_0$, where  we see that $\f_2$ diverges at the boundary of parameter space.}
    \label{fig:parameter_plots}
\end{figure*}

%%%%%%%%%%%%%%%%%%%%%%%
%%%%%%%%%%%%%%%%%%%%%%%
\subsubsection{Background equations }

The background equations, derived from the action  written in six dimensions, and expressed in terms of the holographic coordinate, $\r$, are  the following---see also Appendix~\ref{Sec:sigma-model} and Refs.~\cite{Elander:2010wd,Elander:2010wn,
Bianchi:2003ug,Berg:2005pd,Berg:2006xy,Elander:2009bm,Elander:2014ola,Elander:2018aub,
Elander:2020csd,Elander:2024lir}:
\small{\beqs
\label{eq:phi eom}
4\frac{\6\mathcal{V}_7}{\6\f_1} &=&\f_1''(\r)+[5A'(\r) - \frac{1}{\sqrt{10}}\c'(\r)]\f_1'(\r)+\frac{e^{-4\sqrt{\frac{2}{5}}\c}}{\sqrt{2}}\left(-e^{\sqrt{2}\f_1+\sqrt{\frac{2}{5}}\f_2}{\cA_7^{(1)}}'(\r)^2+e^{-\sqrt{2}\f_1+\sqrt{\frac{2}{5}}\f_2}{\cA_7^{(2)}}'(\r)^2\right)\,, \\
\label{eq:phi2 eom}
4\frac{\6\mathcal{V}_7}{\6\f_2}&=&\f_2''(\r)+[5A'(\r) - \frac{1}{\sqrt{10}}\c'(\r)]\f_2'(\r)+\frac{e^{-4\sqrt{\frac{2}{5}}\c}}{\sqrt{10}}\left(-e^{\sqrt{2}\f_1+\sqrt{\frac{2}{5}}\f_2}{\cA_7^{(1)}}'(\r)^2-e^{-\sqrt{2}\f_1+\sqrt{\frac{2}{5}}\f_2}{\cA_7^{(2)}}'(\r)^2\right) \,,\\
\label{eq:chi eom}
-\frac{2\mathcal{V}_7}{\sqrt{10}}&=&\c''(\r)+[5A'(\r)-\frac{1}{\sqrt{10}}\c'(\r)]\c'(\r)+\frac{e^{-4\sqrt{\frac{2}{5}}\c}}{\sqrt{10}}\left(e^{\sqrt{2}\f_1+\sqrt{\frac{2}{5}}\f_2}{\cA_7^{(1)}}'(\r)^2+e^{-\sqrt{2}\f_1+\sqrt{\frac{2}{5}}\f_2}{\cA_7^{(2)}}'(\r)^2\right) \, ,\\
\label{eq:A71 eom}
0&=&{\cA_7^{(1)}}''(\r)+{\cA_7^{(1)}}'(\r)\left(\sqrt{2}\f_1'(\r)+\sqrt{\frac{2}{5}}\f_2'(\r)+5A'(\r)-\frac{9}{\sqrt{10}}\c'(\r)\right) \, , \\
\label{eq:A72 eom}
0&=&{\cA_7^{(2)}}''(\r)+{\cA_7^{(2)}}'(\r)\left(-\sqrt{2}\f_1'(\r)+\sqrt{\frac{2}{5}}\f_2'(\r)+5A'(\r)-\frac{9}{\sqrt{10}}\c'(\r)\right) \, ,\\
\label{eq:EEom1}
-2\mathcal{V}_7&=&10A'(\r)^2+4A''(\r)-\frac{4A'(\r)\c'(\r)}{\sqrt{10}}+\frac{\f_1'(\r)^2+\f_2'(\r)^2}{4}+\c'(\r)^2 +\nn \\
&&\hspace{12mm} +\frac{e^{-4\sqrt{\frac{2}{5}}\c}}{4}\left(e^{\sqrt{2}\f_1+\sqrt{\frac{2}{5}}\f_2}{\cA_7^{(1)}}'(\r)^2+ e^{-\sqrt{2}\f_1+\sqrt{\frac{2}{5}}\f_2}{\cA_7^{(2)}}'(\r)^2\right)\, ,\\
\label{eq:EEom2}
-2\mathcal{V}_7&=&10A'(\r)^2-\frac{\f_1'(\r)^2+\f_2'(\r)^2}{4}-\c'(\r)^2 -\frac{e^{-4\sqrt{\frac{2}{5}}\c}}{4}\left(e^{\sqrt{2}\f_1+\sqrt{\frac{2}{5}}\f_2}{\cA_7^{(1)}}'(\r)^2+e^{-\sqrt{2}\f_1+\sqrt{\frac{2}{5}}\f_2}{\cA_7^{(2)}}'(\r)^2\right)\,,
\eeqs }
where primed variables denote derivatives in respect to $\rho$, so that $f^{\prime}(\rho)=\frac{\partial}{\partial \rho}f(\rho)$.

We note that  Eqs.~(\ref{eq:A71 eom}) and~(\ref{eq:A72 eom}) may be reformulated in terms of two total derivatives:
\begin{equation}
    \begin{split}
        \6_\r\left(e^{\sqrt{2}\f_1(\r)+\sqrt{\frac{2}{5}}\f_2(\r)+5A(\r)-\frac{9}{\sqrt{10}}\c(\r)}{\cA_7^{(1)}}'(\r)\right)&=0\,, \\
        \6_\r\left(e^{-\sqrt{2}\f_1(\r)+\sqrt{\frac{2}{5}}\f_2(\r)+5A(\r)-\frac{9}{\sqrt{10}}\c(\r)}{\cA_7^{(2)}}'(\r)\right)&=0\,, 
    \end{split}
\end{equation}
and hence we identify two conserved quantities, $\cC_1$ and $\cC_2$, as
\begin{equation}
    \begin{split}
        e^{\sqrt{2}\f_1(\r)+\sqrt{\frac{2}{5}}\f_2(\r)+5A(\r)-\frac{9}{\sqrt{10}}\c(\r)}{\cA_7^{(1)}}'(\r)&\equiv\cC_1\,, \\
        e^{-\sqrt{2}\f_1(\r)+\sqrt{\frac{2}{5}}\f_2(\r)+5A(\r)-\frac{9}{\sqrt{10}}\c(\r)}{\cA_7^{(2)}}'(\r)&\equiv \cC_2\,.
    \end{split}
\end{equation}

By forming the combination  $-2\sqrt{10}\times$(\ref{eq:chi eom})$+$(\ref{eq:EEom1})$+$(\ref{eq:EEom2}), and using the definitions of the two conserved quantities, $\cC_1$ and $\cC_2$, identified above,  we arrive at the equation:
\begin{equation}
    20A'(\r)^2-52\sqrt{\frac{2}{5}}A'(\r)\c'(\r)+2\c'(\r)^2+4A''(\r)-2\sqrt{10}\c''(\r)-2e^{-5A(\r)+\frac{1}{\sqrt{10}}\c(\r)}(\cC_1 {\cA_7^{(1)}}'(\r)+\cC_2 {\cA_7^{(2)}}'(\r))=0\,,
\end{equation}
which can also be recast as a total derivative, and hence
leads us to identify a third conserved quantity, $\cC_3$, defined as
\begin{equation}\label{eq:conserved3}
    -2(\cC_1{\cA_7^{(1)}}(\r)+\cC_2 {\cA_7^{(2)}}(\r))+e^{5A(\r)-\frac{\c(\r)}{\sqrt{10}}}\left(4A'(\r)-2\sqrt{10}\c'(\r)\right)\equiv \cC_3\,.
\end{equation}

%%%%%%%%
%%%%%%%%
\subsubsection{UV Expansions}
\label{Sec:UV}

All the solutions of interest approach the same AdS$_7$ background geometry, asymptotically, for $\rho\rightarrow +\infty$. The dual field theories are all deformations of the same UV fixed point. We find it convenient  to replace the radial direction with $z=e^{-\frac{\r}{2}}$, and expand in powers of small $z$. The asymptotic UV expansion of the background fields in the  whole class  is the following:
\footnotesize{\beqs
\label{eq:UVexp A71}
\cA_7^{(1)}(z)&=&\cA_{7,U}^{(1)}+z^4 \cA^{(1)}_{7,4}-\frac{2}{15} z^6 \left(\sqrt{2} \cA^{(1)}_{7,4} \left(5 \phi _{1,2}+\sqrt{5} \phi
   _{2,2}\right)\right)+\nn \\
   && \frac{\cA^{(1)}_{7,4} }{40} z^8 \left(\log(z)\left(-24 \phi _{1,2}^2 -48 \sqrt{5} \phi _{2,2} \phi
   _{1,2} +72 \phi _{2,2}^2 \right)-19 \phi _{1,2}^2
   \right.\nn \\
   &&\left.
   -2 \sqrt{5} \phi _{2,2} \phi _{1,2}-15 \phi _{2,2}^2+20 \sqrt{2}
   \f_{1,4}+4 \sqrt{10} \f_{2,4}\right)+O\left(z^{10}\right)\, , \\
\label{eq:UVexp A72}
\cA_7^{(2)}(z)&=&\cA_{7,U}^{(2)}+z^4 \cA^{(2)}_{7,4}+\frac{2}{15} \sqrt{2} z^6 \cA^{(2)}_{7,4} \left(5 \phi _{1,2}-\sqrt{5} \phi
   _{2,2}\right)+\nn \\
   &&\frac{\cA^{(2)}_{7,4}}{40} z^8  \left(\log(z)\left(24 \phi _{1,2}^2 -48 \sqrt{5} \phi _{2,2} \phi _{1,2} -72 \phi _{2,2}^2 \right)+19 \phi _{1,2}^2
   \right.\nn \\
   &&\left.
   -2 \sqrt{5} \phi _{2,2} \phi _{1,2}+15 \phi _{2,2}^2+20 \sqrt{2} \phi
   _{1,4}-4 \sqrt{10} \phi _{2,4}\right)+O\left(z^{10}\right)\, ,\\
\label{eq:UVexp phi1}
\f_1(z)&=&z^2 \phi _{1,2}+\frac{1}{5} z^4 \left(5 \phi _{1,4}-6 \sqrt{10} \phi _{1,2} \phi _{2,2} \log (z)\right)+\nn \\
&&\frac{1}{60} z^6
\left(108 \phi _{1,2}^3 \log (z)-108 \phi _{2,2}^2 \phi _{1,2} \log (z)-77 \phi _{1,2}^3+63 \phi _{2,2}^2 \phi _{1,2}-18
\sqrt{10} \phi _{2,4} \phi _{1,2}-18 \sqrt{10} \phi _{1,4} \phi _{2,2}\right)+\nn \\
&& \frac{1}{600} z^8 \left(\log ^2(z)\left(-432 \sqrt{10} \phi
_{2,2} \phi _{1,2}^3 +1296 \sqrt{10} \phi _{2,2}^3 \phi _{1,2}\right)+\right. \nn \\
&&\left.\qquad\log(z)\left(312 \sqrt{10} \phi _{2,2} \phi _{1,2}^3
+360 \phi _{1,4} \phi _{1,2}^2 -2088 \sqrt{10} \phi _{2,2}^3 \phi _{1,2} +720 \phi _{2,2} \phi
_{2,4} \phi _{1,2} -1080 \phi _{1,4} \phi _{2,2}^2 \right) -\right. \nn \\ 
&&\qquad \left.115 \sqrt{10} \phi _{2,2} \phi _{1,2}^3+120 \phi _{1,4}
\phi _{1,2}^2+1321 \sqrt{10} \phi _{2,2}^3 \phi _{1,2}-760 \phi _{2,2} \phi _{2,4} \phi _{1,2}
  \right.  \nn \\
   &&\left.
+600 \phi _{1,4} \phi
_{2,2}^2-60 \sqrt{10} \phi _{1,4} \phi _{2,4}\right)+O\left(z^{10}\right), \\
\label{eq:UVexp phi2}
\f_2(z)&=&z^2 \phi _{2,2}+\frac{1}{5} z^4 \left(-3 \sqrt{10} \phi _{1,2}^2 \log (z)+9
   \sqrt{10} \phi _{2,2}^2 \log (z)+5 \phi _{2,4}\right)+\nn \\
   &&\frac{1}{60} z^6
   \left(972 \phi _{2,2}^3 \log (z)-108 \phi _{1,2}^2 \phi _{2,2} \log (z)-637
   \phi _{2,2}^3+63 \phi _{1,2}^2 \phi _{2,2}+54 \sqrt{10} \phi _{2,4} \phi
   _{2,2}-18 \sqrt{10} \phi _{1,2} \phi _{1,4}\right)+\nn \\
   &&\frac{z^8}{2400} \left(1296
   \sqrt{10} \phi _{1,2}^4 \log ^2(z)-9504 \sqrt{10} \phi _{2,2}^2 \phi
   _{1,2}^2 \log ^2(z)+11664 \sqrt{10} \phi _{2,2}^4 \log ^2(z)-\right.\nn \\
   &&\quad\left.\log(z)\left(1440 \sqrt{10}
   \phi _{1,2}^4 +4992 \sqrt{10} \phi _{2,2}^2 \phi _{1,2}^2 -4320 \phi _{2,4} \phi _{1,2}^2 +2880 \phi _{1,4} \phi _{2,2}
   \phi _{1,2} +8928 \sqrt{10} \phi _{2,2}^4 +12960 \phi
   _{2,2}^2 \phi _{2,4} \right)+ \right. \nn \\
   && \left.\quad 759 \sqrt{10} \phi _{1,2}^4-814 \sqrt{10}
   \phi _{2,2}^2 \phi _{1,2}^2+2400 \phi _{2,4} \phi _{1,2}^2-3040 \phi _{1,4}
   \phi _{2,2} \phi _{1,2}-11921 \sqrt{10} \phi _{2,2}^4-120 \sqrt{10} \phi
   _{1,4}^2+\right. \nn \\
   && \left. \quad 360 \sqrt{10} \phi _{2,4}^2+4960 \phi _{2,2}^2 \phi
   _{2,4}\right)+O\left(z^{10}\right)\, , \\
   \label{eq:UVexp chi}
   \c(z)&=&  \chi_U-\frac{\sqrt{10}}{4} \log (z)-\frac{z^4
   \left(\phi _{1,2}^2+\phi _{2,2}^2\right)}{16 \sqrt{10}}+ \nn \\
   && \frac{z^6}{360}
  \left(
\frac{6 \sqrt{10} \phi _{2,2}^3-6 \sqrt{10} \phi _{1,2}^2 \phi _{2,2}-40 \phi _{2,4} \phi
   _{2,2}-40 \phi _{1,2} \phi _{1,4}+15 \sqrt{10} \chi _6}{\sqrt{10}}+\chi _6+72\log(z)(\phi_{1, 2}^2 \phi_{2, 2}-\phi_{2, 2}^3)\right)+\nn \\
   &&\frac{z^8}{12800} \left(\log^2(z)\left(-288 \sqrt{10} \phi _{1,2}^4 +576 \sqrt{10} \phi _{2,2}^2 \phi _{1,2}^2 -2592 \sqrt{10}
   \phi _{2,2}^4 \right)-\right. \nn \\
   && \left. \quad \log(z)\left(216 \sqrt{10} \phi _{1,2}^4 +432 \sqrt{10}
   \phi _{2,2}^2 \phi _{1,2}^2 +960 \phi _{2,4} \phi _{1,2}^2 +1920 \phi _{1,4} \phi _{2,2} \phi _{1,2} -1944 \sqrt{10} \phi
   _{2,2}^4 -2880 \phi _{2,2}^2 \phi _{2,4} \right)+ \right. \nn \\
   && \left. \quad 163 \sqrt{10}
   \phi _{1,2}^4-270 \sqrt{10} \phi _{2,2}^2 \phi _{1,2}^2+360 \phi _{2,4}
   \phi _{1,2}^2+720 \phi _{1,4} \phi _{2,2} \phi _{1,2}+1355 \sqrt{10} \phi
   _{2,2}^4-80 \sqrt{10} \phi _{1,4}^2- \nn \right. \\
   && \left. \quad 80 \sqrt{10} \phi _{2,4}^2-1080 \phi
   _{2,2}^2 \phi _{2,4}\right)+O\left(z^{10}\right)\, , \\
   \label{eq:UVexp A}
   A(z)&=&A_U-\frac{5}{4}\log(z)-\frac{z^4}{32}(\f_{1, 2}^2+\f_{2, 2}^2)+ \nn \\
   && \frac{z^6}{720}\left(72\sqrt{10}\log(z)(\f_{1, 2}^2\f_{2, 2}-\f_{2, 2}^3)-40(\f_{1, 2}\f_{1, 4}+\f_{2, 2}\f_{2, 4})-6\sqrt{10}(\f_{1, 2}^2\f_{2, 2}-\f_{2, 2}^3)+15\sqrt{10}\c_6\right) \nn \\
   && \frac{z^8}{2560}\bigg(\log(z)^2(-288\f_{1, 2}^4+576\f_{1, 2}^2\f_{2, 2}^2-2592\f_{2, 2}^4) + \nn \\
   &&\log(z)\left(-216\f_{1, 2}^4+192\sqrt{10}\f_{1, 2}\f_{1, 4}\f_{2, 2}+432\f_{1, 2}^2\f_{2, 2}^2-1944\f_{2, 2}^4+96\sqrt{10}\f_{1, 2}^2\f_{2, 4}-288\sqrt{10}\f_{2, 2}^2\f_{2, 4}\right)+\nn \\
   &&163\f_{1, 2}^4-80\f_{1, 4}^2+72\sqrt{10}\f_{1, 2}\f_{1, 4}\f_{2, 2}-270\f_{1, 2}^2\f_{2, 2}^2+1355\f_{2, 2}^4
      \nn \\
   &&
   +36\sqrt{10}\f_{2, 2}^2\f_{2, 4}-108\sqrt{10}\f_{2, 2}^2\f_{2, 4}-80\f_{2, 4}^2 \bigg) +\cO(z^{10})\,.
\eeqs}
We  define $\chi_6$ in such a way that for $\chi_6=0$ we recover the relation $\chi=\frac{\sqrt{10}}{5} A$,
typifying the domain-wall solutions in $D=7$ dimensions.

The UV expansions are determined by $11$ parameters: they correspond to the five sources and five vacuum expectation values of the operators dual to the five sigma-model scalars, with the addition of a harmless additive constant in the warp factor, $A$. In these expressions, we denote the parameters as $\{\f_{1, 2}, \f_{2, 2}, \f_{1, 4}, \f_{2, 4}, \cA_{7, U}^{(1)}, \cA_{7, U}^{(2)}, \cA_{7, 4}^{(1)}, \cA_{7, 4}^{(2)}, \c_U, \c_6, A_U\}$. One recovers domain-wall solutions in $D=7$ dimensions, with AdS$_7$ geometry, by setting to zero all these constants, except for $\cA_{7, U}^{(1)}$,  $\cA_{7, U}^{(2)}$, and the trivial $A_U$.

%%%%%%%%
%%%%%%%%
%%%%%%%%
\subsubsection{Soliton (confining) solutions}

The soliton solutions that are the main  interest for our analysis can be written 
 in closed form, by  first rewriting the metric after a change of radial coordinate
 from $\r$ to the new $\varrho$~\cite{Liu:1999ai, Cvetic:1999xp}, as follows:
\begin{equation}\label{metricAna}
    \dd s_7^2=\frac{(H_1(\vr)H_2(\vr))^{\frac{1}{5}}}{f(\vr)}\dd\vr^2 + \frac{1}{4}(H_1(\vr)H_2(\vr))^{\frac{1}{5}}\vr^2\dd x_{1, 4}^2+(H_1(\vr)H_2(\vr))^{-\frac{4}{5}}f(\vr) \dd \eta^2\,.
\end{equation}
The new functions $H_1(\varrho)$, $H_2(\varrho)$, and $f(\varrho)$ can be  identified as follows:
\beqs
     \c(\r)&\equiv&\frac{\sqrt{10}}{8}\log\left[(H_1H_2)^{-\frac{4}{5}} f\right], \\
    A(\r)&\equiv&\frac{1}{2}\log\left[\frac{1}{4}(H_1H_2)^{\frac{1}{5}}\vr^2\right]+\frac{1}{\sqrt{10}}\c(\vr)\, ,\\
    \frac{\6}{\6 \r} &\equiv& \frac{\sqrt{f}}{(H_1H_2)^{\frac{1}{10}}}\frac{\6}{\6 \vr}\,.
\eeqs
By imposing  the background equations, one arrives to explicit expressions for the background functions:
\beqs
H_i&=&1-\frac{Q_i^2}{\vr^4}\,~~~~~(i=1,\,2) ,\\
f&=&-\frac{\m}{\vr^4}+\frac{1}{4}\vr^2H_1 H_2\, ,\\
\f_1&=&\frac{1}{\sqrt{2}}\log\left[\frac{H_1}{H_2}\right]\, ,\\
\f_2&=&\frac{1}{\sqrt{10}}\log\left[H_1 H_2\right]\, ,\\
\cA_7^{(i)}&=&\pm\left(\frac{\sqrt{\m}(1-H_i^{-1})}{Q_i}-\frac{\sqrt{\m}(1-H_i(\vr_0)^{-1})}{Q_i}\right)\,.
\eeqs
They depends on the parameters $Q_1$, $Q_2$, and $\mu$.
To avoid a singularity,  the conditions $\vr^2_0\geq {\rm Max}[|Q_i|]$  and $\vr_0\geq 0$ must  satisfied, where $\vr_0$ is defined as the largest  root of $f(\vr)$. Notice that $\cA_7^{(i)}$ vanishes at the end of space, when $\varrho\rightarrow\varrho_0$.\footnote{The algebraic equation $f(\vr_0)=0$  admits multiple solutions, which one might be tempted to interpret in terms of multiple branches of solutions to the background equations. Unfortunately, only one such algebraic solution gives rise to a gravity background ending at $\vr_0$. By replacing in the analytic expressions, in particular Eq.~(\ref{metricAna}),  solutions  that violates the requirement $\vr^2_0\geq {\rm Max}[|Q_i|]$,  results in the metric developing a singularity  at $\vr={\rm Max}[|Q_i|]>\vr_0$, before the end of space. Such solutions to the algebraic relations are discarded.} The scalars $\phi_1$ and $\phi_2$, that have no charge under the two Abelian gauge groups, develop profiles that are controlled by the fluxes, the magnitude of which are given by $\cA_7^{(i)}$, and break the $SO(5)$ symmetry to its $SO(2)\times SO(2)$ subgroup.

In order to characterise these solutions in the large class identified in Sect.~\ref{Sec:UV}, we  perform a UV expansion of the solutions in the  
parameter $z=e^{-\frac{\r}{2}}$. We do so by first defining a new coordinate $\mathfrak{z}=\frac{1}{\vr}$ and writing this as a series in $z$:
\beqs \label{eq:UVexp confining sol}
\mathfrak{z}&=&\frac{z}{2}-\frac{1}{320}(Q_1^2+Q_2^2)z^5-\frac{\m}{384}z^7 + \frac{Q_1^4+7Q_1^2Q_2^2+Q_2^4}{51200}z^9+\frac{13(Q_1^2+Q_2^2)}{307200}z^{11}+\mathcal{O}(z^{13})\, , 
\eeqs
The soliton solutions then read as follows:
\beqs 
\label{eq:UVexp A71 conf}
\cA_7^{(1)}(z)&=&\frac{Q_1\sqrt{\m}}{Q_1^2-\vr_0^4}
+\frac{Q_1\sqrt{\m}}{16}z^4+\frac{Q_1(3Q_1^2-2Q_2^2)\sqrt{\m}}{1280}z^8+\cO(z^{10})\, , \\
\label{eq:UVexp A72 conf}
\cA_7^{(2)}(z)&=&\frac{Q_2\sqrt{\m}}{Q_2^2-\vr_0^4}
+\frac{Q_2\sqrt{\m}}{16}z^4+\frac{2Q_2(-Q_1^2+3Q_2^2)\sqrt{\m}}{1280}z^8+\cO(z^{10})\, , \\
\label{eq:UVexp phi1 conf}
\f_1(z)&=&\frac{-Q_1^2+Q_2^2}{16\sqrt{2}}z^4+\frac{-Q_1^4+Q_2^4}{2560\sqrt{2}}z^8+\cO(z^{10})\, ,\\
\label{eq:UVexp phi2 conf}
\f_2(z)&=&-\frac{Q_1^2+Q_2^2}{16\sqrt{10}}z^4-\frac{Q_1^4-8Q_1^2Q_2^2+Q_2^4}{2560\sqrt{10}}z^8+\cO(z^{10})\, ,\\
\label{eq:UVexp chi conf}
\c(z)&=&-\frac{\sqrt{10}}{4}\log(z)-\frac{\m}{16\sqrt{10}}z^6-\frac{3Q_1^4-4Q_1^2Q_2^2+3Q_2^4}{20480\sqrt{10}}z^8 +\cO(z^{10})\, , \\
\label{eq:UVexp A conf}
A(z)&=&-\frac{5}{4}\log(z)-\frac{\m}{768}z^6-\frac{3Q_1^4-4Q_1^2Q_2^2+3Q_2^4}{40960}z^8 +\cO(z^{10})\,.
\eeqs
We hence arrive to the identifications:
\beqs
\f_{1, 2}&=&\f_{2, 2}=\c_U=A_U=0\, \\
\f_{1, 4}&=&\frac{-Q_1^2+Q_2^2}{16\sqrt{2}}\, ,\\
\f_{2, 4}&=&\frac{-Q_1^2-Q_2^2}{16\sqrt{10}}\, ,\\
\cA_{7, 4}^{(i)}&=&\pm \frac{Q_i\sqrt{\m}}{16}\, , \\
\label{eq:A7Ui}
\cA_{7, U}^{(i)}&=&\pm\frac{Q_i\sqrt{\m}}{Q_i^2-\vr_0^4}\, ,\\
\c_6&=&-\frac{\m}{16\sqrt{10}}\, .
\eeqs
The three parameters $Q_1$, $Q_2$, and $\mu$ can be traded for $\cA_{7, U}^{(i)}$, $\cA_{7, U}^{(i)}$, $\chi_6$. We illustrate the properties of the soliton solutions in Fig.~\ref{fig:parameter_plots}, by plotting the value of the parameters as a function of $\varrho_0$, for representative choices of other parameters.

Absence of conical singularity near the end of space, at $\vr\rightarrow \vr_0$, is ensured by assessing the expansion of the induced two-dimensional metric in the plane described by polar coordinates $(\eta,\vr)$,  near $\vr_0$:
\beqs
    \dd s_2^2&=& \di \rho^2 +e^{4\sqrt{\frac{5}{2}}\chi}\di\eta^2
    \nonumber\\
    &=&
    \frac{(H_2H_2)^{\frac{1}{5}}}{f}\dd \vr^2 +(H_1H_2)^{(-\frac{4}{5})}f \dd \eta^2 \approx \frac{(H_1(\vr_0)H_2(\vr_0))^{\frac{1}{5}}}{f'(\vr_0)(\vr-\vr_0)}\dd \vr^2 +(H_1(\vr_0)H_2(\vr_0))^{(-\frac{4}{5})}f'(\vr_0)(\vr-\vr_0) \dd \eta^2\,.
\eeqs
Making the change of variable $\frac{(H_2H_2)^{\frac{1}{5}}}{f'(\r_0)(\vr-\vr_0)}\dd \vr^2= \di \rho^2$, such that $\rho=\sqrt{\frac{4(H_1(\vr_0)H_2(\vr_0))^{\frac{1}{5}}}{f'(\vr_0)}(\vr-\vr_0)}$, the induced metric  described by the polar coordinates, $(\rho,\,\eta)$, reads:
\beqs
 \dd s_2^2 = \dd \rho^2 + \left(\frac{f'(\vr_0)}{2\sqrt{H_1(\vr)H_2(\vr_0)}}\right)^2\rho^2 \dd \eta^2\,,
\eeqs
which requires  to set $f'(\vr_0)=\pm2\sqrt{H_1(\vr)H_2(\vr_0)}$, hence avoiding the arising of a conical singularity. This expression translates into the following additional constraint on the parameters characterising the solution:
\beqs\label{constraint}
 0&=&   Q_2^2\vr_0^4+\vr_0^8+Q_1^2(\vr_0^4-3Q_2^2)\pm 4\vr_0^3\sqrt{(Q_1^2-\vr_0^4)(Q_2^2-\vr_0^4)}+8\vr_0^2\m\,. 
\eeqs
This constraint must be satisfied at  the same time as the defining equation for $\vr_0$:
\beqs
f(\vr_0)&=&0\,=\,
\label{Eq:f0}
\frac{1}{4} \vr_0^2 \left(1-\frac{{Q_1}^2}{\vr_0^4}\right)
   \left(1-\frac{{Q_2}^2}{\vr_0^4}\right)-\frac{\mu }{\vr_0^4}\,.
\eeqs
This constraint further reduces the number of free parameters to two, which can be fixed in more than one way: for example, we can fix
$\cA_{7, U}^{(1)}$ and $\cA_{7, U}^{(1)}$, which we will generically refer to as the sources in later parts of the paper, and use the constraints to set $\varrho_0$ and $\mu$, and ultimately $Q_1$ and $Q_2$. By inspection, we find that  non-singular, smooth soliton solutions exist only for  $\varrho_0<\frac{4}{3}$.

By making use of Eq.~(\ref{Eq:f0}), and of the definition of the leading-order coefficients in the UV expansions, in Eq.~(\ref{eq:A7Ui}), we can rewrite 
\beqs
\label{Eq:Q1}
(Q_1^2-\vr_0^4)(Q_2^2-\vr_0^4)&=&4\mu\vr_0^2\,,\\
\label{Eq:Q2}
Q_1^2Q_2^2&=&\left(4\vr_0^2\cA_{7, U}^{(1)}\cA_{7, U}^{(2)}\right)^2\,,\\
\label{Eq:Q3}
Q_1^2+Q_2^2&=&-4\frac{\mu}{\vr_0^2}+\vr_0^4+\left(4\cA_{7, U}^{(1)}\cA_{7, U}^{(2)}\right)^2\,,
\eeqs
and hence the constraint derived from the absence of conical singularities reads
\beqs
\left(2\mu+\vr_0^6-16\vr_0^2\left(\cA_{7, U}^{(1)}\right)^2 \left(\cA_{7, U}^{(2)}\right)^2\right)^2 &=&16 \mu \vr_0^4\,.
\eeqs
The relation to the source terms, $\cA_{7, U}^{(i)}$, is completed by the following, obtained by combining the definition of  
$\cA_{7, U}^{(i)}$ with the three expressions in Eqs.~(\ref{Eq:f0}),  (\ref{Eq:Q1}), and~(\ref{Eq:Q3}):
\beqs
\left(\cA_{7, U}^{(1)}\right)^2+\left(\cA_{7, U}^{(2)}\right)^2&=&
\frac{256 \vr_0^2 \left(\cA_{7, U}^{(1)}\right)^4 \left(\cA_{7, U}^{(2)}\right)^4
-32 \left(\cA_{7, U}^{(1)}\right)^2 \left(\cA_{7, U}^{(2)}\right)^2 (\vr_0^6+2\mu)+\vr_0^{4}\left(\vr_0^6-4\mu\right)}{16 \mu \vr_0^2}\,.
\eeqs

Finally, we find it convenient to define an alternative parameterisation of the magnetic fluxes by introducing polar coordinates $\cA_{7, U}^{(1)}=\cA_U\cos(\theta)$ and $\cA_{7, U}^{(2)}=\cA_U\sin(\theta)$,  so that the relations between 
the four variables read:
\beqs
\mu&=&2 \vr_0^4 \left(2-2 \cA_U^2-\vr_0^2\right)\,,\\
\sin^2(2\theta)&=&\pm\left(
\frac{\cA_U^4 \vr_0^2 \left(-8 \cA_U^2-3 \vr_0^2+8\right)-4 \sqrt{2} \sqrt{-\cA_U^8 \vr_0^4 \left(2
   \cA_U^2+\vr_0^2-2\right)}}{4 \cA_U^8}
   \right)\,.
\eeqs

We conclude this section by noticing that we have identified two widely different classes of solutions, that can be both parameterised in terms of the leading coefficients, $\cA_{7, U}^{(1)}$ and $\cA_{7, U}^{(2)}$, of the expansion of the seventh component of the two Abelian gauge fields.
Because of the compact nature of the seventh dimension, $\eta$, the value of such gauge fields corresponds to non-trivial magnetic fluxes in the dual field theory language.
The AdS$_7$ solutions have geometries that are independent of $\cA_{7, U}^{(1)}$, and $\cA_{7, U}^{(2)}$, are interpreted in terms of conformal field theories in $D-1=6$ dimensions, and exist for any choices of these two parameters. 

The soliton solutions obtained by the aforementioned restrictions 
among the parameters $\mu$, $\rho_0$, $Q_1$, and $Q_2$ have a more sophisticated interpretation in field theory terms. The compacification of the seventh dimension is causing it to shrink, and the dual field theory ultimately
reduces to a five-dimensional one that confines. The presence of magnetic flux is no longer harmless, as when the circle shrinks to zero size the magnetic flux must  vanish as well, resulting in  restrictions to the admissible values of $\cA_{7, U}^{(1)}$, and $\cA_{7, U}^{(2)}$.
As shown explicitly in Appendix~\ref{Sec:gravityinvariants}, these solutions are smooth and regular except, except for the ones with $\mu=0$, that are singular when $\vr \rightarrow \vr_0$.
In the next sections, we will perform a stability analysis aimed at understanding which of these two classes of solutions is physically realised, as a function of the two sources, $\cA_{7, U}^{(1)}$, and $\cA_{7, U}^{(2)}$.

%%%%%%%%%%%%%%%%%%%%%%%
%%%%%%%%%%%%%%%%%%%%%%%
\section{Global stability analysis: Free energy} 
\label{Sec:FreeEnergy}

\begin{figure*}[t]
    \centering
    \subfloat[Phase Diagram \label{Fig:PhaseDiagram}]{\includegraphics[width=0.45\linewidth]{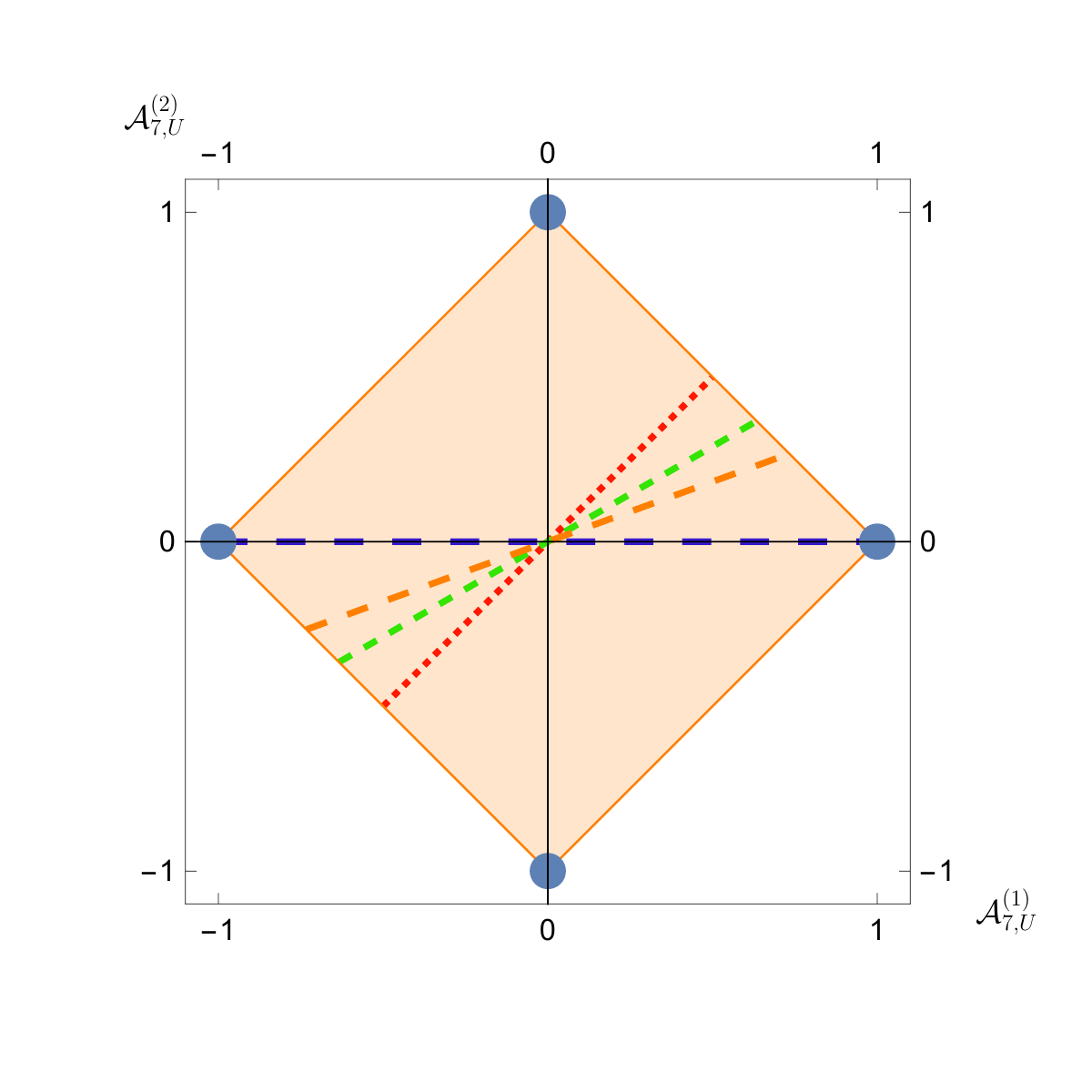}} 
    \subfloat[Free Energy \label{Fig:FreeEnergy}]{\includegraphics[width=0.45\linewidth]{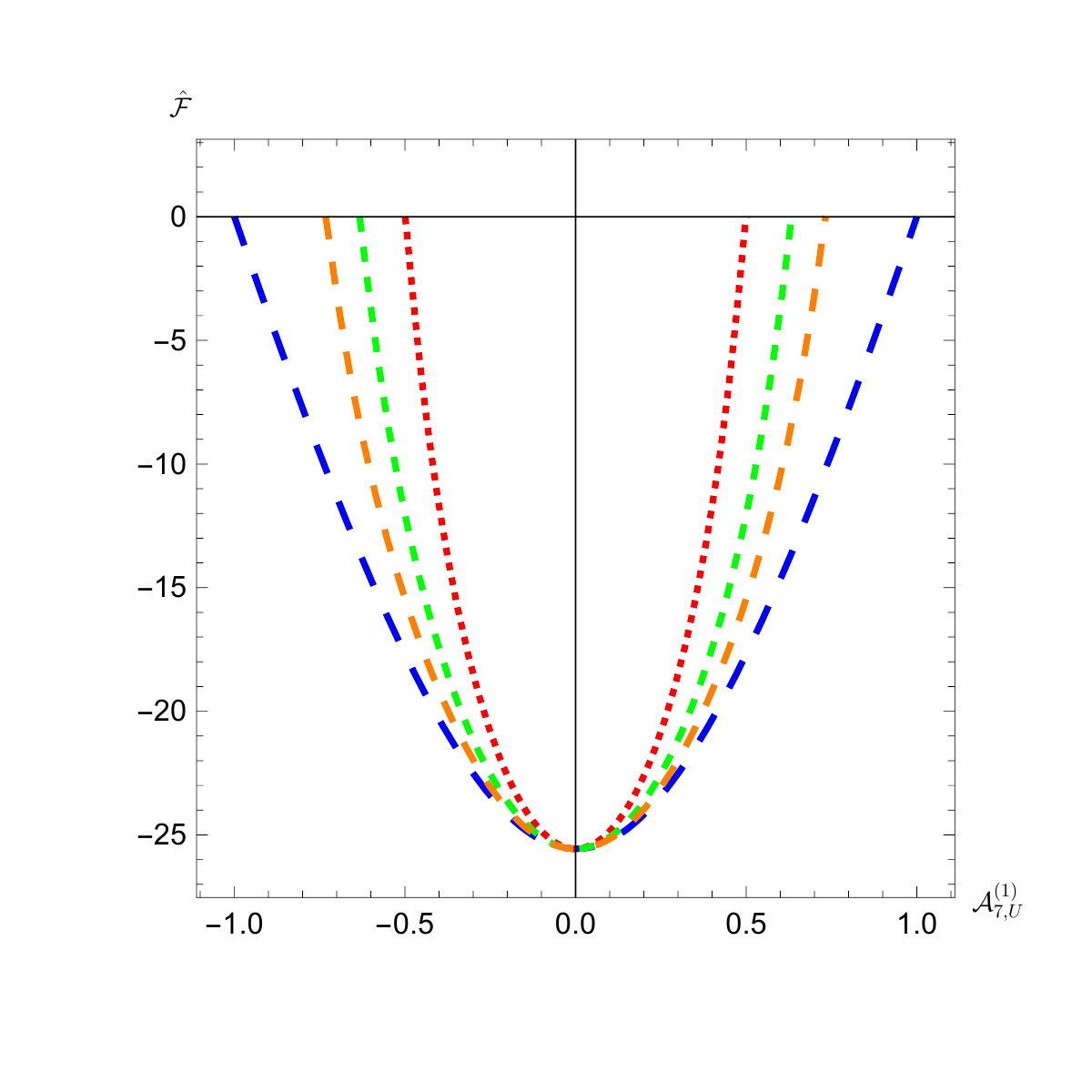}}
    \caption{Left panel: phase diagram of the model, in the plane defined by the sources  $\cA_{7, U}^{(1)}$ and $\cA_{7, U}^{(2)}$. Inside the shaded area, the vacuum is given by regular soliton solutions, corresponding to confining solutions with magnetic flux in the dual theory. Outside the shaded region, the vacuum is given by domain wall solutions leading to AdS$_7$ geometry. Along the sides of the square one has first-order phase transitions. The dashed line correspond to lines of constant ratio $\cA_{7, U}^{(2)}/\cA_{7, U}^{(1)}=\left(0,\tan\left(\frac{\pi}{9}\right),\tan\left(\frac{\pi}{6}\right),\tan\left(\frac{\pi}{4}\right)\right)$ (long dashing to short dashing). Right panel: the free energy density, $\hat{\cal F}$  of the dual, strongly coupled  field theory living in six dimensions, expressed in units of the scale $\Lambda$, as a function of the parameter $\cA_{7, U}^{(1)}$, and for values of $\cA_{7, U}^{(2)}$ chosen along the four straight lines in the phase diagram in the left panel, with constant ratio $\cA_{7, U}^{(2)}/\cA_{7, U}^{(1)}=\left(0,\tan\left(\frac{\pi}{9}\right),\tan\left(\frac{\pi}{6}\right),\tan\left(\frac{\pi}{4}\right)\right)$---long dashed to short dashed lines in blue, orange, green, and red, respectively.}
    \label{fig:F}
\end{figure*}

The holographic renormalisation prescription for the calculation of the free energy of the dual field theory starts with the evaluation of the on-shell action in the gravity dual~\cite{Bianchi:2001kw,Skenderis:2002wp,Papadimitriou:2004ap},
which  in seven dimensions reads
\beqs
{\cal S}_7^{\rm on-shell}
&=&\frac{1}{2\pi}\int \di^5 x\di\eta\di\rho\partial_{\rho}\left[
e^{5A(\rho)-\frac{1}{\sqrt{10}}\chi(\rho)}\left(-\frac{1}{2}\partial_{\rho}A(\rho)+\frac{1}{2\sqrt{10}}\partial_{\rho}\chi(\rho)\right)
\right]\,,
\eeqs
having made use of Einstein's equations, Eqs.~(\ref{eq:EEom1}) and~(\ref{eq:EEom2}). To this, one adds boundary localised terms, required by holographic renormalisation. The Gibbons-Hawking-York (GHY),  boundary localised contribution to the action is
 \beqs
 {\cal S}_{\rm GHY}^{\rm on-shell}&=&
\frac{1}{2\pi} \int\di^5x \di \eta\di \rho\sum_{i=1}^2\delta(\rho-\rho_i)(-)^i\left[ \sqrt{-\tilde{g}}\frac{\cal \kappa}{2}\right]
\nonumber\\
&=&
\frac{1}{2\pi} \int\di^5x \di \eta\di \rho\sum_{i=1}^2\delta(\rho-\rho_i)(-)^i\left[
e^{5A(\rho)-\frac{1}{\sqrt{10}}\chi(\rho)}\left(\frac{5}{2}\partial_{\rho}A(\rho)-\frac{1}{2\sqrt{10}}\partial_{\rho}\chi(\rho)
\right)\right]
\,,
 \eeqs
 where we introduce the vector orthonormal to the boundary, $N_{\hat{M}}=(0,\,0,\,0,\,0,\,0,\,1,\,0)$, the induced metric 
 $\tilde{g}_{\hat{M}\hat{N}}=g_{\hat{M}\hat{N}}-N_{\hat{M}}N_{\hat{N}}$, and 
 the extrinsic curvature, 
 ${\cal \kappa}\equiv\tilde{g}^{\hat{M}\hat{N}}\nabla_{\hat{M}}N_{\hat{N}}=-\tilde{g}^{\hat{M}\hat{N}}\Gamma^{\hat{P}}_{\,\,\,\hat{M}\hat{N}}N_{\hat{P}}$.
Finally, there are boundary localised potentials, of the form
 \beqs
 {\cal S}_{\lambda}^{\rm on-shell}&=&
\frac{1}{2\pi} \int\di^5x \di \eta\di \rho\sum_{i=1}^2\delta(\rho-\rho_i)(-)^i\left[ \sqrt{-\tilde{g}}\frac{}{}\lambda_i\right]
\nonumber\\
&=&
\frac{1}{2\pi} \int\di^5x \di \eta\di \rho\sum_{i=1}^2\delta(\rho-\rho_i)(-)^i\left[
e^{5A(\rho)-\frac{1}{\sqrt{10}}\chi(\rho)}\frac{}{}\lambda_i\right]
\,.
 \eeqs
Putting the three contributions together yields:
\beqs
{\cal S}&=&{\cal S}_7^{\rm on-shell}+ {\cal S}_{\rm GHY}^{\rm on-shell}+ {\cal S}_{\lambda}^{\rm on-shell}\\
&=&
\frac{1}{2\pi} \int\di^5x \di \eta\di \rho\sum_{i=1}^2\delta(\rho-\rho_i)(-)^i\left[
e^{5A(\rho)-\frac{1}{\sqrt{10}}\chi(\rho)}\left({2}\partial_{\rho}A(\rho)\frac{}{}+\frac{}{}\lambda_i\right)
\right]\,.
\eeqs

In order to apply holographic renormalisation~\cite{Bianchi:2001kw,Skenderis:2002wp,Papadimitriou:2004ap} to the reduction onto a circle,  as in Refs.~\cite{Elander:2020ial,Elander:2020fmv,Elander:2021wkc,Elander:2022ebt,Fatemiabhari:2024lct}, we impose
the requirements $\lambda_1=-2\partial_{\rho}A(\rho)$,  that is necessary for the consistency of the variational problem giving the background solutions, and $\lambda_2={\cal W}_2$ (see Appendix~\ref{Sec:super} for the complete form of ${\cal W}_2$ required in general classes of solutions), that automatically cancels the divergences. We ultimately arrive to the final expression
\beqs
{\cal S}^{\rm ren}&=&{\cal S}\,=\,\int\di^5x \left[
e^{5A(\rho)-\frac{1}{\sqrt{10}}\chi(\rho)}\left({2}\partial_{\rho}A(\rho)\frac{}{}+\frac{}{}{\cal W}_2\right)
\right]_{\rho=\rho_2}\,,
\eeqs
where we performed the trivial integral $\int \di \eta=2\pi$.
Finally, using the general relation
\begin{equation}
    F=-\lim_{\vr_1 \rightarrow \vr_0} \lim _{\vr_2 \rightarrow \infty} \mathcal{S} \equiv \int dx^6 d\eta \mathcal{F}\,,
\end{equation} 
we arrive at renormalised result  for free energy density, 
valid for all the solutions in Sec.~\ref{Sec:UV},
\beqs
\mathcal{F}&=&
\frac{1}{120} e^{5 A_U-\frac{\chi _U}{\sqrt{10}}} \left(-10 \phi _{1,2} \phi _{1,4}+3
   \sqrt{10} \phi _{2,2}^3 \left(3 \log \left(\phi _{2,2}\right)-2\right)+\phi _{2,2}
   \left(3 \sqrt{10} \phi _{1,2}^2 \left(2-3 \log \left(\phi _{2,2}\right)\right)-10 \phi
   _{2,4}\right)+15 \sqrt{10} \chi _6\right)\,,
\eeqs
which in the case in which both the sources for the operators associated with the scalars $\phi_i$ vanish, with $\phi _{1,2}=\phi _{2,2}=0$, simplifies to the expression:
\beqs
\mathcal{F}&=&
\frac{\sqrt{10}}{8} e^{5 A_U-\frac{\chi _U}{\sqrt{10}}} \chi _6\,.
\eeqs

In the case of the soliton solutions discussed in this paper, using the expansion for small $z$ of the analytical solutions,
 we find that  $\chi_6=-\frac{\mu}{16\sqrt{10}}$, and hence
\beqs
\mathcal{F}&=&-\frac{\mu}{128}\,.
\eeqs
From the relation in Eq.~(\ref{Eq:Q1}) one knows that $\mu\geq0$, and hence $\mathcal{F}\leq 0$.
Conversely, the domain-wall AdS$_7$ solutions have $\chi_6=0$, and hence ${\cal F}=0$.
Hence, the soliton solutions have lower free energy than the domain-wall solutions, as long as they exist,
and phase coexistence may arise when $\mu=0=\chi_6={\cal F}$. We therefore find a zero-temperature phase transition between two types of solution. The first is the soliton solution given by the metric in Eq.~(\ref{metricAna}),  dual to a confining gauge theory. The second is the domain-wall, $AdS_7$ solution with metric in Eq.~(\ref{eq:metric}), which has $\c=\frac{\sqrt{10}}{8}\r$ and $A=\frac{5}{8}\r$, and is dual to a conformal field theory.

The critical line is defined by the requirement $\mu=0$, which is satisfied when $\vr_0^2=2-2 \cA_U^2$. In this case the metric can be 
 brought into the domain-wall form:
\beqs
\di s_7^2(\mu=0) &=&\frac{4}{\vr^2(H_1(\vr)H_2(\vr))^{4/5}} \di \vr^2 +\frac{1}{4}\vr^2(H_1(\vr)H_2(\vr))^{1/5}\left(\di x_{1,4}^2+\di\eta^2\right)\,,
\eeqs
 and furthermore
$\sin^2(2\theta)=\frac{(\cA_U^2-1)^2}{\cA_U^4}$. One then finds that 
$\vr_0^4=Q_1^2=Q_2^2=4(1-\cA_U^2)^2$.
In this case one arrives to the conclusion that the critical line 
in the first octant of the $\left(\cA_{7, U}^{(1)},\,\cA_{7, U}^{(2)}\right)$ plane, is defined by relation
\beqs
\sin(\theta)=-\frac{\sqrt{2 \cA_{ U}^2-1}-1}{2 \cA_{ U}}\,,
\eeqs
which can be recast  as $\cA_{7, U}^{(1)}+\cA_{7, U}^{(2)}=1$. Extending to the other quadrants by symmetry, one concludes that the condition
$\mu=0$ defines a square, with sides $\sqrt{2}$ and vertex on the axis. This is illustrated in Fig.~\ref{fig:F}, to which we will return shortly.

By exploiting the symmetries of the system, it is most insightful to look at the solutions in the two limiting cases in which
$\cA_{7, U}^{(1)}=\cA_{U}$ and $\cA_{7, U}^{(2)}=0$ and contrast it with the case
$\cA_{7, U}^{(2)}=\cA_{7, U}^{(1)}=\cA_{U}/\sqrt{2}>0$, as representative of the whole space of solutions.
In the former case, the two constraints can be rewritten as follows:
\beqs
\left(2\mu+\vr_0^6\right)^2 &=&16 \mu \vr_0^4\,,\\
\left(\cA_{7, U}^{(1)}\right)^2&=&
\frac{\vr_0^{4}\left(\vr_0^6-4\mu\right)}{16 \mu \vr_0^2}\,,
\eeqs
These admit four solutions, but, by imposing the reality constraints, $\vr_0^2>0$ and $\vr_0>0$,  one identifies uniquely the free energy.
It is a negative definite quantity for $\cA_{7, U}^{(1)}<1$, that vanishes when $\cA_{7, U}^{(1)}=1$.
For larger values of $\cA_{7, U}^{(1)}$ the globally stable solutions are domain-wall solutions with constant background field
$\cA_{7}^{(1)}$, vanishing $\phi_i$, and vanishing ${\cal F}$.

By contrast, setting $\sin^2(2\theta)=1$ yields the somewhat more complicated relations
\beqs
\left(2\mu+\vr_0^6-16\vr_0^2\left(\cA_{7, U}^{(1)}\right)^4\right)^2 &=&16 \mu \vr_0^4\,,\\
2\left(\cA_{7, U}^{(1)}\right)^2&=&
\frac{256 \vr_0^2 \left(\cA_{7, U}^{(1)}\right)^8
-32 \left(\cA_{7, U}^{(1)}\right)^4 (\vr_0^6+2\mu)+\vr_0^{4}\left(\vr_0^6-4\mu\right)}{16 \mu \vr_0^2}\,.
\eeqs
This system admits six solutions. Four of them have complex $\vr_0$, which is unphysical. Of the remaining two, we must choose the one with largest $\vr_0$, and ensure that it satisfies both $\vr_0>Q_1$ and $\vr_0>Q_2$.
This second condition restrict the space to  $\cA_U<1/\sqrt{2}$, as outside the square defined by $\mu=0$ one finds that neither of the two values of $\vr_0$ is larger than both $Q_1$ and $Q_2$.
As a result the qualitative shape of the free energy is closely resembling the aforementioned $\theta=0$ case, with
${\cal F}<0$ for $\cA_{7, U}^{(2)}=\cA_{7, U}^{(1)}<\frac{1}{2}$, and vanishing when $\cA_{7, U}^{(2)}=\cA_{7, U}^{(1)}=\frac{1}{2}$.
For larger values of $\cA_{7, U}^{(2)}=\cA_{7, U}^{(1)}$ the globally stable solutions are domain-wall solutions with constant background fields
$\cA_{7}^{(2)}=\cA_{7}^{(1)}$, vanishing $\phi_i$, and vanishing ${\cal F}$.

To reinstate physical units, it is convenient to define the following scale, given by the time it takes for light to reach the end of space from infinity, travelling along the holographic direction, $\vr$~\cite{Csaki:2000cx}:
\beqs
\Lambda^{-1} &\equiv&
\int_{\rho_0+\r_{1}}^{\infty} \di\rho\, e^{\frac{\chi}{\sqrt{10}}-A}
\,=\,
 \int_{\vr_0+\vr_{1}}^{+\infty} \di \vr\,\frac{2}{\vr\sqrt{f(\vr)}}\,,
\eeqs
where we require the inclusion of an IR cutoff, $\vr_{1}$, in the lower limit of the integration to avoid a divergence at the end of space. We numerically intergrate and take $\vr_{1}=10^{-6}.$

We illustrate the behavior of the free energy in Fig.~\ref{fig:F}.
The left-hand panel shows the phase diagram, in the  $\left(\cA_{7, U}^{(1)},\,\cA_{7, U}^{(2)}\right)$ plane, together with a selection of lines obtained by holding fixed the ratio $\left(\cA_{7, U}^{(1)}/\cA_{7, U}^{(2)}\right)$.
The phase diagram is characterised by lines of first-order phase transitions forming a polygon (a square),
as expected on the basis of the symmetries of the system.
The right-hand panel shows the free energy density expressed in terms of the scale $\Lambda$, as $\hat{\cal F}={\cal F} \Lambda^{-6}$. The resulting plot displays a continuous curve, that is not differentiable at the transition.
As expected, the endpoints of the confining branches obtained at fixed source ratio, $\tan(\q)$, take the form of domain-wall solutions. Unlike their $AdS_7$ counterparts, these solutions are singular at the end of space. This distinction results in the coexistence of two separate configurations, a hallmark of first-order phase transitions, as is the non-differentiable character of the free energy at the transition.~\footnote{First-order phase transitions are often accompanied also by the existence of metastable and unstable branches of solutions, that are not realised at equilibrium, but can be  produced under controlled conditions studied in details. This is not the case in this example: there are other solutions to the algebraic constraint defining $\vr_0$, as discussed earlier in the manuscript, but as we anticipated these additional solutions cannot be interpreted as background geometries, and hence do not have a dual field theory interpretation. This is at odds with what found for other theories---see, e.g., Ref.~\cite{Elander:2025fpk}---in which the presence of  metastable and unstable branches  leads to the typical swallow-tail diagrams in sketches of the free energy as a function of the sources.}
The presence of a singularity indicates the incompleteness of the classical supergravity treatment, large curvature effects emerging in close proximity of the  phase transitions---see Appendix~\ref{Sec:gravityinvariants}.

%%%%%%%%%%%%%%%%%%%%%%%
%%%%%%%%%%%%%%%%%%%%%%%
\section{Local stability analysis: fluctuation Spectra} 
\label{Sec:MassSpectra}

\begin{figure*}[t] 
    \centering
    \subfloat[$\th=0$ \label{Fig:Spectrum 0}]{\includegraphics[width=0.45\linewidth]{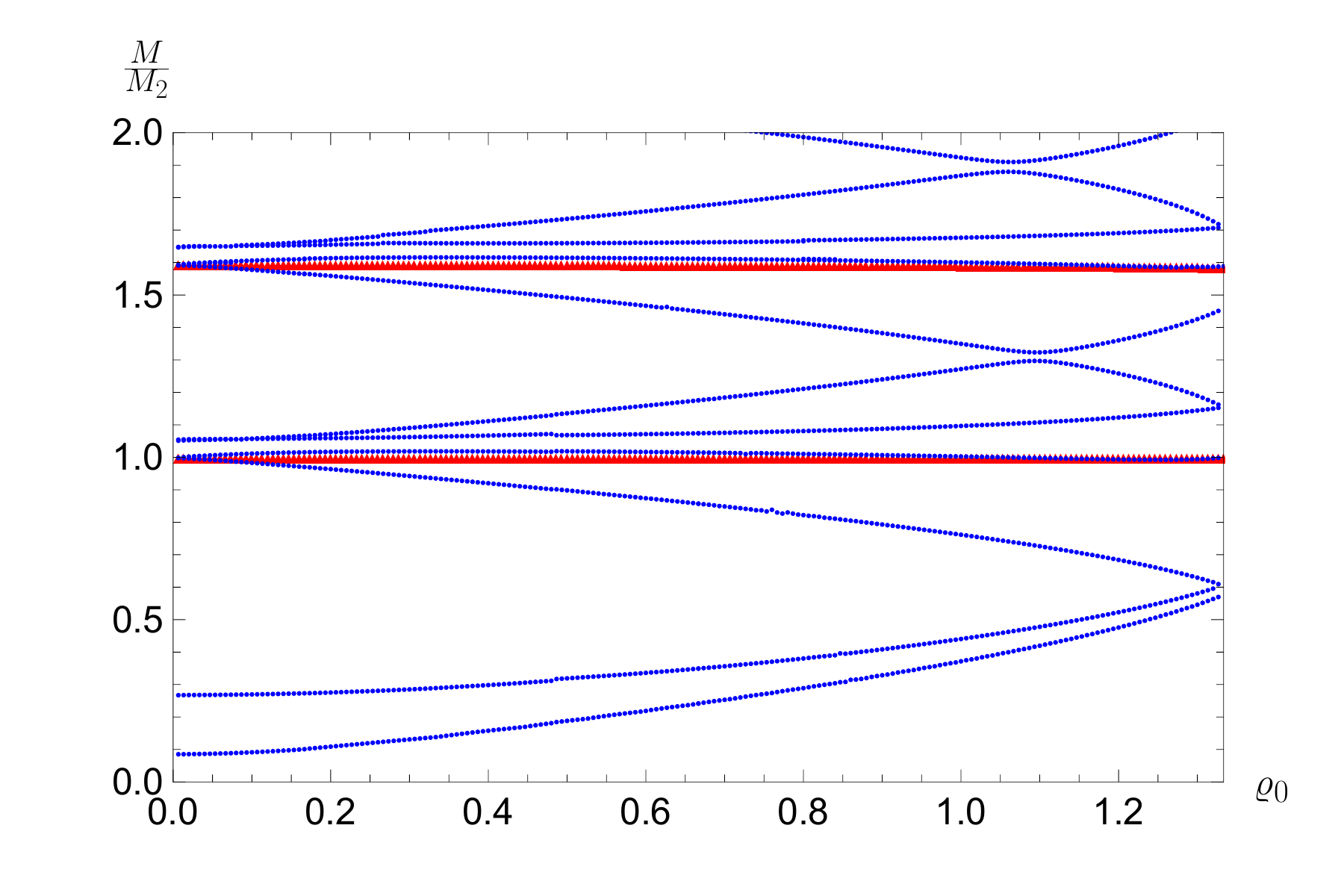}}
    \subfloat[$\th=\frac{\p}{9}$ \label{Fig:Spectrum pi/9}]{\includegraphics[width=0.45\linewidth]{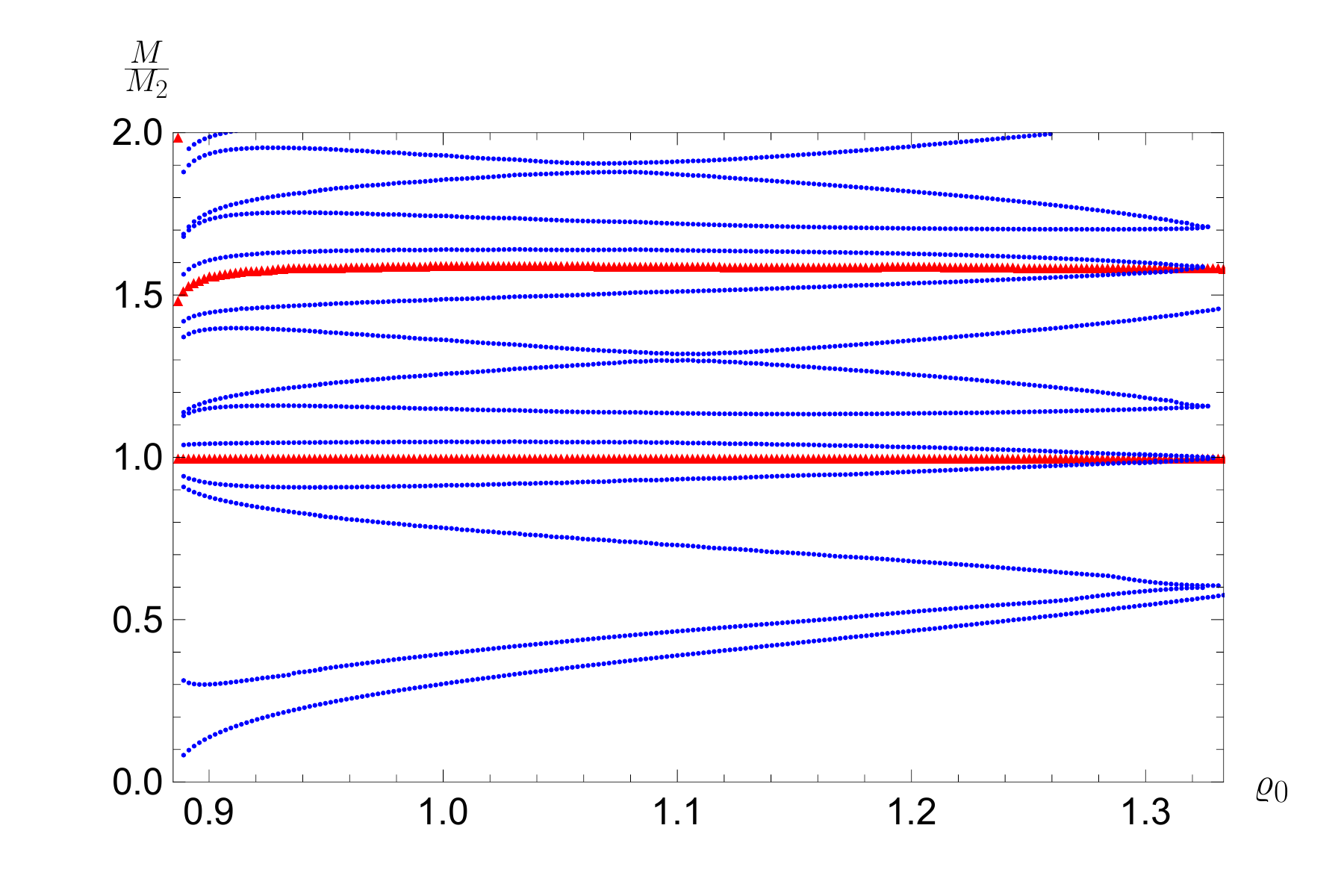}}\\
    \subfloat[$\th=\frac{\p}{6}$ \label{Fig:Spectrum pi/6}]{\includegraphics[width=0.45\linewidth]{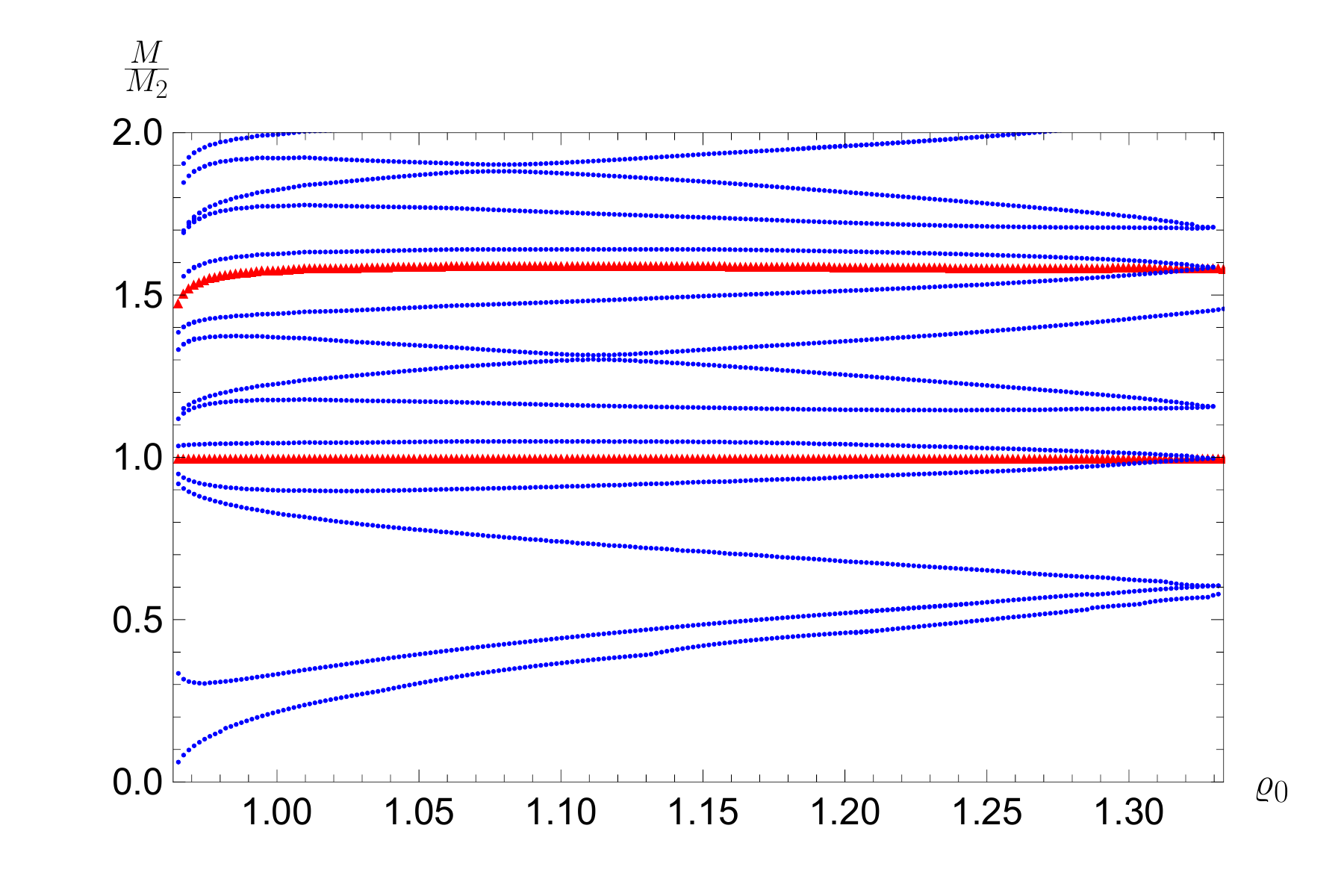}}
    \subfloat[$\th=\frac{\p}{4}$ \label{Fig:Spectrum pi/4}]{\includegraphics[width=0.45\linewidth]{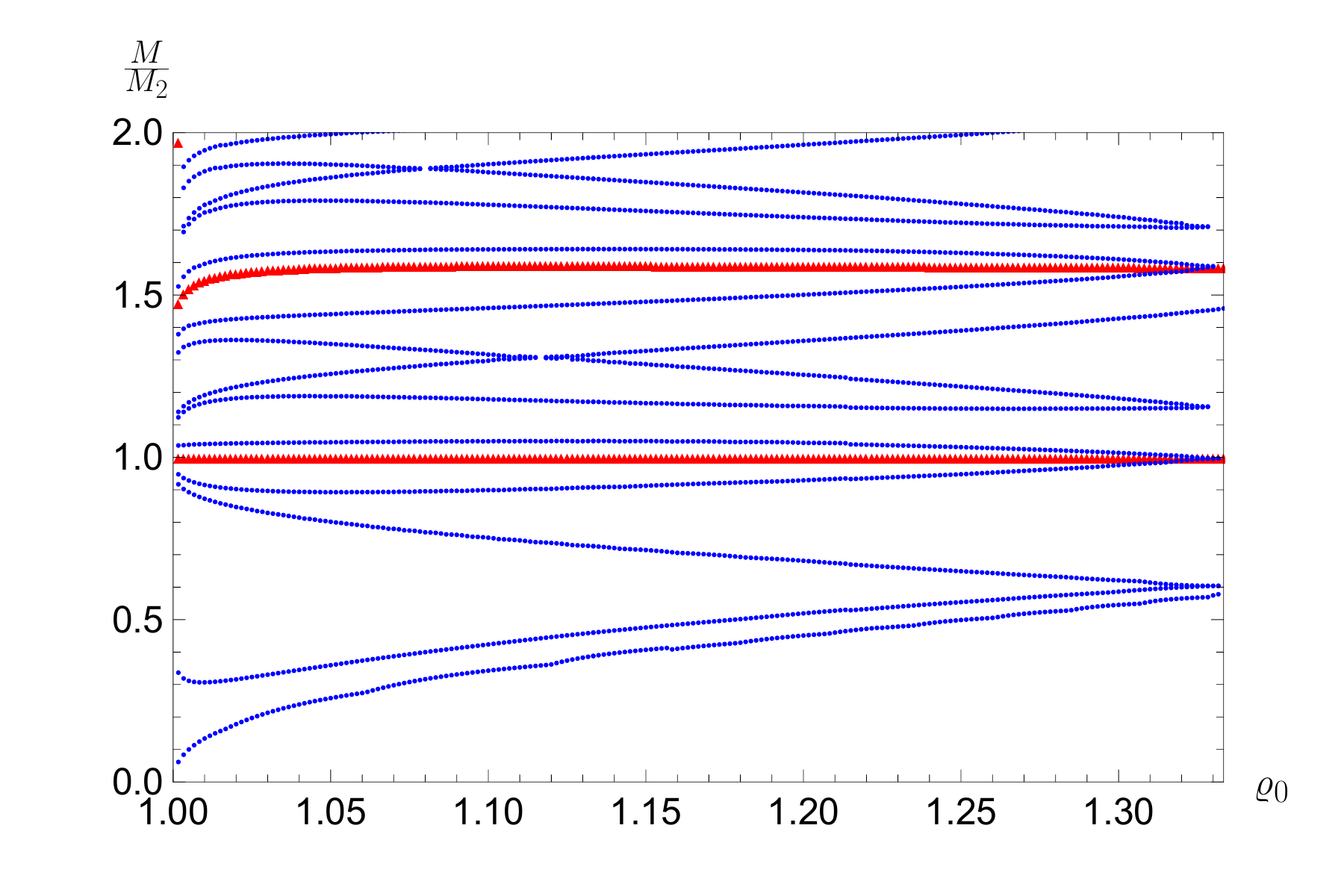}}\\
    \subfloat[Probe approximation, $\th=0$ \label{Fig:Spectrum probe}]{\includegraphics[width=0.45\linewidth]{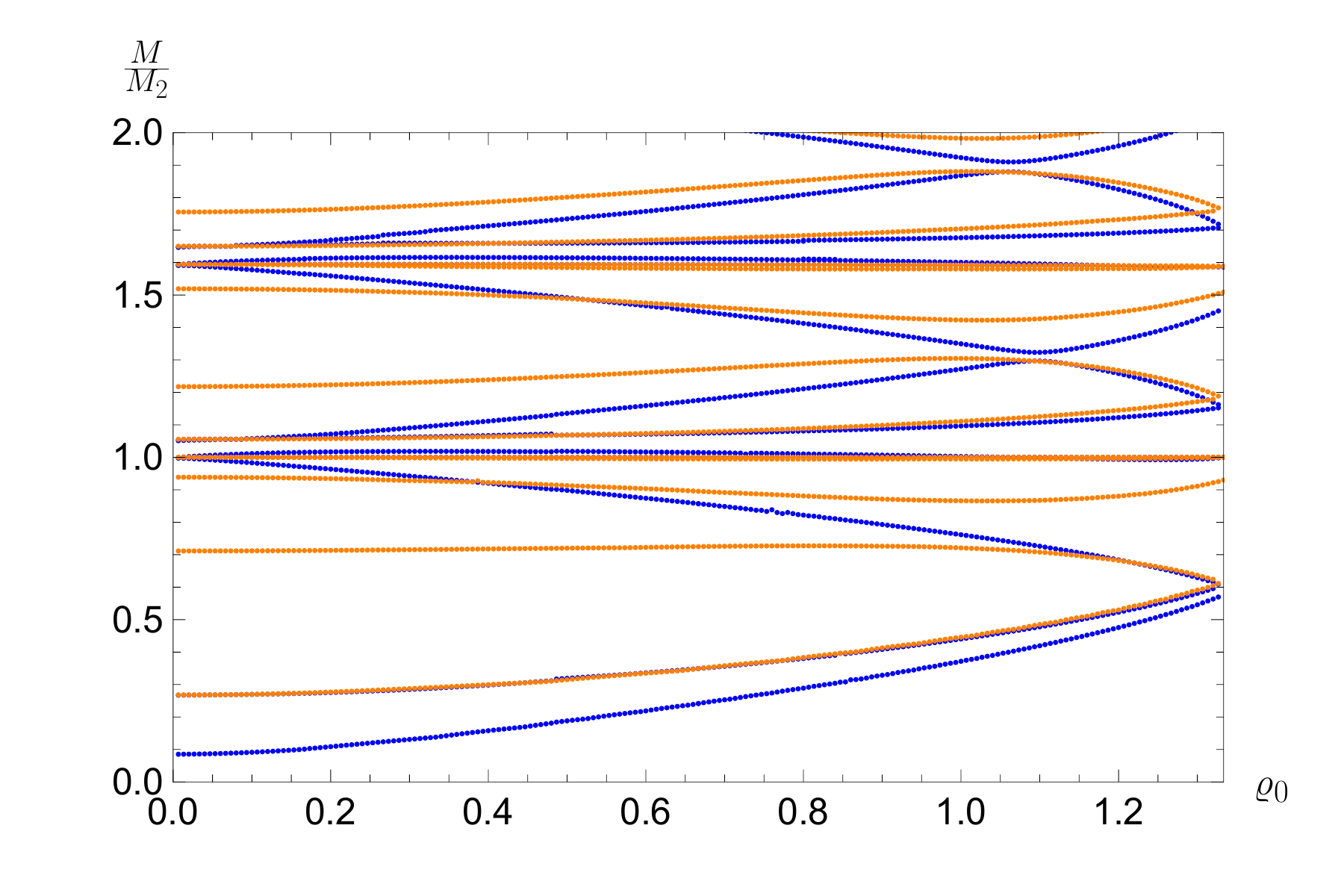}}
    \subfloat[Universal energy scale\label{Fig:energy scale}]{\includegraphics[width=0.45\linewidth]{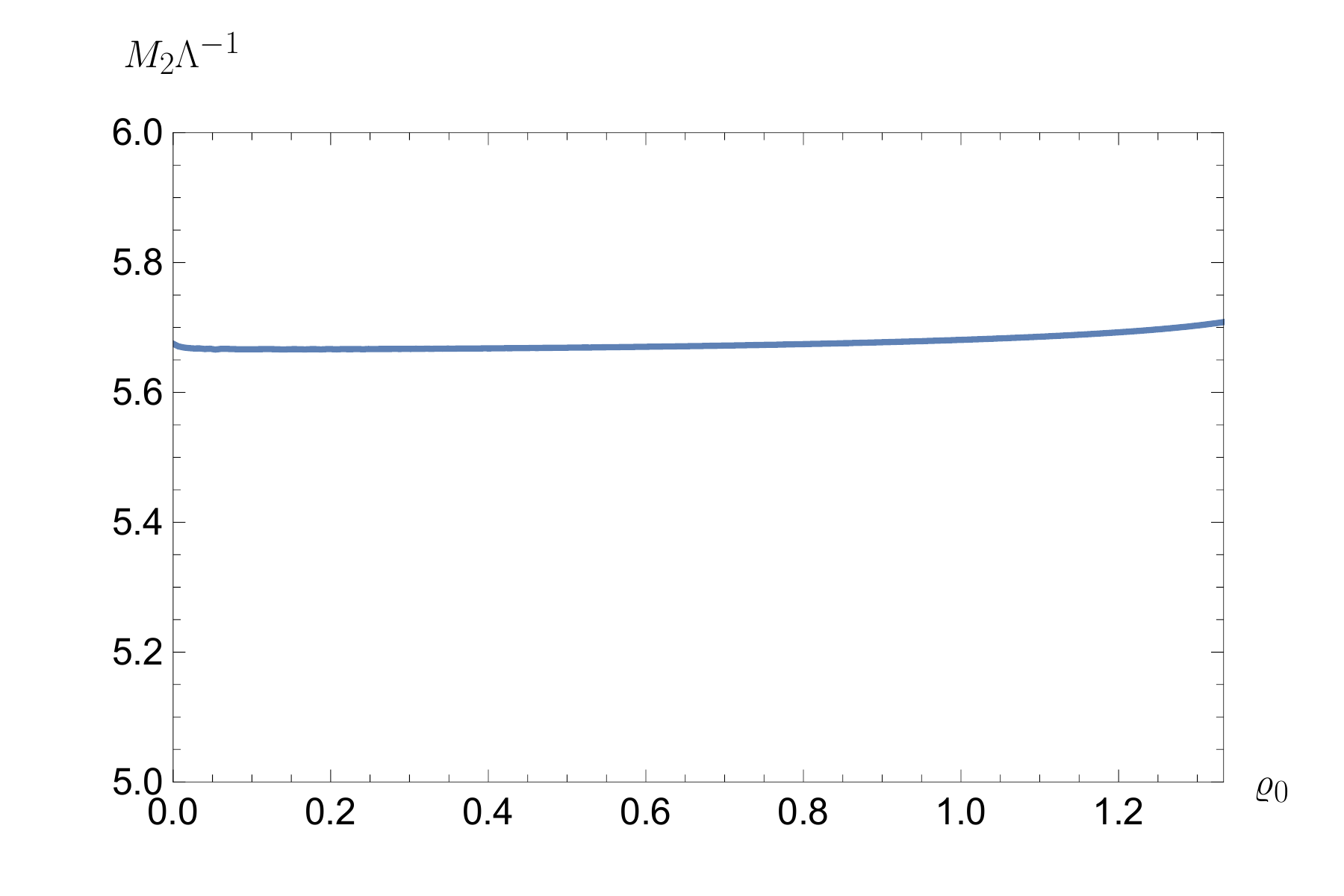}}
    
    \caption{Top four panels: mass spectra, normalised to the mass of the lightest spin-2 fluctuation, $M_2$, as a function of $\varrho_0$, for four  examples of one-parameter subclasses of soliton (confining) solutions, obtained by fixing the ratio of the two magnetic fluxes---see the free energy of the same 
    solutions  in Fig.~\ref{Fig:PhaseDiagram}.
 Spin-2 states are shown as (red) triangles and spin-0 as (blue) disks.  The far left of each plot corresponds to points along the line of first order phase transitions in Fig.~\ref{Fig:PhaseDiagram}. Fifth panel: comparison of the spin-0 states with the result of using the probe approximation, shown with the orange rectangles, for solutions with $\theta=0$. The probe approximation entirely misses the lightest state, suggesting this state contains a substantial contribution from the dilaton. The numerical results are obtained for 
 $\varrho_{IR}=10^{-6} \varrho_0$. Sixth panel: ratio of the mass of the lightest spin-2 states, $M_2$, to the universal energy scale, $\L$,  for the solutions in which $\q=0$; the ratio is roughly constant across the parameter space, suggesting either of the two observables could be used to set the physical scale, and compare between theories of this class.
 }
    \label{fig:spectra}
\end{figure*}

We perform  a local stability study of the soliton solutions, restricted to the region inside the square in Fig.~\ref{fig:F}, 
in which they are regular and energetically favored. 
To this purpose, we compute  the mass spectrum 
of fluctuations of the scalar and tensor fields in the gravity description, along the soliton solutions, which we interpret as bound states of the dual field theory (in Yang-Mills theories, these would be identified with glueballs).
We adopt the gauge invariant formalism developed in Refs.~\cite{Bianchi:2003ug,Berg:2005pd,Berg:2006xy,
Elander:2009bm,Elander:2010wd,Elander:2010wn}, for both spin-0 and spin-2 cases---see also Refs.~\cite{Elander:2014ola,Elander:2018aub,
Elander:2020csd,Elander:2024lir}, and Appendix~\ref{Sec:gaugeinvariantformalism}, in which we fix and explain the notation.

Using the holographic direction $r$, the gauge-invariant combinations of the scalar fluctuations obey the following coupled  equations:
\begin{equation}\label{eq:Scalar EoM}
	0 = \left[ \mathcal{D}^2_r + 5\partial_r A \mathcal{D}_r  + e^{-2A}M^2 \right] \mathfrak{a}^a
	- \left[ V^a _{|c} - R_{bcd}^a \partial_r {\Phi}^b \partial_r {\Phi}^d + \frac{(\partial_r {\Phi}^a V^b + V^a \partial_r {\Phi}^b)G_{bc}}{\partial_r A} + \frac{ V \partial_r {\Phi}^a \partial_r {\Phi}^b G_{bc}}{(\partial_r A)^2} \right] \mathfrak{a}^c,
\end{equation}
subject to the boundary condition:
\begin{equation} \label{eq:fluctuationboundarycond}
    \6_r\Phi^a\6_r\Phi_b\mathcal{D}_r\mathfrak{a}^b|_{r_i}=\left[-2\6_rAe^{-2A}M^2\delta_b^a+\6_r\Phi^a\left(\frac{V}{\6_rA}\6_r\Phi_b+V_b\right)\right]\mathfrak{a}^b|_{r_i},
\end{equation}
where $\mathfrak{a}^a$ are the five scalar fluctuations, and $r_i=\{r_{1}, r_{2}\}$ are the cutoffs we impose along the holographic direction,  in the IR, near the end of space, $r_0$, and in the far UV. It is understood that at the end of the calculation the results are extrapolated to the limits $r_1\rightarrow r_0$ and $r_2\rightarrow +\infty$, respectively. We denote as $M^2$ the mass squared of the fluctuation, in the dual five-dimensional confining theory. The full expressions for the fluctuation equations and boundary conditions written in terms of holographic variable $\r$ are given in Appendix~\ref{Sec:Fluctuations}. 

In our numerical study, we use Eq.~(\ref{eq:fluctuationboundarycond}) to set the IR boundary conditions for the evolution of the linearised equations. In order to improve the convergence of the numerical results we find it useful to impose the UV boundary conditions on asymptotic expansions of the scalar fluctuations, and then match the resulting, constrained expansions to the numerical solutions of the fluctuation equations at the holographic coordinate $\rho=\rho_2$. The UV expansions are presented explicitly in  Appendix~\ref{Sec:UVexpansions}.

The spin-2, tensor fluctuations, $\mathfrak e^\mu{}_\nu$,  obey the simpler differential equations
\beqs
\label{eq:tensoreom}
	0&=& \left[ \partial_r^2 + (D-1) \partial_r {A} \partial_r + e^{-2{ A}(r)} M^2 \right] \mathfrak e^\mu_{\,\,\,\nu}, \nn \\
    &=& \left[e^{2A(\r)-\sqrt{\frac{2}{5}}\c(\r)}\6_{\r}^2+5e^{2A(\r)-\sqrt{\frac{2}{5}}\c(\r)}\6_{\r}A\6_{\r}+M^2 \right] \mathfrak e^\mu_{\,\,\,\nu} \, ,
\eeqs
with boundary conditions given by 
\beq
\label{eq:tensorbc}
\left.\frac{}{}	\partial_\r \mathfrak e ^\mu_{\,\,\,\nu} \right|_{\r=\r_i}= 0 \,.
\eeq

We performed an extensive numerical study of spectra, focusing on four representative branches of soliton solutions,
characterised by four choices of the angle controlling the ratio between the two Abelian fluxes, $\theta=0,\,\pi/9,\,\pi/6,\,\pi/4$,
 as illustrated in Fig.~\ref{fig:spectra}.
 In all cases, we repeated the calculations for several choices of the IR and UV cutoff, looking to minimise residual numerical artefacts, within the limits of numerical precision of the calculations. We notice that the improvement makes the convergence in the UV so fast that no discernible corrections could be found by changing the UV cutoff within reasonable ranges.
We did not have a similar process available for the IR, yet we found that  by solving the equation rewritten in the variable $\vr$, and choosing
 $\varrho_{IR}=10^{-6} \varrho_0$ as the IR cutoff, the results are a good approximation to the exact ones, for the purposes of this publication. We present in Appendix~\ref{Sec:ircutoffdependence} examples demonstrating the degree of convergence of the results, as a function of the  IR cutoff. 

The results, shown in Fig.~\ref{fig:spectra}, can be summarised as follows. Firstly, we do not find evidence of local instabilities, in the form of tachyons, anywhere in the relevant portion of parameter space. This is to be contrasted with the results of Refs.~\cite{Elander:2020ial,Elander:2020fmv,Elander:2021wkc}, in which the arising of a first-order phase transition is always accompanied by the presence of a tachyonic state that emerges along the same branch of solutions, but for choices of tunable parameters past the phase transition.
In the case at hand,  the transition is happening at the end of the physical branches of solutions of interest, beyond  which background solutions  of this class do not exist.

The five scalar and one tensor gauge-invariant fluctuations all lead to discrete mass spectra, with typical spacing and scale  well represented by the lightest of the spin-2 states.\footnote{As displayed by the last panel of Fig.~\ref{fig:spectra}, we could has well have normalised the masses with the scale $\Lambda$, with no significant qualitative change to the overall emerging picture, as the ratio $M_2/\Lambda$ is effectively a constant.}  The two lightest scalar modes are the only ones that are substantially lighter than the lightest spin-2 particle, ranging from $0.6$ down to $0.05$ and $0.3$ for the lightest and next-to-lightest state, in units of the tensor mass, $M_2$. This is happening over a sizeable portion of parameter space, not just in proximity of the transition, and deserves further discussion, after we present the results 
we obtained with the probe approximation, in the next section. We observe here that this behavior is also in sharp contrast to Refs.~\cite{Elander:2020ial,Elander:2020fmv,Elander:2021wkc}, that show no particular suppression of the mass of the scalar bound states along the physically realised portion of the branches of solutions, before the inset of the phase transition---in the cases in the literature, a suppression of the mass of the lightest scalar is observed only along the metastable portion of the branches of solutions, past the transition itself.

%%%%%%%%%%%%%%%%%%%%%%%
%%%%%%%%%%%%%%%%%%%%%%%
\subsection{Probe approximation} 
\label{Sec:probe}

We now turn to examining the extent to which the scalar fluctuations discussed in the previous section couple as the dilaton,  with particular interest to the lightest scalars and to the regions of parameter space close to the phase transition. To investigate this point, we adopt the probe approximation, following the approach and notation of Ref.~\cite{Elander:2020csd}.
The key insight underlying this approximation, and the results presented in this section, stems from the definition of gauge-invariant scalar combinations given in Eq.~(\ref{eq:scalarfluc}). These combinations consist of two additive terms: the first, $\varphi^a$, represents small fluctuations of the sigma-model scalar fields around their background values; the second is proportional to $h$, the trace of the five-dimensional metric fluctuations which couples to the trace of the boundary stress-energy tensor, the dilatation operator.
The probe approximation refers to computing the spectrum while deliberately neglecting the contribution of $h$ to the gauge-invariant combinations. In doing so, the resulting spectrum reflects only the dynamics of the scalar fluctuations, $\varphi^a$. 

While this is an approximation, and therefore not physically complete (in particular, we are explicitly breaking gauge invariance) it is nonetheless useful for diagnostic purposes. If the probe approximation yields a spectrum that closely matches the result from the full gauge-invariant analysis, we can infer that the metric fluctuations, $h$, play a negligible role. This would imply that the scalar fluctuations are not significantly coupled to the dilaton operator of the dual field theory (the trace of the stress-energy tensor). Conversely, the emergence of substantial differences between the two results for the  spectra would indicate that the affected scalar modes are the result of substantial mixing with the dilaton.

In the probe approximation, the equations of motion for the fluctuations simplify considerably and take the following form:
\begin{equation}\label{eq:Scalar Probe EoM}
	0 = \left[ \mathcal{D}^2_r + 5\partial_r A \mathcal{D}_r  + e^{-2A}M^2 \right] \mathfrak{p}^a
	- \left[ V^a _{|c} - R_{bcd}^a \partial_r {\Phi}^b \partial_r {\Phi}^d \right] \mathfrak{p}^c,
\end{equation}
whilst the boundary conditions reduce to Dirichlet:
\begin{equation}
    0=\mathfrak{p}^a|_{r_i}.
\end{equation}
We report the expression for the equations in the $\rho$ variable in Appendix~\ref{Sec:probeeqs}.
The results of our analysis are displayed in the fifth panel of Fig.~\ref{fig:spectra}, for the choice $\theta=0$ in which only one of the two Abelian fields is non-trivial in the background. The comparison between the probe approximation and the result of the complete calculation with gauge-invariant variables is overall quite good: particularly for the largest values of $\varrho_0$ available, and for the mass of the heaviest fluctuations, the probe approximation works well, demonstrating that the majority of states we identified have little contamination from the dilaton.

Focusing on the two lightest states shows a striking contrast. The next-to-lightest state is always well captured by the probe approximation, demonstrating that it has nothing to do with dilaton and scale invariance. The lightest state is entirely missed, demonstrating its nature as mostly an approximate dilaton. This is possibly the most striking example of the probe approximation failing completely to identify the mass of a state---see the catalogue of examples in Ref.~\cite{Elander:2020csd}---a phenomenon that persists throughout the whole parameter space, not just near the phase transition.

This observation adds to the distinctive peculiarity of this theory.
We find evidence of first-oder phase transitions, but not of the transition becoming weaker, or second-order (as was the case in the results discussed in Refs.~\cite{Faedo:2024zib,Elander:2025fpk}), anywhere in the region of parameter space that is open to exploration. Similarly to what emerged in the theories in Refs.~\cite{Elander:2020ial,Elander:2020fmv,Elander:2021wkc} and the bottom-up models in Refs.~\cite{Elander:2022ebt,Fatemiabhari:2024lct}, we find that there is no sense in which a parametrically light dilaton emerges near the transition---in which region the geometry exhibits large curvature, undermining the classical supergravity treatment.
And yet, we find that, for $\theta=0$, in a large portion of parameter space that  extends to the whole branch of confining solutions, the lightest state is a dilaton, and its mass can be as much as one order of magnitude suppress in respect to the typical scale set by confinement. To emphasise this finding we can compare the fifth panel of Fig.~\ref{fig:spectra} to the figures in Ref.~\cite{Elander:2022ebt}, particularly the summary Figure~10. In this earlier publication, a whole class of bottom-up models were demonstrated to exhibit a first-order phase transition, accompanied by the existence of a parametrically light dilaton, along the metastable branches of solutions. Yet, along the physically realised, stable branches of solution, the  dilaton is never parametrically light. Its mass is minimised in closest proximity to
the transition. The ratio between mass of the dilaton and of lightest spin-2 state is found to be $M_d/M_2\gsim 0.36$. In sharp contrast, in the theory discussed in this paper, this ratio is much smaller, $M_d/M_2\sim 1/10$, over large parts of parameter space. Furthermore, this interesting behaviour emerges without having to tune any of the parameters, contrary to earlier examples in the literature.

%%
%%%%%%%%%%%%%%%%%%%%%%%
%%%%%%%%%%%%%%%%%%%%%%%
\section{Outlook} 
\label{Sec:outlook}

This research sits at the confluence of two ongoing research programmes. On the one hand, it is part of the systematic exploration, by means of holography, of confining field theories for which it is possible to compute the spectrum of bound states in regions of parameter space near phase transitions, as formulated in the outlook of Ref.~\cite{Elander:2020ial}---see also Refs.~\cite{Elander:2020fmv,Elander:2021wkc,Elander:2025fpk} in the top-down approach to holography, and Refs.~\cite{Elander:2022ebt,Fatemiabhari:2024lct,Faedo:2024zib}. The aim of this programme is to understand under what circumstances a light dilaton emerges as a bound state, and hence to provide the foundations for dilaton effective field theory and its applications.
On the other hand, the work in Ref.~\cite{Anabalon:2021tua} identified a new solution-generating technique within top-down supergravity---see also Refs~\cite{Nunez:2023xgl, Nunez:2023nnl,Fatemiabhari:2024aua,Chatzis:2024top,Chatzis:2024kdu,Chatzis:2025dnu,Chatzis:2025hek,Anabalon:2025sok}---giving access to new families of non-trivial holographic descriptions of confining field theories. We are here using this solution-generating technique as part of the tools deployed for the aforementioned exploration of the relation between dilaton and phase transitions.

We have presented the results of an extensive global and local stability analysis of a two-parameter family of regular classical solutions of maximal supergravity in $D=7$ dimensions with $SO(2)\times SO(2)$ gauge symmetry. Solutions in this class take the soliton form, and admit an elegant interpretation in terms of a  strongly coupled confining  field theory obtained by circle compactification of the ${\cal N}=(2,0)$ superconformal theory in $D-1=6$ dimensions. The global stability analysis makes use of the holographically renormalised free energy, and confirms the presence of a first-order phase transition in the space of solutions. In a compact region of  parameter space, inside a square, the soliton/confining family is energetically favoured, in respect to a competing family of AdS$_7$ supergravity solutions with constant gauge field that take the domain-wall form and are physically realised outside of the square. The local stability analysis, performed by computing gauge invariant fluctuations of the spin-0 and spin-2 supergravity fields, confirms that there are no further instabilities, all the accessible states having non-zero, positive mass squared.

The main element of novelty emerging from this work is that the spectrum of bound states of the dual field theory  (the dilaton in particular) behaves in an unusual way, in respect to former examples found in the literature. In the two known cases in which a parametric suppression of the dilaton mass has been found~\cite{Faedo:2024zib,Elander:2025fpk}, there is evidence of a nearby critical point at which a phase transition becames of second order, which is not the case for the theories discussed here.
The dilaton we identify is not parametrically light.
Yet, the suppression of its mass, with respect to the mass of the lightest spin-2 bound state, $M_d/M_2\sim 1/10$, is strong enough that if it could be engineered to take place in a realistic new physics model with composite dynamics it would be sufficient to avoid most exclusion bounds---see the discussions of the misalignment angle in Refs.~\cite{Panico:2015jxa,Cacciapaglia:2020kgq} and references therein.
Furthermore, the suppression of the dilaton mass persists over a large portion of parameter space, extending far away from the first-order phase transition characterising the system, without requiring to fine-tune the parameters.

This finding marks important progress towards building a complete picture of the conditions upon which strongly coupled, confining gauge theories can give rise to a light dilaton in their spectrum, separated in mass from the other bound states of the theory. It also highlights how our current understanding of the underlying mechanism is still incomplete, warranting a systematic exploration of other holographic models, in other dimensions, and in theories  with less symmetry. The solution generating technique we borrowed from Ref.~\cite{Anabalon:2021tua} is proving a useful tool in this direction. But our study opens fundamental questions about the relation of the magnetic fluxes to the emergence of a light dilaton, and its relation to approximate scale invariance.   It would be desirable to have at disposal other such techniques to explore the broader space of relevant top-down holographic theories, and we envision starting an extensive exploration of the space of similar theories, to address the aforementioned open questions.

%%%%%%%%%%%%%%%%%%%%%%%
%%%%%%%%%%%%%%%%%%%%%%%

\begin{acknowledgments}

We would like to thank Ali Fatemiabhari and Carlos Nunez for useful discussions.

The work of  MP and JR has been supported in part by the STFC  Consolidated Grants 
No. ST/P00055X/1, No. ST/T000813/1, and ST/X000648.
MP received funding from the European Research Council (ERC) under the European 
Union’s Horizon 2020 research and innovation program under Grant Agreement No.~813942. 

JR is supported by the STFC DTP with contract No. ST/Y509644/1.

\vspace{1.0cm}
{\bf Research Data Access Statement}---The data generated for this manuscript can be downloaded from  Ref.~\cite{piai_2026_18633529}. 
\vspace{1.0cm}

{\bf Open Access Statement}---For the purpose of open access, the authors have applied a Creative Commons 
Attribution (CC BY) licence  to any Author Accepted Manuscript version arising.

\end{acknowledgments}

%%%%%%%%%%%%%%%%%%%%%%%
%%%%%%%%%%%%%%%%%%%%%%%
\appendix

%%%%%%%%%%%%%%%%%%%%%%%
%%%%%%%%%%%%%%%%%%%%%%%
\section{Sigma model coupled to gravity, generalities}
\label{Sec:sigma-model}

We report here the general equations defining a sigma-model coupled to gravity, by adopting the notation of Refs.~\cite{Elander:2010wd,Elander:2010wn}---see also Refs.~\cite{Bianchi:2003ug,Berg:2005pd,Berg:2006xy,Elander:2009bm,Elander:2014ola,Elander:2018aub,
Elander:2020csd,Elander:2024lir}. In general $D$ dimensions, we write the two-derivative sigma-model
action, for real field $\Phi^a$, with $a=1,\,\cdots,\,n$,  as follows:
\beqs
\label{eq:sigmamodelaction}
	\mathcal S_D &=& \int \dd^D x \sqrt{-g_D} \, \left\{ \frac{\cal R}{4} - \frac{1}{2} g^{MN} G_{ab} \partial_M \Phi^a \partial_N \Phi^b - \mathcal V_D(\Phi^a) \right\}\nonumber \,,
\eeqs
where $G_{ab}$ is the  metric in the sigma-model space, and $\mathcal V_D(\Phi^a)$ is the potential. 
We use the following conventions for gravity. Firstly the Christoffel symbols are defined as follows
\begin{equation}
\Gamma^{P}{}_{MN}
= \tfrac12\, g^{PQ}
\left(
\partial_{M} g_{NQ}
+ \partial_{N} g_{MQ}
- \partial_{Q} g_{MN}
\right)\,.
\end{equation}
The Riemann tensor is
\begin{equation}
\cR_{MNP}{}^Q \equiv \partial_{N}\Gamma^{Q}{}_{MP}
- \partial_{M}\Gamma^{Q}{}_{NP}
+ \Gamma^{S}{}_{MP}\Gamma^{Q}{}_{SN}
- \Gamma^{S}{}_{NP}\Gamma^{Q}{}_{SM}\,,
\end{equation}
and the Ricci tensor and scalar are obtained by contracting the indices, with
$
\cR_{MN}
= \cR_{MPN}{}^P
$, and $
\cR = \cR_{MN} g^{MN}$.

We assume that the background geometry is characterised by the domain wall ansatz for the metric:
\beqs
\di s_D^2 &=& e^{2A} \di x_{1,D-2}^2 + \di r^2\,,
\eeqs
where $A=A(r)$. Similarly, we allow the scalars to have a non-trivial profile depending only on the holigraphic direction, 
with $\Phi^a =  \Phi^a(r)$ in the background.
In the calculations, we assume the holographic direction to be bounded, as in $r_1<r<r_2$. The two boundaries act, respectively, as an infrared (IR) and ultraviolet (UV) regulator, with it understood that the limits $r_1\rightarrow r_0$ and $r_2\rightarrow +\infty$ have to be taken at the end of the calculation, to recover physical results that do not depend on the regulators.
The general expression for the equations of motions, satisfied by the background scalars in general $D$ dimensions,
 are the following:
\beqs
\label{eq:backgroundEOM1}
\partial_r^2\Phi^a\,+\,(D-1)\partial_r {A}\partial_r\Phi^a\,+\, {\cal G}^a_{\,\,\,\,bc}\partial_r\Phi^b\partial_r\Phi^c\,-\,\mathcal V^a
&=&0\,,
\eeqs
where the sigma-model derivatives are given by $\mathcal V^a\equiv  G^{ab}\partial_b \mathcal V$, and  
$\partial_b \mathcal V\equiv \frac{\partial \mathcal V}{\partial \Phi^b}$. The Einstein equations reduce to
\beqs
\label{eq:backgroundEOM2}
(D-1)(\partial_r { A})^2\,+\,\partial_r^2 { A}\,+\,\frac{4}{D-2} \mathcal V &=&
0\,,\\
(D-1)(D-2)(\partial_r { A})^2\,-\,2 G_{ab}\partial_r\Phi^a\partial_r\Phi^b\,+\,4 \mathcal V&=&0\,.
\eeqs

%%%%%%%%%%%%%%%%%%%%%%%
\section{Superpotential and supersymmetric solutions}\label{Sec:super}

In this Appendix, we report and discuss  the superpotential of the theory in $D=7$ dimensions, and its role in the process of holographic renormalisation for field-theory observables computed in general, dual gravity background solutions. 
We start from the domain-wall ansatz for the metric, which here we write as
\beqs
\di s_{D}^2&=&\di \rho^2+e^{2{\cal A}}\di x_{1,D-2}^2\,.
\eeqs
If the scalar  potential, ${\cal V}_D$, can be written as
\beqs
\label{Eq:super}
{\cal V}_D&=& \frac{1}{2} G^{ab}\frac{\partial {\cal W}_D}{\partial \Phi^a} \frac{\partial {\cal W}_D}{\partial \Phi^a} -\frac{D-1}{D-2}\left( {\cal W}_D\right)^2\,,
\eeqs
for a superpotential, ${\cal W}_D={\cal W}_D(\Phi^a)$, that depends only on the scalar fields, then any solution of the 
following differential equations
\beqs
\partial_{\rho} {\cal A} &=&-\frac{2}{D-2}{\cal W}_D\,,\\
\partial_{\rho} \Phi^a&=&G^{ab}\frac{\partial{\cal W}_D}{\partial \Phi^b}\,
\eeqs
is also a solution to the classical equations derived from the potential.
In these equations, the background value of both ${\cal A}$ and the scalars, $\Phi^a$,  depend only on $\rho$.

In the case at hand, with $D=7$, the potential admits the following approximation, for small fields:
\beqs
{\cal V}_7
   &\simeq & -\frac{15}{8} \,-\,\frac{1}{4} \left(\phi_1^2+\phi_2^2\right)\,+\,\cdots\,,
   \eeqs
   where we expanded in powers of the (small) scalar fields, 
   $\phi_{1,2}$, starting from the AdS$_7$ solutions with $\phi_i=0$ and ${\cal A}={\cal A}_0+\frac{1}{2}\rho$. 
  There are  four possible superpotentials that satisfy the defining relation in Eq.~(\ref{Eq:super}):
 \beqs
 {\cal W}_7&=&-\frac{5}{4}-\frac{3\pm 1}{16}\phi_1^2-\frac{3\pm 1}{16}\phi_2^2\,+\cdots\,.
 \eeqs
 By choosing the minus sign in both quadratic terms, so that 
$ {\cal W}_7=-\frac{5}{4}-\frac{1}{8}(\phi_1^2+\phi_2^2)\,+\cdots$, one finds that the first-order equations are
 \beqs
 \partial_{\rho} {\cal A}&=&\frac{1}{2} \,+\,\cdots\,,\\
 \partial_{\rho} \phi_{1,2} &=&-\phi_{1,2}\,+\,\cdots\,.
 \eeqs
By setting $z=e^{-\rho/2}$, one finds that $e^{\cal A}\propto \frac{1}{z}$ and $\phi_{1,2} \propto z^2$,
 indicating that gravity solutions obtained as small perturbations of the AdS$_7$ ones have a dual description in terms of deformations of a six-dimensional CFT, in the presence of two operators of dimensions $\Delta=4$ (so that $6-\Delta=2$ is the scaling dimension of their deforming coupling). 
When computing the free energy of a general solution, the superpotential plays the role of a counterterm in our prescription. We must require to cancel all divergences, which in this case can be done by retaining in the superpotential terms up to cubic order in the scalars:
 \beqs
  \label{Eq:W2}
 {\cal W}_2&=&-\frac{5}{4}-\frac{\phi_1^2+\phi_2^2}{8} +\frac{3}{4\sqrt{10}}\phi_2\ln\left({\phi_2}\right)\left(\phi_1^2-\phi_2^2\right)\,.
 \eeqs
The free energy is determined by the action evaluated on shell. Its UV-divergent terms match those arising from the superpotential, and take the form $\frac{1}{z^6}[W_0 + W_2\f_i^2+\cdots]$. 
The UV expansion of scalar fields takes the form $\f_i \approx z^2 \f_{i,2}+z^4\f_{i,4}$ where $\f_{i,2}$ encodes the sources and $\f_{i,4}$ is proportional to the vacuum expectation values (VEVs) of the dual operators. In the body of the paper, we restrict attention to solutions in which the source is zero, but with non-zero VEV. Hence,  because all terms in the UV expansion of the superpotential, beyond the constant term, scale as $z^8$ or higher, they do not contribute to the divergence in the free energy---yet, we report the general result, for completeness.

We conclude this Appendix by showing an additional class of solutions, obtained using the first order equations.
These domain-wall solutions are singular, have vanishing free energy, and correspond to field theories in $D=6$ dimensions in which all deformations are trivial (though the VEVs of the two operators are not), hence they have no effect on the stability analysis discussed in the body of this paper.
The superpotential conjugate to $ {\cal W}_2$ in Eq.~(\ref{Eq:W2}) admits a superpotential that can be written in closed form:
\beqs
{\cal W}_{\rm susy}&=&-e^{-\frac{1}{\sqrt{10}}\phi_2}\left(\frac{1}{4}e^{\sqrt{\frac{5}{2}}\phi_2}+\cosh\left(\frac{\phi_1}{\sqrt{2}}\right)\right)\,.
\eeqs
This superpotential has the following small-field expansion:
\beqs
{\cal W}_{\rm susy}&=&-\frac{5}{4}-\frac{1}{4}\left(\phi_1^2+\phi_2^2\right)\,+\,\cdots\,,
\eeqs
hence the solutions of the first order equations derived from it  amount to turning on the sub-leading terms in the UV expansions of $\phi_{1}$ and $\phi_2$, corresponding to the emergence of VEVs for field theory operators in the dual language. One then finds that for all solutions solutions one has 
$\phi_{1,2}=\phi_{2,2}=\chi_6=0$, hence the free energy vanishes exactly, ${\cal F}=0$.\footnote{
Also $-{\cal W}_{\rm susy}$ is an admissible superpotential.
Its adoption would require a change of sign in the radial variable, $\tau$.}
Due to the symmetries of the system, for any solution of the first-order equations equations, there exists another equivalent one with  $\phi_1\rightarrow -\phi_1$ and the other background fields unchanged.

The resulting equations of motion are the following:
\beqs
\partial_{\rho}\phi_1&=&-2\sqrt{2}e^{-\frac{\phi_2}{\sqrt{10}}}\sinh\left(\frac{\phi_1}{\sqrt{2}}\right)\,,\\
\partial_{\rho}\phi_2&=&-2\sqrt{\frac{2}{5}}e^{-\frac{\phi_2}{\sqrt{10}}}
\left(e^{\sqrt{\frac{5}{2}}\phi_2}-\cosh\left(\frac{\phi_1}{\sqrt{2}}\right)\right)\,,\\
\partial_{\rho}{\cal A}&=&\frac{1}{10}e^{-\frac{\phi_2}{\sqrt{10}}}
\left(e^{\sqrt{\frac{5}{2}}\phi_2}+4\cosh\left(\frac{\phi_1}{\sqrt{2}}\right)\right)\,.
\eeqs
We can solve these equations, by applying the change of variable $\partial_{\rho}=e^{-\frac{\phi_2}{\sqrt{10}}}\partial_{\tau}$, to rewrite them as follows:
 \beqs
\partial_{\tau}\phi_1&=&-2\sqrt{2}\sinh\left(\frac{\phi_1}{\sqrt{2}}\right)\,,\\
\partial_{\tau}\phi_2&=&-2\sqrt{\frac{2}{5}}
\left(e^{\sqrt{\frac{5}{2}}\phi_2}-\cosh\left(\frac{\phi_1}{\sqrt{2}}\right)\right)\,,\\
\partial_{\tau}{\cal A}&=&\frac{1}{10}
\left(e^{\sqrt{\frac{5}{2}}\phi_2}+4\cosh\left(\frac{\phi_1}{\sqrt{2}}\right)\right)\,.
\eeqs
The general solution is the following:
\beqs
\phi_1^{\pm}(\tau)&=&\pm{2}{\sqrt{2}}\,{\rm arctanh}\left[e^{-2(\tau-\tau_o)}\right]\,,\\
\phi_2^{\pm}(\tau)&=&\sqrt{\frac{2}{5}}\log\left[\frac{\sinh(2(\tau-\tau_o))}{\cosh(2(\tau-\tau_o))-1\pm\left(1-\cosh(2(\bar{\t}_o-\t_0))\right)}\right]\,,\\
{\cal A}(\tau)&=&{\cal A}_o -\frac{1}{2}(\tau-\t_o)+\\
&&\nonumber
+\log\left[\left(e^{4(\tau-\tau_o)}-1\right)^{\frac{1}{5}}
\left(e^{4(\tau-\tau_o)}+1+2\left(-1\pm\left(1-\cosh(2(\bar{\t}_o-\t_o)\right) \right) e^{2(\tau-\tau_o)}\right)^{\frac{1}{20}}\right]\,,
\eeqs
with $\tau_o$, $\bar{\t}_o$,  and ${\cal A}_o$ integration constants.

%%%%%%%%%%%%%%%%%%%%%%%
\section{Lift to $D=11$ dimensions}\label{Sec:lift}

We report here the general form of the background metric, by lifting the $SO(2)\times SO(2)$ truncation back to $D=11$ dimensions . We identify the seven dimensional action in Eq.~(\ref{eq:7Daction}) and the potential  in Eq.~(\ref{Eq:pot7})
with Eqs.~(4.6) and~(4.7) of Ref.~\cite{Cvetic:1999xp}, respectively. Doing so requires setting the gauge coupling $g^2\equiv 1$ in the latter equation. The lift is based on defining
\beqs
X_1&\equiv&e^{-\frac{1}{\sqrt{2}}\f_1-\frac{1}{\sqrt{10}}\f_2}\,,\\
X_2&\equiv&e^{+\frac{1}{\sqrt{2}}\f_1-\frac{1}{\sqrt{10}}\f_2}\,,\\
X_0&\equiv&(X_1X_2)^{-2}\,=\,e^{\frac{4}{\sqrt{10}}\f_2}\,.
\eeqs 
We choose the ranges of the four angles describing the manifold to be the following:
\beqs
0\leq &\tilde{\psi}& \leq\frac{\pi}{2}\,, \\
0 \leq &\tilde{\xi}& \leq \pi\,,\\
0\leq & \tilde{\varphi}_i& < 2\pi\,.
\eeqs
We solve the constraint $\mu_0^2+\mu_1^2+\mu_2^2=1$ by defining 
\beqs
\mu_0&\equiv&\cos \tilde{ \xi}\,,\\
\mu_1&\equiv&\sin \tilde{\xi} \cos \tilde{\psi}\,,\\
\mu_2&\equiv&\sin \tilde{\xi} \sin \tilde{\psi}\,,
\eeqs
and introduce the variable
\beqs
\tilde{\Delta}&\equiv&\sum_{i=0}^2 X_i \mu_i^2\,,
\eeqs
which depends on the two angles, $\tilde{\varphi}_i$, along the two circles corresponding to the  $SO(2)\times SO(2)$ directions.
The metric in $11$ dimensions takes the following form~\cite{Cvetic:1999xp}:
\beqs
\di s_{11}^2&=&\tilde{\Delta}^{1/3}\di s_7^2\,+\,\tilde{\Delta}^{-2/3}\left(X_0^{-1}\di \mu_0^2
+\sum_{i=1,2}X_i^{-1}\left(\di \mu_i^2+\mu_i^2\left(\di \tilde{\varphi}_i+{\cal A}^{(i)}_{\hat{M}}\di x^{\hat{M}}\right)^2\right)\right)\,.
\eeqs

As a consistency check,  notice that if $\phi_1=\phi_2=0={\cal A}^{i}_{\hat{M}}$, then $X_1=X_2=X_0=1=\tilde{\Delta}$,
in which  case:
\beqs
\di s_{11}^2(0)&=&\di s_7^2\,+\,\left[\di \mu_0^2
+\sum_{i=1,2}\left(\di \mu_i^2+\mu_i^2\di \tilde{\varphi}_i^2\right)\right]\nonumber\\
&=& \di s_7^2\,+\,\left[\frac{}{}\di \tilde{\xi}^2 +\sin^2 \tilde{\xi} \left(\di \tilde{\psi}^2 +\cos^2\tilde{\psi} \di \tilde{\varphi}_1^2
+\sin^2\tilde{\psi} \di \tilde{\varphi}_2^2\right)\frac{}{}\right]\,.
\eeqs
The four dimensional internal submanifold has indeed ${\rm vol}({\cal S}^4)=\frac{8\pi^2}{3}$, and furthermore the round bracket in the last form of the metric describes the 3-sphere, with angular coordinates $(\tilde{\psi},\tilde{\varphi}_1,\tilde{\varphi}_2)$.

\begin{figure*}[b]
    \centering
    \subfloat[Ricci Scalar \label{Fig:Ricci}]{\includegraphics[width=0.45\linewidth]{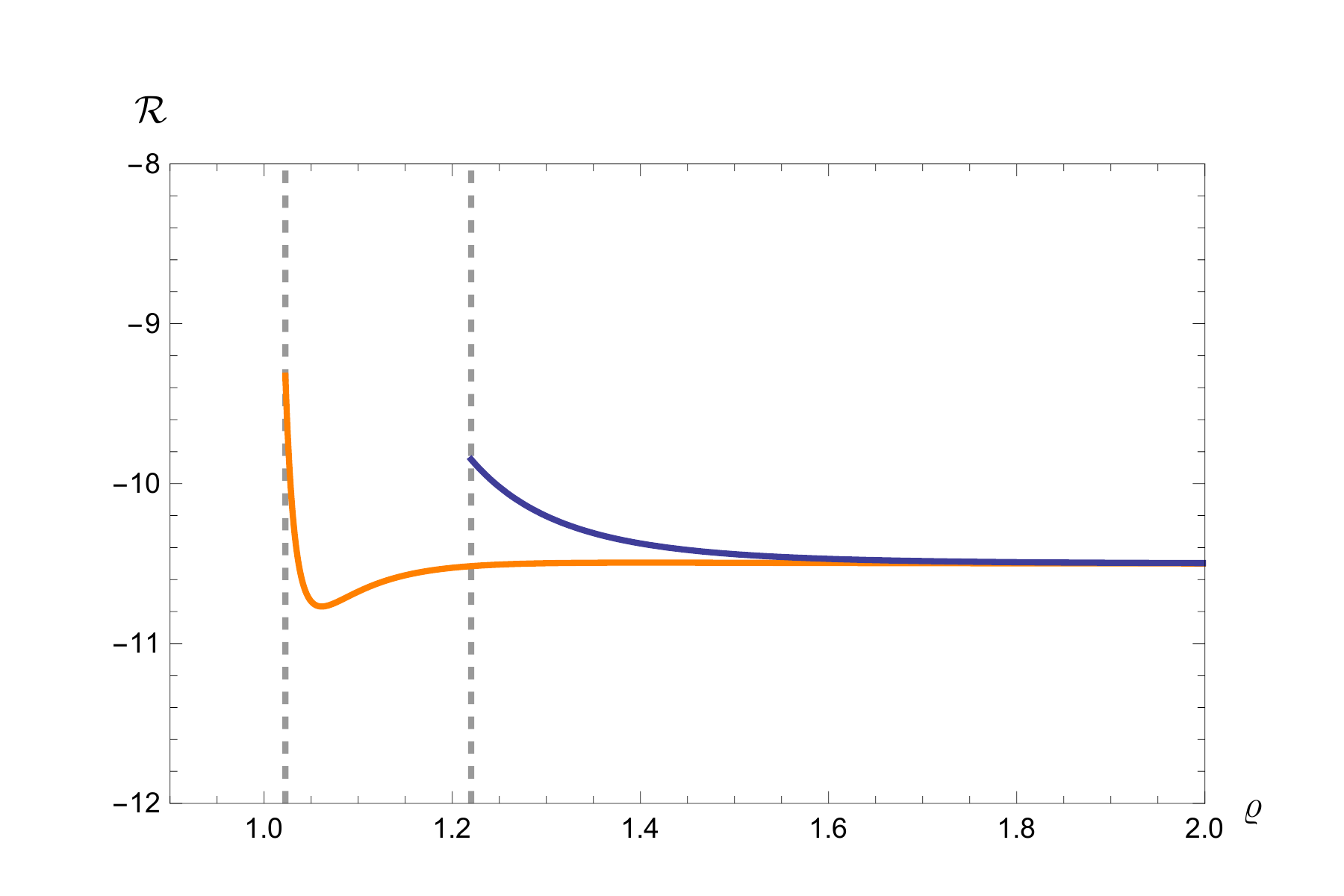}} 
    \subfloat[$\cR_{MN}\cR^{MN}$ \label{Fig:R2}]{\includegraphics[width=0.45\linewidth]{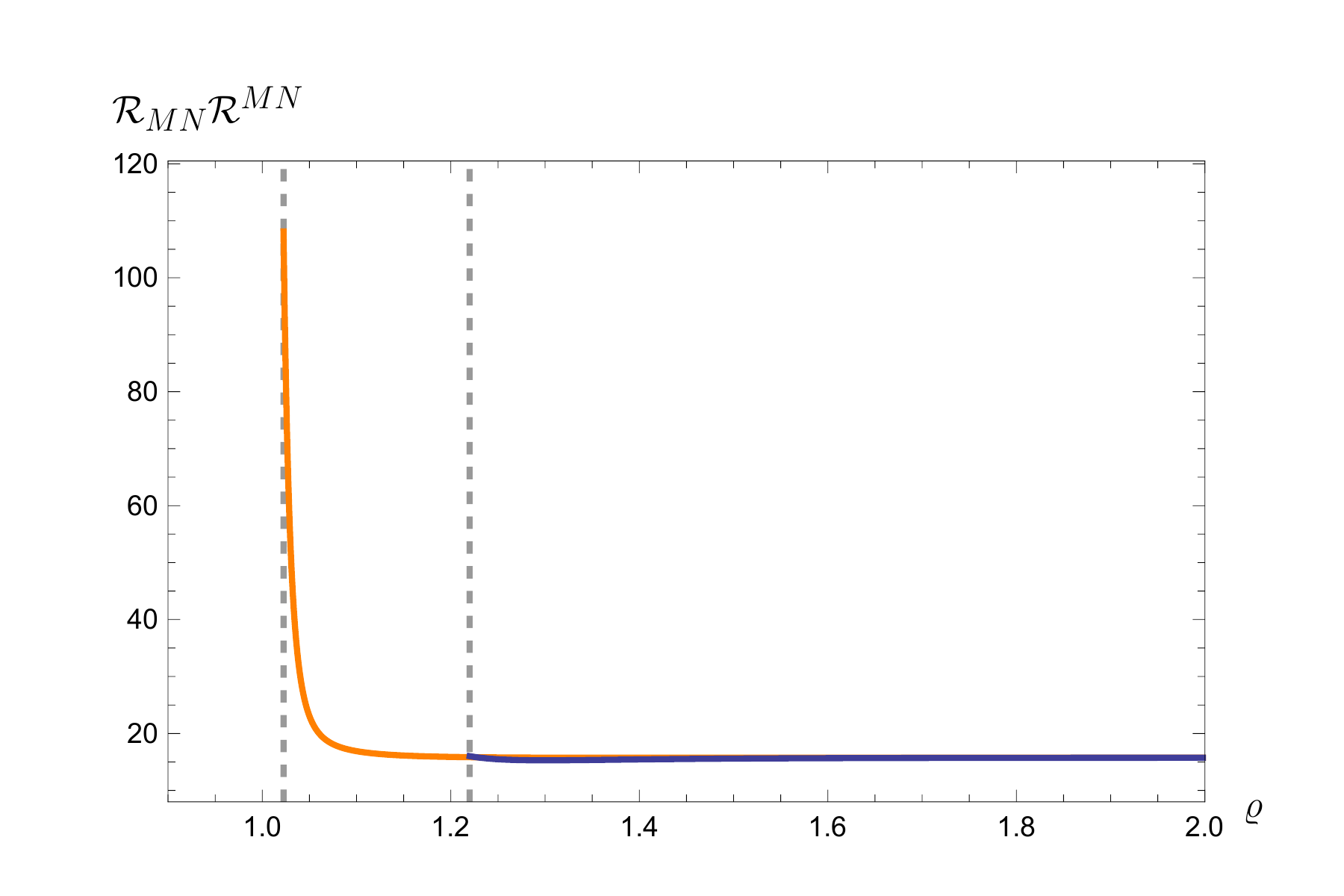}}
    
    \subfloat[$\cR_{MNPQ}\cR^{MNPQ}$ \label{Fig:R4}]{\includegraphics[width=0.45\linewidth]{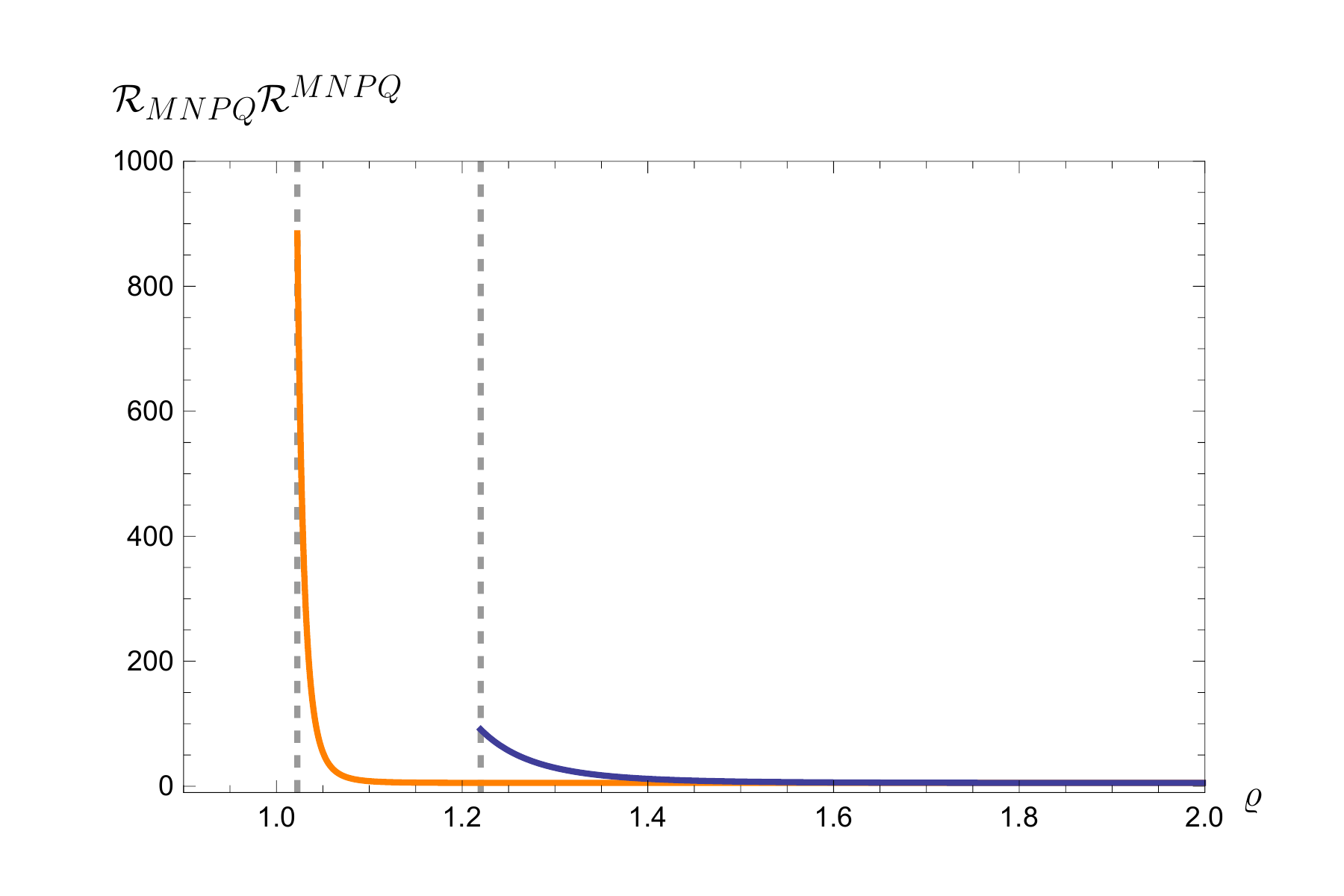}}
    \subfloat[Ricci Scalar, $\m=0$ \label{Fig:RicciBoundary}]{\includegraphics[width=0.45\linewidth]{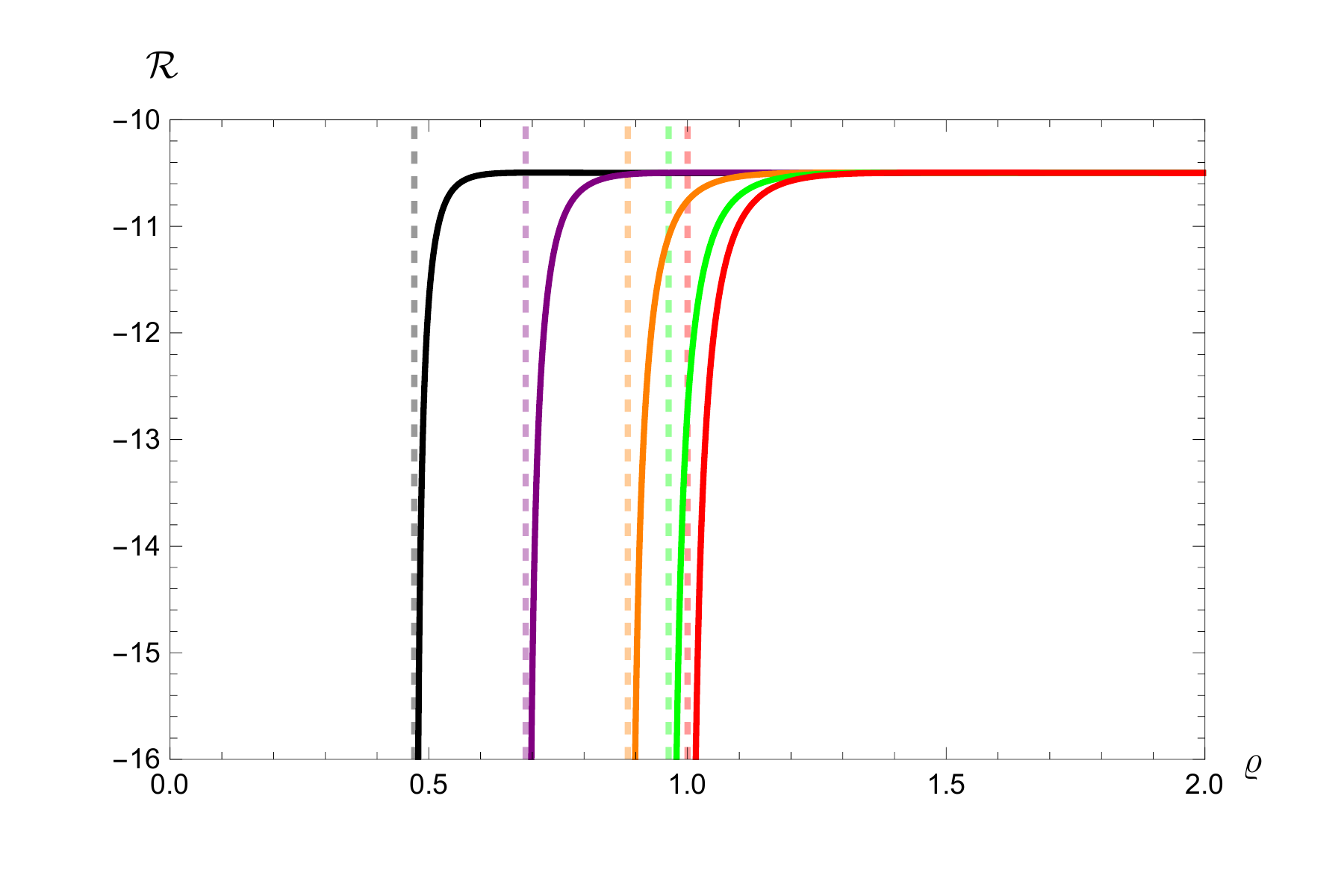}}
    \caption{Gravitational invariants for a selection of soliton (confining) solutions, as a function of the holographic direction, $\varrho$. Panels~\ref{Fig:Ricci}, \ref{Fig:R2}, and \ref{Fig:R4}, show three gravitational invariants for two particular choices of regular confining solutions, parametrized by $\{\cA_{7,U}^{(1)}, \cA_{7,U}^{(2)}\}=\{0.396,0.216\}$ and $\{0.796,0.565\}$, in blue and orange respectively. The grey dashed vertical lines correspond to the value of  $\vr_0$, where the space ends for each solution. All invariants are found to be smooth and regular over the entire space. The bottom right panel, \ref{Fig:RicciBoundary}, shows the Ricci scalar calculated for a selection of solutions  along the line of phase transitions, along which $\m=0$ and the free energy vanishes (the sides of the square in Fig.~\ref{Fig:PhaseDiagram}). The red, green, orange, purple and black lines correspond to soliton solutions for which the value of the angle $\theta={\frac{\p}{4}, \frac{\pi}{6}, \frac{\pi}{9}, \frac{\pi}{20}}$ and $ \frac{\pi}{25}$ respectively. The dashed lines show the corresponding end of space for each solution, 
    at which point the Ricci scalar diverges.}
    \label{fig:Gravitational_Invariants}
\end{figure*}

In principle,  one could compactify one of the external dimensions on a circle, and rewrite the background solutions in $D=10$ dimensions, within type-IIA supergravity.  The standard prescription to extract the field-theory behaviour of the Wilson loop of the (lower dimensional) dual field theory, and the related static potential between heavy quarks~\cite{Maldacena:1998im,Rey:1998ik}, would require to  compute the effect of the curved background on a probe string, allowed to develop extended configurations stretching in the space directions.   We observe that in the presence of non-trivial solutions, for which the scalar fields have a non-vanishing bulk profile, $\phi_i\neq 0$, the metric of the internal, four-dimensional manifold is characterised by a warp factor, $\tilde{\Delta}^{-2/3}$, that depends on the internal angles parametrising it, and is not a simple, smooth function. Because  of the presence of the factor of $\tilde{\Delta}^{-2/3}$, performing this exercise would require to solve non-linear, coupled  equations involving the internal angles, besides the radial direction,  and doing so exceeds the scopes of this study. Furthermore, special singularities emerge in the internal part of the geometry of the solutions of interest in the body of this paper, and rendering  the field-theory interpretation of the results less transparent than in Ref.~\cite{Witten:1998zw}.

%%%%%%%%%%%%%%%%%%%%%%%
%%%%%%%%%%%%%%%%%%%%%%%
\section{Gravitational Invariants}
\label{Sec:gravityinvariants}

The space-time geometry in $D=7$ dimensions for the backgrounds we study in the body of the paper, which correspond to confining field theories in $D-1=6$ dimensions compactified on a shrinking cirle, is characterised by curvature invariants that can be written in the following form:
\beqs
{\cal R}&=&
-10 A''(\rho )+2 \sqrt{10} A'(\rho ) \chi '(\rho )-30 A'(\rho )^2+\sqrt{\frac{2}{5}} \chi ''(\rho
   )-\frac{11}{5} \chi '(\rho )^2\\
   &=&
   20 A'(\rho )^2-\frac{1}{5} e^{\sqrt{\frac{2}{5}} (\phi_2(\rho )-4 \chi (\rho ))-\sqrt{2} \phi_1(\rho )} \left(e^{2 \sqrt{2} \phi_1(\rho )} \cA_7^{(1)\prime}(\rho )^2+\cA_7^{(2)\prime}(\rho )^2\right)-2
   \chi '(\rho )^2+\frac{48\, {\cal V}_7}{5}\\
   &=&
   \frac{1}{10 \varrho ^{32/5} \left(\left(\varrho
   ^4-Q_1^2\right) \left(\varrho ^4-Q_2^2\right)\right)^{6/5}}
   \left[\frac{}{}15 Q_1^4 Q_2^4+48 \mu  \varrho ^6 \left(Q_1^2+Q_2^2\right)-32 \mu  Q_1^2
   Q_2^2 \varrho ^2+126 \varrho ^{12} \left(Q_1^2+Q_2^2\right)\right.+\nonumber\\
   &&\left.\frac{}{}+6 Q_1^2 Q_2^2
   \varrho ^4 \left(Q_1^2+Q_2^2\right)-3 \varrho ^8 \left(3 Q_1^4+52 Q_1^2
   Q_2^2+3 Q_2^4\right)-105 \varrho ^{16}\frac{}{}\right]\,.
\eeqs
Here, the first expression is valid as long as the vector fields all vanish in the background, the second has been obtained by making use of the equations of motion,  and the third by making use of the exact soliton solutions, having changed variable in the holographic direction, but without imposing the constraint derived from the absence of conical singularities.

Upon imposing the constraint to avoid a conical singularity, the Ricci Scalar calculated for solutions along the line of phase transition, where $\m=0$ (and $Q_1^2=Q_2^2=\vr_0^4$) is given by
\begin{equation}
    \cR^{(\m=0)}=-\frac{3(35\vr^8-14\vr^4 \vr_0^4-5\vr_0^8)}{10\vr^8\left(\frac{(\vr^4-\vr_0^4)^2}{\vr^8}\right)^{\frac{1}{5}}}\,.
\end{equation}
This  expression diverges at the end of space, when $\vr=\vr_0$. Conversely,  in the case of the AdS$_7$, domain-wall solutions, the Ricci Scalar takes on the constant value $-\frac{21}{2}$.

We calculate three gravitational invariants, $\mathcal{R}, \cR_{MN}\cR^{MN}$, and $\cR_{MNPQ}\cR^{MNPQ}$, for the confining solutions. In Fig.~\ref{fig:Gravitational_Invariants} we display some representative examples for solutions close to  (but not at) the phase transition. We also display the Ricci scalar, $\cR$, for solutions with $\m=0$, which sit along the line of phase-transition. We find that the confining solutions are regular, with all three invariants finite, away from the phase transition. However, the Ricci scalar diverges for solutions at the end of these branches, corresponding to the phase transition. The classical  supergravity approximation in the limiting case of $\m=0$ is not reliable at this point, due to the large curvature.

%%%%%%%%%%%%%%%%%%%%%%%
%%%%%%%%%%%%%%%%%%%%%%%
\section{Gauge invariant formalism for the fluctuations}
\label{Sec:gaugeinvariantformalism}

We summarise here the main equations entering the gauge-invariant treatment of the fluctuations of the background, as developed in Refs.~\cite{Bianchi:2003ug,Berg:2005pd,Berg:2006xy,Elander:2009bm,Elander:2010wd,Elander:2010wn,Elander:2014ola,Elander:2018aub,Elander:2020csd}.  The sigma-model scalar fields, $\Phi^a$, are rewritten 
as follows~\cite{Bianchi:2003ug,Berg:2006xy,Berg:2005pd,Elander:2009bm,Elander:2010wd}:
\beq
	\Phi^a(x^\mu,r) =  \Phi^a(r) + \varphi^a(x^\mu,r) \,,
\eeq
where $\varphi^a(x^\mu,r)$ are small fluctuations around the background 
solutions, $\Phi^a(r)$.
The metric fluctuations are decomposed by foliating the holographic direction as in the Arnowitt-Deser-Misner  (ADM) formalism~\cite{Arnowitt:1962hi}, as follows:
\beqs
	\dd s_D^2 &=& \left( (1 + \nu)^2 + \nu_\sigma \nu^\sigma \right) \dd r^2 + 2 \nu_\mu \dd x^\mu \dd r 
	+ e^{2 {A}(r)} \left( \eta_{\mu\nu} + h_{\mu\nu} \right) \dd x^\mu \dd x^\nu \,,
	\eeqs
	where 
	\beqs
	h^\mu{}_\nu &=& \mathfrak e^\mu{}_\nu + i q^\mu \epsilon_\nu + i q_\nu \epsilon^\mu 
	+ \frac{q^\mu q_\nu}{q^2} H + \frac{1}{D-2} \delta^\mu{}_\nu h\,.
\eeqs
In these expressions, $\nu(x^\mu,r)$, $\nu^\mu(x^\mu,r)$,
 $\mathfrak e^\mu{}_\nu(x^\mu,r)$, $\epsilon^\mu(x^\mu,r)$, $H(x^\mu,r)$, and $h(x^\mu,r)$ are small
  fluctuations around the background metric. The gauge-invariant  $\epsilon^\mu(x^\mu,r)$  and the traceless $\mathfrak e^\mu{}_\nu(x^\mu,r)$
 are both transverse. The set of  gauge-invariant (under diffeomorphism) combinations  of the fluctuations can be written as follows:
\beqs
\label{eq:scalarfluc}
	\mathfrak a^a &=& \varphi^a - \frac{\partial_r  \Phi^a}{2(D-2)\partial_r { A}} h \,, \\
	\mathfrak b &=& \nu - \partial_r \left( \frac{h}{2(D-2)\partial_r {A}} \right) \,, \\
	\mathfrak c &=& e^{-2{ A}} \partial_\mu \nu^\mu - \frac{e^{-2{ A}} q^2 h}{2(D-2) \partial_r A} 
	- \frac{1}{2} \partial_r H \,, \\
	\mathfrak d^\mu &=& e^{-2{A}} P^\mu{}_\nu \nu^\nu - \partial_r \epsilon^\mu \,.
\eeqs
The equations of motion obeyed by
$\mathfrak{b}$ and $\mathfrak{c}$ are algebraic, and  can be decoupled in the treatment of the scalar fluctuations.
 The equation of motion for $\mathfrak d^\mu$ is also algebraic and does not lead to a spectrum of states.

The treatment of the fluctuations of the scalars is suitable to the adoption of a language that
captures the curvature of  the internal space, and makes use of its symmetries,  in  parallel to the space-time. 
One hence introduces the sigma-model connection, that 
 descends from the sigma-model metric, $ G_{ab}$, and the
 sigma-model derivative,  $\partial_b=\frac{\partial}{\partial \Phi^b}$, to read 
\beqs
 {\cal G}^d_{\,\,\,\,ab}&\equiv& \frac{1}{2} G^{dc}\left(\frac{}{}\partial_a G_{cb}
+\partial_b G_{ca}-\partial_c G_{ab}\right)\,.
\eeqs
The sigma-model Riemann tensor captures the curvature of the internal sigma-model space. We use conventions in which we write 
\beqs
 {\cal R}^a_{\,\,\,\,bcd}
&\equiv& \partial_c {\cal G}^a_{\,\,\,\,bd}-\partial_d {\cal G}^a_{\,\,\,\,bc}
+ {\cal G}^e_{\,\,\,\,bd} {\cal G}^a_{\,\,\,\,ce}- {\cal G}^e_{\,\,\,\,bc} {\cal G}^a_{\,\,\,\,de}\,,
\eeqs
and hence the sigma-model covariant derivative is
\beqs
D_b X^d_{\,\,\,\,a}&\equiv& \partial_b X^d_{\,\,\,\,a}+{\cal G}^d_{\,\,\,\,cb}X^c_{\,\,\,\,a}
- {\cal G}^c_{\,\,\,\,ab}X^d_{\,\,\,\,c}\,.
\eeqs

Adopting this formalism, the equations of motion for $\mathfrak a^a$ are the following:
\beqs
\label{eq:scalareom}
	0 &=& \Big[ {\cal D}_r^2 + (D-1) \partial_{r}{A} {\cal D}_r - e^{-2{A}} q^2 \Big] \mathfrak{a}^a \,\,\\ \nonumber
	&& - \Big[  {\mathcal V}^{\,\,a}{}_{\,|c} - \mathcal{R}^a{}_{bcd} \partial_{r}\Phi^b \partial_{r}\Phi^d + 
	\frac{4 (\partial_{r}\Phi^a  {\mathcal V}^{\,b} +  {\mathcal V}^{\,a} 
	\partial_{r}\Phi^b) G_{bc}}{(D-2) \partial_{r} {A}} + 
	\frac{16  {\mathcal V} \partial_{r}\Phi^a \partial_{r}\Phi^b G_{bc}}{(D-2)^2 (\partial_{r}{A})^2} \Big] \mathfrak{a}^c\,,
\eeqs
where $q^2=\eta_{\mu\nu}q_{\mu}q_{\nu}=-M^2$, while the boundary conditions are given by
\beqs
\label{eq:scalarbc}
 \frac{2  e^{2A}\partial_{r} \Phi^a }{(D-2)q^2 \partial_{r}{ A}}
	\left[ \partial_{r} \Phi^b{\cal D}_r -\frac{4  {\cal V} \partial_{r} \Phi^b}{(D-2) 
	\partial_r { A}} - {\cal V}^b \right] \mathfrak a_b - \mathfrak a^a\Big|_{r_i} = 0 \, .
\eeqs
In all these expressions, the background covariant derivative is  
$\mathcal D_r \mathfrak a^a \equiv \partial_r \mathfrak a^a +
 \mathcal G^a_{\ bc} \partial_r  \Phi^b \mathfrak a^c$,
 while we use the shorthand ${\mathcal V}^a{}_{|b} \equiv \frac{\partial {\mathcal V}^a}{\partial \Phi^b} + \mathcal G^a_{\ bc} {\mathcal V}^c$.

The tensor fluctuations, $\mathfrak e^\mu{}_\nu$, must  obey the simpler differential equations
\beq
\label{eq:tensor fluc}
	\left[ \partial_r^2 + (D-1) \partial_r {A} \partial_r - e^{-2{ A}(r)} q^2 \right] \mathfrak e^\mu_{\,\,\,\nu} = 0 \,,
\eeq
with boundary conditions given by 
\beq
\label{eq:tensor fluc bc}
\left.\frac{}{}	\partial_r \mathfrak e ^\mu_{\,\,\,\nu} \right|_{r=r_i}= 0 \,.
\eeq

%%%%%%%%%%%%%%%%%%%%%%%
%%%%%%%%%%%%%%%%%%%%%%%
\subsection{Fluctuation equations for the scalars}
\label{Sec:Fluctuations}

We report here the explicit form of the fluctuation equations used for the numerical study the results of which are reported in the body of the paper, written in terms of the variable $\rho$.  Primed variables stand for derivatives in respect to $\rho$, so that $g^{\prime}=\partial_{\rho} g(\rho)$.
The five gauge-invariant scalar fluctuations, $\left\{\mathfrak{a}^{\phi_1},\,\mathfrak{a}^{\phi_2},\,\mathfrak{a}^{\chi},\,
\mathfrak{a}^{{\cal A}_7^{(1)}},\,\mathfrak{a}^{{\cal A}_7^{(2)}} \right\}$, obey the following  equations.
\beqs
\label{eq:phi1 fluc}
0&=& \Bigg[160 e^{2A+\sqrt{2}\f_1+\sqrt{\frac{2}{5}}(\f_2+4\c)}A'^2\6_{\r}^2+16e^{2A+\sqrt{2}\f_1+\sqrt{\frac{2}{5}}(\f_2+4\c)}A'^2(50A'-\sqrt{10}\c')\6_{\r}+  \\
&& 5\Big(32A'^2\left(e^{2A+\frac{\f_1}{\sqrt{2}}+\frac{5\f_2+8\c}{\sqrt{10}}}(1+e^{\sqrt{2}\f_1})-e^{2A+2\sqrt{\frac{2}{5}}\f_2}(e^{2\sqrt{2}\f1}\left({\cA_7^{(1)}}'\right)^2+\left({\cA_7^{(2)}}'\right)^2)+e^{\sqrt{2}\f_1+\sqrt{\frac{2}{5}}(\f_2+5\c)}M^2\right)+ \nn \\
&& 16\sqrt{2}e^{2A+\frac{\f_1}{\sqrt{2}}+\frac{5\f_2+8\c}{\sqrt{10}}}(-1+e^{\sqrt{2}\f_1})A'\f_1'+e^{2A+\frac{\f_1}{\sqrt{2}}+4\sqrt{\frac{2}{5}}\c}\left(8e^{\frac{\f_1}{\sqrt{2}}}-e^{\frac{\f_2}{\sqrt{2}}+\sqrt{10}\f_2}+4e^{\sqrt{\frac{5}{2}}\f_2}(1+e^{\sqrt{2}\f_1})\right)\f_1^2\Big)\Bigg]\fa^{\f_1}+ \nn \\
&&\Bigg[32\sqrt{5}e^{2\sqrt{\frac{2}{5}}\f_2}A'^2\Bigg(3e^{\frac{\f_1}{\sqrt{2}}+\frac{\f_2+8\c}{\sqrt{10}}}(-1+e^{\sqrt{2}\f_1})-e^{2\sqrt{2}\f_1}\left({\cA_7^{(1)}}'\right)^2+\left({\cA_7^{(2)}}'\right)\Bigg)+ \nn \\
&& 5 e^{4 \sqrt{\frac{2}{5}} \chi (\rho )+\frac{\phi_1(\rho )}{\sqrt{2}}} \left(4 \left(e^{\sqrt{2} \f_1}+1\right) e^{\sqrt{\frac{5}{2}} \f_2}-e^{\frac{\f_1}{\sqrt{2}}+\sqrt{10} \f_2}+8 e^{\frac{\f_1}{\sqrt{2}}}\right) \f_1' \f_2'+ \nn \\
&& 16 \sqrt{2} A' e^{\sqrt{2} \f_1+\frac{8 \chi +5 \f_2}{\sqrt{10}}} \left(\sqrt{5} \f_1' \left(3 \cosh \left(\frac{\f_1}{\sqrt{2}}\right)+\sinh \left(\sqrt{\frac{5}{2}} \f_2\right)-3 \cosh\left(\sqrt{\frac{5}{2}} \f_2\right)\right)+5 \sinh \left(\frac{\f_1}{\sqrt{2}}\right) \f_2'\right)\Bigg]e^{2A}\fa^{\f_2} + \nn \\
&& \Bigg[-160 \sqrt{2} e^{2 A+2 \sqrt{2} \phi _1+2 \sqrt{\frac{2}{5}} \phi _2} \left(A'\right)^2 \cA_7^{(1)}\6_{\r}+ \nn \\
&&e^{2 A+\frac{3 \phi _1}{\sqrt{2}}+\sqrt{\frac{2}{5}} \phi _2}\Bigg(20 e^{\sqrt{\frac{5}{2}} \phi _2} (2 \sqrt{2} (e^{\sqrt{2} \phi _1}-1) A'+(e^{\sqrt{2} \phi _1}+1) \phi _1')
{\cA_7^{(1)}}'-  \nn \\
&& 16 e^{\frac{\phi _1}{\sqrt{2}}+\sqrt{\frac{2}{5}} \phi _2} (A')^2 {\cA_7^{(1)}}'(25 \sqrt{2}
A'+\sqrt{5} \left(2 \phi _2'-9 \chi '\right)+10 \phi _1')+5 \sqrt{2} {\cA_7^{(1)}}'')+40 e^{\frac{\phi _1}{\sqrt{2}}} \phi _1'
{\cA_7^{(1)}}'-5 e^{\frac{\phi _1}{\sqrt{2}}+\sqrt{10} \phi _2} \phi _1' {\cA_7^{(1)}}'\Bigg)\Bigg]\fa^{\cA_7^{(1)}} \nn \\
&&\Bigg[160 \sqrt{2} e^{2 A+2 \sqrt{\frac{2}{5}} \f_2} (A')^2 {\cA_7^{(2)}}'\6_{\r}\nn + \\
&& e^{2 A+\sqrt{\frac{2}{5}} \phi _2} \Bigg(40 e^{\sqrt{\frac{5}{2}} \phi _2} {\cA_7^{(2)}}' \left(2 \sqrt{2} A' \sinh \left(\frac{\phi
   _1}{\sqrt{2}}\right)+\phi _1' \cosh \left(\frac{\phi _1}{\sqrt{2}}\right)\right)+ \nn \\
   && 16 e^{\sqrt{\frac{2}{5}} \phi _2} (A')^2
   {\cA_7^{(2)}}'\left((25 \sqrt{2} A'+\sqrt{5} (2 \phi _2'-9 \chi')-10 \phi _1')+5 \sqrt{2} {\cA_7^{(2)}}''\right)-5e^{\sqrt{10} \phi _2} \phi _1' {\cA_7^{(2)}}'+40 \phi _1' {\cA_7^{(2)}}'\Bigg)\Bigg]\fa^{\cA_7^{(2)}}+ \nn \\
&& \Bigg[-16 \sqrt{5} e^{2 \sqrt{\frac{2}{5}} \phi _2} \left(A'\right)^2 \left(-2 e^{2 \sqrt{2} \phi _1} \left({\cA_7^{(1)}}'\right)^2+2 \left({\cA_7^{(2)}}'\right)^2+\left(e^{\sqrt{2} \phi _1}-1\right) e^{\frac{8 \chi +\phi _2}{\sqrt{10}}+\frac{\phi _1}{\sqrt{2}}}\right)+\nn \\
   && \sqrt{2} A' e^{4 \sqrt{\frac{2}{5}}
   \chi +\frac{\phi _1}{\sqrt{2}}} \left(40 e^{\sqrt{\frac{5}{2}} \phi _2} \left(e^{\sqrt{2} \phi _1}-1\right) \chi '-\sqrt{5} \left(4
   e^{\sqrt{\frac{5}{2}} \phi _2} \left(e^{\sqrt{2} \phi _1}+1\right)+8 e^{\frac{\phi _1}{\sqrt{2}}}-e^{\frac{\phi _1}{\sqrt{2}}+\sqrt{10} \phi
   _2}\right) \phi _1'\right)+\nn \\ 
   &&5 \left(4 e^{\sqrt{\frac{5}{2}} \phi _2} \left(e^{\sqrt{2} \phi _1}+1\right)+8 e^{\frac{\phi
   _1}{\sqrt{2}}}-e^{\frac{\phi _1}{\sqrt{2}}+\sqrt{10} \phi _2}\right) \chi ' e^{4 \sqrt{\frac{2}{5}} \chi +\frac{\phi _1}{\sqrt{2}}} \phi _1'\Bigg]4e^{2A}\fa^{\c}\nn \, ,
\eeqs
\beqs
\label{eq: phi2 fluc}
0&=&\Bigg[32 \sqrt{5} e^{2 \sqrt{\frac{2}{5}} \phi _2} \left(A'\right)^2 \left(-e^{2 \sqrt{2} \phi _1} \left({\cA_7^{(1)}}'\right)^2+\left({\cA_7^{(2)}}'\right)^2+3
   \left(e^{\sqrt{2} \phi _1}-1\right) e^{\frac{8 \chi +\phi _2}{\sqrt{10}}+\frac{\phi _1}{\sqrt{2}}}\right)+ \\
   && 16 \sqrt{2} A' e^{\frac{8 \chi +5 \phi
   _2}{\sqrt{10}}+\sqrt{2} \phi _1} \left(5 \phi _2' \sinh \left(\frac{\phi _1}{\sqrt{2}}\right)+\sqrt{5} \phi _1' \left(\sinh
   \left(\sqrt{\frac{5}{2}} \phi _2\right)+3 \cosh \left(\frac{\phi _1}{\sqrt{2}}\right)-3 \cosh \left(\sqrt{\frac{5}{2}} \phi
   _2\right)\right)\right)+ \nn \\
   && 5 \left(4 e^{\sqrt{\frac{5}{2}} \phi _2} \left(e^{\sqrt{2} \phi _1}+1\right)+8 e^{\frac{\phi _1}{\sqrt{2}}}-e^{\frac{\phi
   _1}{\sqrt{2}}+\sqrt{10} \phi _2}\right) e^{4 \sqrt{\frac{2}{5}} \chi +\frac{\phi _1}{\sqrt{2}}} \phi _1' \phi _2'\Bigg]e^{2A}\fa^{\f_1}+ \nn \\
&& \Bigg[160 \left(A'\right)^2 e^{2 A+\sqrt{\frac{2}{5}} \left(4 \chi +\phi _2\right)+\sqrt{2} \phi _1} \6_{\r}^2 + 16 \left(A'\right)^2 \left(50 A'-\sqrt{10} \chi '\right) e^{2 A+\sqrt{\frac{2}{5}} \left(4 \chi +\phi _2\right)+\sqrt{2} \phi _1} \6_{\r} +\nn \\
&& 32 \left(A'\right)^2 \Bigg(-e^{2 A+2 \sqrt{2} \phi _1+2 \sqrt{\frac{2}{5}} \phi _2} \left({\cA_7^{(1)}}'\right)^2-e^{2 A+2 \sqrt{\frac{2}{5}} \phi _2}
  \left({\cA_7^{(2)}}'\right)^2+e^{4 \sqrt{\frac{2}{5}} \chi +\sqrt{2} \phi _1} \left(5 M^2 e^{\sqrt{\frac{2}{5}} \left(\chi +\phi _2\right)}-8 e^{2 A}
   \left(2 e^{\sqrt{10} \phi _2}-1\right)\right)+\nn \\
   && 9 \left(e^{\sqrt{2} \phi _1}+1\right) e^{2 A+\frac{8 \chi +5 \phi _2}{\sqrt{10}}+\frac{\phi
   _1}{\sqrt{2}}}\Bigg)+16 \sqrt{10} \left(3 e^{\sqrt{\frac{5}{2}} \phi _2} \left(e^{\sqrt{2} \phi _1}+1\right)-4 e^{\frac{\phi _1}{\sqrt{2}}}-2
   e^{\frac{\phi _1}{\sqrt{2}}+\sqrt{10} \phi _2}\right) A' \phi _2' e^{2 A+4 \sqrt{\frac{2}{5}} \chi +\frac{\phi _1}{\sqrt{2}}}+\nn \\
   && 5 \left(4
   e^{\sqrt{\frac{5}{2}} \phi _2} \left(e^{\sqrt{2} \phi _1}+1\right)+8 e^{\frac{\phi _1}{\sqrt{2}}}-e^{\frac{\phi _1}{\sqrt{2}}+\sqrt{10} \phi
   _2}\right) \left(\phi _2'\right){}^2 e^{2 A+4 \sqrt{\frac{2}{5}} \chi +\frac{\phi _1}{\sqrt{2}}}\Bigg]\fa^{\f_2}+ \nn \\
   && \Bigg[32 \sqrt{10} e^{2 A+2 \sqrt{2} \phi _1+2 \sqrt{\frac{2}{5}} \phi _2} \left(A'\right)^2 {\cA_7^{(1)}}'\6_{\r}+\nn \\
   &&e^{2 A+\frac{3 \phi _1}{\sqrt{2}}+\sqrt{\frac{2}{5}} \phi _2} \Bigg(-8 e^{\frac{\phi _1}{\sqrt{2}}} \left(4 \sqrt{10} A'-5 \phi _2'\right)
  {\cA_7^{(1)}}'+4 e^{\sqrt{\frac{5}{2}} \phi _2} \left(e^{\sqrt{2} \phi _1}+1\right) \left(6 \sqrt{10} A'+5 \phi _2'\right) {\cA_7^{(1)}}'-\nn \\
   &&e^{\frac{\phi _1}{\sqrt{2}}+\sqrt{10} \phi _2} \left(16 \sqrt{10} A'+5 \phi _2'\right) {\cA_7^{(1)}}'-16 e^{\frac{\phi
   _1}{\sqrt{2}}+\sqrt{\frac{2}{5}} \phi _2} \left(A'\right)^2 \left({\cA_7^{(1)}}' \left(5 \sqrt{10} A'-9 \chi '+2 \sqrt{5} \phi _1'+2 \phi
   _2'\right)+\sqrt{10}{\cA_7^{(1)}}''\right)\Bigg)\Bigg]\fa^{\cA_7^{(1)}} + \nn \\
  && \Bigg[-32 \sqrt{10} \left(A'\right)^2 {\cA_7^{(2)}}'\6_{\r}+8 \left(5 \phi _2'-4 \sqrt{10} A'\right) {\cA_7^{(2)}}'(\rho )-e^{\sqrt{10} \phi _2} \left(16 \sqrt{10} A'+5 \phi _2'\right) {\cA_7^{(2)}}'+\nn \\
   && 8 e^{\sqrt{\frac{5}{2}} \phi _2} \cosh \left(\frac{\phi _1}{\sqrt{2}}\right) \left(6 \sqrt{10} A'+5 \phi _2'\right){\cA_7^{(2)}}'-16
   e^{\sqrt{\frac{2}{5}} \phi _2} \left(A'\right)^2 \left({\cA_7^{(2)}}' \left(5 \sqrt{10} A'-9 \chi '-2 \sqrt{5} \phi _1'+2 \phi
   _2'\right)+ \right.\nn \\
   && \left. \sqrt{10} {\cA_7^{(2)}}''\right)\Bigg]e^{2 A+2 \sqrt{\frac{2}{5}} \phi _2} \fa^{\cA_7^{(2)}}+\nn \\
   &&\Bigg[32 e^{2 \sqrt{\frac{2}{5}} \phi _2} \left(A'\right)^2 \left(e^{2 \sqrt{2} \phi _1} \left({\cA_7^{(1)}}'\right)^2+\left({\cA_7^{(2)}}'\right)^2\right)+e^{\sqrt{\frac{2}{5}} \left(4 \chi +5 \phi _2\right)+\sqrt{2} \phi _1} \left(\sqrt{10} A' \left(\phi _2'-16 \chi '\right)+32
   \left(A'\right)^2-5 \chi ' \phi _2'\right)-\nn \\
   && 4 \left(e^{\sqrt{2} \phi _1}+1\right) e^{\frac{8 \chi +5 \phi _2}{\sqrt{10}}+\frac{\phi _1}{\sqrt{2}}}
   \left(\sqrt{10} A' \left(\phi _2'-6 \chi '\right)+12 \left(A'\right)^2-5 \chi ' \phi _2'\right)+\nn \\
   &&8 e^{4 \sqrt{\frac{2}{5}} \chi +\sqrt{2} \phi _1}
   \left(-\sqrt{10} A' \left(4 \chi '+\phi _2'\right)+8 \left(A'\right)^2+5 \chi ' \phi _2'\right)\Bigg]4e^{2A}\fa^{\c} \nn \, ,
\eeqs

\beqs
0&=&\Bigg[16 \sqrt{5} e^{2 \sqrt{\frac{2}{5}} \phi _2} \left(A'\right)^2 \left(-2 e^{2 \sqrt{2} \phi _1} \left({\cA_7^{(1)}}'\right)^2+2 \left({\cA_7^{(2)}}'\right)^2+\left(e^{\sqrt{2} \phi _1}-1\right) e^{\frac{8 \chi +\phi _2}{\sqrt{10}}+\frac{\phi _1}{\sqrt{2}}}\right)+ \\
&& \sqrt{2} A' e^{\frac{8 \chi +5 \phi_2}{\sqrt{10}}+\sqrt{2} \phi _1} \left(\sqrt{5} \phi _1' \left(-9 \sinh \left(\sqrt{\frac{5}{2}} \phi _2\right)+8 \cosh \left(\frac{\phi_1}{\sqrt{2}}\right)+7 \cosh \left(\sqrt{\frac{5}{2}} \phi _2\right)\right)-80 \chi ' \sinh \left(\frac{\phi _1}{\sqrt{2}}\right)\right)- \nn \\
&&5\left(4 e^{\sqrt{\frac{5}{2}} \phi _2} \left(e^{\sqrt{2} \phi _1}+1\right)+8 e^{\frac{\phi _1}{\sqrt{2}}}-e^{\frac{\phi _1}{\sqrt{2}}+\sqrt{10}\phi _2}\right) \chi ' e^{4 \sqrt{\frac{2}{5}} \chi +\frac{\phi _1}{\sqrt{2}}} \phi _1'\Bigg]e^{2A}\fa^{\f_1} +\nn \\
&&\Bigg[32 e^{2 \sqrt{\frac{2}{5}} \phi _2} \left(A'\right)^2 \left(e^{2 \sqrt{2} \phi _1} \left({\cA_7^{(1)}}'\right)^2+\left({\cA_7^{(2)}}'\right)^2\right)+e^{\sqrt{\frac{2}{5}} \left(4 \chi +5 \phi _2\right)+\sqrt{2} \phi _1} \left(\sqrt{10} A' \left(\phi _2'-16 \chi '\right)+32\left(A'\right)^2-5 \chi ' \phi _2'\right)-\nn \\
&& 4 \left(e^{\sqrt{2} \phi _1}+1\right) e^{\frac{8 \chi +5 \phi _2}{\sqrt{10}}+\frac{\phi _1}{\sqrt{2}}}
\left(\sqrt{10} A' \left(\phi _2'-6 \chi '\right)+12 \left(A'\right)^2-5 \chi ' \phi _2'\right)+\nn \\
&& 8 e^{4 \sqrt{\frac{2}{5}} \chi +\sqrt{2} \phi _1}
\left(-\sqrt{10} A' \left(4 \chi '+\phi _2'\right)+8 \left(A'\right)^2+5 \chi ' \phi _2'\right)\Bigg]e^{2A}\fa^{\f_2}+ \nn \\
&&\Bigg[32 \sqrt{10} e^{2 A+2 \sqrt{2} \phi _1+2 \sqrt{\frac{2}{5}} \phi _2} \left(A'\right)^2 {\cA_7^{(1)}}'\6_{\r}+ e^{2 A+\frac{3 \phi _1}{\sqrt{2}}+\sqrt{\frac{2}{5}} \phi _2} \Bigg(-8 e^{\frac{\phi _1}{\sqrt{2}}} \left(\sqrt{10} A'-5 \chi '\right)
{\cA_7^{(1)}}'+e^{\frac{\phi _1}{\sqrt{2}}+\sqrt{10} \phi _2} \left(\sqrt{10} A'-5 \chi '\right) {\cA_7^{(1)}}'-\nn \\
&& 4 e^{\sqrt{\frac{5}{2}} \phi _2} \left(e^{\sqrt{2} \phi _1}+1\right) \left(\sqrt{10} A'-5 \chi '\right) {\cA_7^{(1)}}'+16 e^{\frac{\phi _1}{\sqrt{2}}+\sqrt{\frac{2}{5}} \phi _2} \left(A'\right)^2 \left({\cA_7^{(1)}}' \left(5 \sqrt{10} A'-9 \chi '+2 \sqrt{5} \phi _1'+2 \phi_2'\right)+\sqrt{10} {\cA_7^{(1)}}''\right)\Bigg)\Bigg]\fa^{\cA_7^{(1)}}+ \nn \\
&& \Bigg[32 \sqrt{10} \phi _2 (A')^2 {\cA_7^{(2)}}'\6_{\r}-8 (\sqrt{10} A'-5 \chi ') {\cA_7^{(2)}}'+e^{\sqrt{10} \phi _2} (\sqrt{10} A'-5 \chi ') {\cA_7^{(2)}}'- 8 e^{\sqrt{\frac{5}{2}} \phi _2} \cosh \left(\frac{\phi _1}{\sqrt{2}}\right) (\sqrt{10} A'-5 \chi ') {\cA_7^{(2)}}'+ \nn \\
&& 16 e^{\sqrt{\frac{2}{5}} \phi _2} (A')^2 \left({\cA_7^{(2)}}' (5 \sqrt{10} A'-9 \chi '-2 \sqrt{5} \phi _1'+2 \phi _2')+\sqrt{10} {\cA_7^{(2)}}''\right)\Bigg]e^{2 A+2 \sqrt{\frac{2}{5}}} \fa^{\cA_7^{(2)}}+ \nn \\
&&\Bigg[160 \left(A'\right)^2 e^{2 A+\sqrt{\frac{2}{5}} \left(4 \chi +\phi _2\right)+\sqrt{2} \phi _1}\6_{\r}^2+16 \left(A'\right)^2 \left(50 A'-\sqrt{10} \chi '\right) e^{2 A+\sqrt{\frac{2}{5}} \left(4 \chi +\phi _2\right)+\sqrt{2} \phi _1}\6_{\r}+ \nn \\
&& 8 \left(A'\right)^2 \left(-16 e^{2 A+2 \sqrt{2} \phi _1+2 \sqrt{\frac{2}{5}} \phi _2}\left({\cA_7^{(1)}}'\right)^2-16 e^{2 A+2 \sqrt{\frac{2}{5}} \phi _2}
   \left({\cA_7^{(2)}}'\right)^2+e^{4 \sqrt{\frac{2}{5}} \chi +\sqrt{2} \phi _1} \left(20 M^2 e^{\sqrt{\frac{2}{5}} \left(\chi +\phi _2\right)}-e^{2 A}
   \left(e^{\sqrt{10} \phi _2}-8\right)\right)+\right.\nn \\
   &&\left. 4 \left(e^{\sqrt{2} \phi _1}+1\right) e^{2 A+\frac{8 \chi +5 \phi _2}{\sqrt{10}}+\frac{\phi
   _1}{\sqrt{2}}}\right)-8 \sqrt{10} \left(4 e^{\sqrt{\frac{5}{2}} \phi _2} \left(e^{\sqrt{2} \phi _1}+1\right)+8 e^{\frac{\phi
   _1}{\sqrt{2}}}-e^{\frac{\phi _1}{\sqrt{2}}+\sqrt{10} \phi _2}\right) A' \chi ' e^{2 A+4 \sqrt{\frac{2}{5}} \chi +\frac{\phi _1}{\sqrt{2}}}+ \nn \\
   && 20
   \left(4 e^{\sqrt{\frac{5}{2}} \phi _2} \left(e^{\sqrt{2} \phi _1}+1\right)+8 e^{\frac{\phi _1}{\sqrt{2}}}-e^{\frac{\phi _1}{\sqrt{2}}+\sqrt{10}
   \phi _2}\right) \left(\chi '\right)^2 e^{2 A+4 \sqrt{\frac{2}{5}} \chi +\frac{\phi _1}{\sqrt{2}}}\Bigg]\fa^{\c} \nn \, ,
\eeqs

\beqs
0&=&\Bigg[160 \sqrt{2} \left(A'\right)^2 e^{2 A+\sqrt{\frac{2}{5}} \left(4 \chi +\phi _2\right)+\frac{3 \phi _1}{\sqrt{2}}}{\cA_7^{(1)}}'\6_{\r}+4 e^{2 A+4 \sqrt{\frac{2}{5}} \chi +\sqrt{2} \phi _1} \Bigg(-8 e^{\frac{\phi _1}{\sqrt{2}}} \left(\sqrt{10} A'-5 \chi '\right) {\cA_7^{(1)}}'+ \\
   && e^{\frac{\phi _1}{\sqrt{2}}+\sqrt{10} \phi _2} \left(\sqrt{10} A'-5 \chi '\right) {\cA_7^{(1)}}'-4 e^{\sqrt{\frac{5}{2}} \phi _2}
   \left(e^{\sqrt{2} \phi _1}+1\right) \left(\sqrt{10} A'-5 \chi '\right) {\cA_7^{(1)}}'-\nn \\
   &&16 e^{\frac{\phi _1}{\sqrt{2}}+\sqrt{\frac{2}{5}} \phi
   _2} \left(A'\right)^2 \left({\cA_7^{(1)}}' \left(5 \sqrt{10} A'-9 \chi '+2 \sqrt{5} \phi _1'+2 \phi _2'\right)+\sqrt{10} {\cA_7^{(1)}}''\right)\Bigg)\Bigg]\fa^{\f_1}+ \nn \\
  && \Bigg[32 \sqrt{10} \left(A'\right)^2 e^{2 A+\sqrt{\frac{2}{5}} \left(4 \chi +\phi _2\right)+\frac{3 \phi _1}{\sqrt{2}}} {\cA_7^{(1)}}'\6_{\r}+
  \mathfrak{a}^{\phi_2}(\rho ) e^{2 A+4 \sqrt{\frac{2}{5}} \chi +\sqrt{2} \phi _1} \Bigg(-8 e^{\frac{\phi _1}{\sqrt{2}}} \left(4 \sqrt{10} A'-5 \phi
   _2'\right) {\cA_7^{(1)}}'+\nn \\
   && 4 e^{\sqrt{\frac{5}{2}} \phi _2} \left(e^{\sqrt{2} \phi _1}+1\right) \left(6 \sqrt{10} A'+5 \phi _2'\right)
  {\cA_7^{(1)}}'-e^{\frac{\phi _1}{\sqrt{2}}+\sqrt{10} \phi _2} \left(16 \sqrt{10} A'+5 \phi _2'\right) {\cA_7^{(1)}}'+\nn \\
   &&16 e^{\frac{\phi
   _1}{\sqrt{2}}+\sqrt{\frac{2}{5}} \phi _2} \left(A'\right)^2 \left({\cA_7^{(1)}}'\left(5 \sqrt{10} A'-9 \chi '+2 \sqrt{5} \phi _1'+2 \phi
   _2'\right)+\sqrt{10} {\cA_7^{(1)}}''\right)\Bigg)\Bigg]\fa^{\f_2}+ \nn \\
   &&\Bigg[16 \left(A'\right)^2 e^{2 A+\sqrt{\frac{2}{5}} \left(4 \chi +\phi _2\right)+\frac{3 \phi _1}{\sqrt{2}}}\Bigg(10 \6_{\r}^2+ \left(50 A'+\sqrt{10} \left(2 \phi _2'-9 \chi
   '\right)+10 \sqrt{2} \phi _1'\right)\6_{\r}\Bigg)+\nn \\
   &&e^{\frac{3 \phi _1}{\sqrt{2}}} \Bigg(16 e^{\sqrt{\frac{2}{5}} \phi _2} \left(A'\right)^2 \left(-10 e^{2 A+\sqrt{2} \phi _1+\sqrt{\frac{2}{5}} \phi
   _2} \left(\cA_7^{(1)\prime}(\rho )\right)^2
   +\right.\nn\\
   &&
   +\left.e^{2 A+4 \sqrt{\frac{2}{5}} \chi } \left(-4 \sqrt{10} \chi ''(\rho )+5 \sqrt{2} \phi_1''(\rho )+\sqrt{10}
   \phi_2''(\rho )+\chi ' \left(4 \chi '-\sqrt{5} \phi _1'-\phi _2'\right)- \right. \right. \nn \\
   && \left. \left. 5 e^{4 \sqrt{\frac{2}{5}} \phi _2}+20 e^{\frac{\phi
   _1}{\sqrt{2}}+\frac{3 \phi _2}{\sqrt{10}}}\right)+10 M^2 e^{\sqrt{10} \chi }\right)+80 \sqrt{2} \left(A'\right)^3 e^{2 A+\sqrt{\frac{2}{5}}
   \left(4 \chi +\phi _2\right)} \left(\sqrt{5} \left(\phi _2'-4 \chi '\right)+5 \phi _1'\right)+\nn \\
   && 5 e^{2 A+\sqrt{2} \phi _1+\frac{7 \phi
   _2}{\sqrt{10}}} \left({\cA_7^{(1)}}\right)'^2 \left(-9 \sinh \left(\sqrt{\frac{5}{2}} \phi _2\right)+8 \cosh \left(\frac{\phi _1}{\sqrt{2}}\right)+7 \cosh
   \left(\sqrt{\frac{5}{2}} \phi _2\right)\right)\Bigg)\Bigg]\fa^{\cA_7^{(1)}} +\nn \\
   &&\Bigg[-5 \left(-4 e^{\sqrt{\frac{5}{2}} \phi _2} \left(e^{\sqrt{2} \phi _1}+1\right)-8 e^{\frac{\phi _1}{\sqrt{2}}}+e^{\frac{\phi _1}{\sqrt{2}}+\sqrt{10}
   \phi _2}\right) e^{2 A+\sqrt{\frac{2}{5}} \phi _2} {\cA_7^{(1)}}' {\cA_7^{(2)}}' \Bigg]\fa^{\cA_7^{(2)}}+ \nn \\
   &&\Bigg[128 \sqrt{10} \left(A'\right)^2 e^{2 A+\sqrt{\frac{2}{5}} \left(4 \chi +\phi _2\right)+\frac{3 \phi _1}{\sqrt{2}}}{\cA_7^{(1)}}'\6_{\r}+4 e^{2 A+4 \sqrt{\frac{2}{5}} \chi +\sqrt{2} \phi _1} \Bigg(-8 e^{\frac{\phi _1}{\sqrt{2}}} \left(\sqrt{10} A'-5 \chi '\right){\cA_7^{(1)}}'+\nn \\
   &&e^{\frac{\phi _1}{\sqrt{2}}+\sqrt{10} \phi _2} \left(\sqrt{10} A'-5 \chi '\right) {\cA_7^{(1)}}'-4 e^{\sqrt{\frac{5}{2}} \phi _2}
   \left(e^{\sqrt{2} \phi _1}+1\right) \left(\sqrt{10} A'-5 \chi '\right) {\cA_7^{(1)}}'-\nn \\
   && 16 e^{\frac{\phi _1}{\sqrt{2}}+\sqrt{\frac{2}{5}} \phi
   _2} \left(A'\right)^2 \left({\cA_7^{(1)}}' \left(5 \sqrt{10} A'-9 \chi '+2 \sqrt{5} \phi _1'+2 \phi _2'\right)+\sqrt{10}{\cA_7^{(1)}}''\right)\Bigg)\Bigg]\fa^{\c}\nn  \, ,
\eeqs

\beqs
0&=&\Bigg[-160 \sqrt{2} \left(A'\right)^2 e^{2 A+\sqrt{\frac{2}{5}} \left(4 \chi +\phi _2\right)+\sqrt{2} \phi _1} {\cA_7^{(2)}}'\6_{\r}+e^{2 A+4 \sqrt{\frac{2}{5}} \chi +\frac{\phi _1}{\sqrt{2}}} \Bigg(-400 \sqrt{2} e^{\frac{\phi _1}{\sqrt{2}}+\sqrt{\frac{2}{5}} \phi _2}
   \left(A'\right)^3 {\cA_7^{(2)}}'+\\
   && 40 \sqrt{2} e^{\sqrt{\frac{5}{2}} \phi _2} \left(e^{\sqrt{2} \phi _1}-1\right) A'{\cA_7^{(2)}}'+16
   e^{\frac{\phi _1}{\sqrt{2}}+\sqrt{\frac{2}{5}} \phi _2} \left(A'\right)^2 \left({\cA_7^{(2)}}'\left(\sqrt{5} \left(9 \chi '-2 \phi
   _2'\right)+10 \phi _1'\right)-5 \sqrt{2}{\cA_7^{(2)}}''\right)+\nn \\
   && 5 \left(4 e^{\sqrt{\frac{5}{2}} \phi _2} \left(e^{\sqrt{2} \phi _1}+1\right)+8
   e^{\frac{\phi _1}{\sqrt{2}}}-e^{\frac{\phi _1}{\sqrt{2}}+\sqrt{10} \phi _2}\right) \phi _1'{\cA_7^{(2)}}'\Bigg)\Bigg]\fa^{\f_1} + \nn \\
   && \Bigg[32 \sqrt{10} \left(A'\right)^2 e^{2 A+\sqrt{\frac{2}{5}} \left(4 \chi +\phi _2\right)+\sqrt{2} \phi _1}{\cA_7^{(2)}}'\6_{\r}+e^{2 A+4 \sqrt{\frac{2}{5}} \chi +\frac{\phi _1}{\sqrt{2}}} \Bigg(-8 e^{\frac{\phi _1}{\sqrt{2}}} \left(4 \sqrt{10} A'-5 \phi
   _2'\right){\cA_7^{(2)}}'+\nn \\
   &&4 e^{\sqrt{\frac{5}{2}} \phi _2} \left(e^{\sqrt{2} \phi _1}+1\right) \left(6 \sqrt{10} A'+5 \phi _2'\right)
   {\cA_7^{(2)}}'-e^{\frac{\phi _1}{\sqrt{2}}+\sqrt{10} \phi _2} \left(16 \sqrt{10} A'+5 \phi _2'\right) {\cA_7^{(2)}}'+\nn \\
   && 16 e^{\frac{\phi
   _1}{\sqrt{2}}+\sqrt{\frac{2}{5}} \phi _2} \left(A'\right)^2 \left({\cA_7^{(2)}}' \left(5 \sqrt{10} A'-9 \chi '-2 \sqrt{5} \phi _1'+2 \phi
   _2'\right)+\sqrt{10} {\cA_7^{(2)}}''\right)\Bigg)\Bigg]\fa^{\f_2}+ \nn \\
   &&\Bigg[5 e^{2 A+2 \sqrt{2} \phi _1+\frac{7 \phi _2}{\sqrt{10}}}  {\cA_7^{(1)}}' {\cA_7^{(2)}}' \left(-9 \sinh \left(\sqrt{\frac{5}{2}} \phi
   _2\right)+8 \cosh \left(\frac{\phi _1}{\sqrt{2}}\right)+7 \cosh \left(\sqrt{\frac{5}{2}} \phi _2\right)\right)\Bigg]\fa^{\cA_7^{(1)}}+ \nn \\
   && \Bigg[16 e^{2 A+\sqrt{\frac{2}{5}} \left(4 \chi +\phi _2\right)+\sqrt{2} \phi _1} \Bigg(10 \left(A'\right)^2 \6_{\r}^2+\left(A'\right)^2 \left(50
   A'+\sqrt{10} \left(2 \phi _2'-9 \chi '\right)-10 \sqrt{2} \phi _1'\right)\6_{\r}\Bigg)
     +\nn\\
   &&16 e^{\sqrt{\frac{2}{5}} \phi _2} \left(A'\right)^2 \Bigg(-10 e^{\sqrt{\frac{2}{5}} \phi _{2 A+2}} \left({\cA_7^{(2)}}'\right)^2+\nn \\
   &&e^{2 A+4 \sqrt{\frac{2}{5}}
   \chi +\sqrt{2} \phi _1} \left(-5 \sqrt{2} \phi_1''(\rho )+\sqrt{10} \left(\phi_2''(\rho )-4 \chi ''(\rho )\right)+\chi ' \left(4
   \chi '+\sqrt{5} \phi _1'-\phi _2'\right)-5 e^{4 \sqrt{\frac{2}{5}} \phi _2}\right)+\nn \\
   && 20 e^{2 A+4 \sqrt{\frac{2}{5}} \chi +\frac{\phi
   _1}{\sqrt{2}}+\frac{3 \phi _2}{\sqrt{10}}}+10 M^2 e^{\sqrt{2} \left(\sqrt{5} \chi +\phi _1\right)}\Bigg)-80 \sqrt{2} \left(A'\right)^3 e^{2
   A+\sqrt{\frac{2}{5}} \left(4 \chi +\phi _2\right)+\sqrt{2} \phi _1} \left(5 \phi _1'-\sqrt{5} \left(\phi _2'-4 \chi '\right)\right)-\nn \\
   && 5 e^{2
   A+\sqrt{\frac{2}{5}} \phi _2} \left(e^{\sqrt{10} \phi _2}-8 e^{\sqrt{\frac{5}{2}} \phi _2} \cosh \left(\frac{\phi _1}{\sqrt{2}}\right)-8\right)
   \left({\cA_7^{(2)}}'\right)^2\Bigg]\fa^{\cA_7^{(2)}} +\nn \\
   && \Bigg[-128 \sqrt{10} \left(A'\right)^2 e^{2 A+\sqrt{\frac{2}{5}} \left(4 \chi +\phi _2\right)+\sqrt{2} \phi _1} {\cA_7^{(2)}}'\6_{\r}+4 e^{2 A+4 \sqrt{\frac{2}{5}} \chi +\frac{\phi _1}{\sqrt{2}}} \Bigg(-8 e^{\frac{\phi _1}{\sqrt{2}}} \left(\sqrt{10} A'-5 \chi '\right)
   {\cA_7^{(2)}}'+\nn \\
   && e^{\frac{\phi _1}{\sqrt{2}}+\sqrt{10} \phi _2} \left(\sqrt{10} A'-5 \chi '\right){\cA_7^{(2)}}'-4 e^{\sqrt{\frac{5}{2}}
   \phi _2} \left(e^{\sqrt{2} \phi _1}+1\right) \left(\sqrt{10} A'-5 \chi '\right){\cA_7^{(2)}}'-\nn \\
   && 16 e^{\frac{\phi
   _1}{\sqrt{2}}+\sqrt{\frac{2}{5}} \phi _2} \left(A'\right)^2 \left({\cA_7^{(2)}}' \left(5 \sqrt{10} A'-9 \chi '-2 \sqrt{5} \phi _1'+2 \phi
   _2'\right)+\sqrt{10}{\cA_7^{(2)}}''\right)\Bigg)\Bigg]\fa^{\c} \nn .
\eeqs

The boundary conditions for the five gauge-invariant, scalar fluctuations are given by the following expressions.
\beqs
\label{eq: BCphi1}
0&=&\Bigg[40 A' \left(\phi _1'\right){}^2 e^{2 A+\sqrt{\frac{2}{5}} \left(4 \chi +\phi _2\right)+\sqrt{2} \phi _1}\6_{\r}+\\
&&20 \sqrt{2} e^{2 A+2 \sqrt{\frac{2}{5}} \phi _2} A' \phi _1' \left(e^{2 \sqrt{2} \phi _1} \left({\cA_7^{(1)}}'\right)^2-\left({\cA_7^{(2)}}'\right)^2+2
   \left(e^{\sqrt{2} \phi _1}-1\right) e^{\frac{8 \chi +\phi _2}{\sqrt{10}}+\frac{\phi _1}{\sqrt{2}}}\right)+ \nn \\
   && 320 M^2 \left(A'\right)^2
   e^{\sqrt{\frac{2}{5}} \left(5 \chi +\phi _2\right)+\sqrt{2} \phi _1}+5 \left(4 e^{\sqrt{\frac{5}{2}} \phi _2} \left(e^{\sqrt{2} \phi
   _1}+1\right)+8 e^{\frac{\phi _1}{\sqrt{2}}}-e^{\frac{\phi _1}{\sqrt{2}}+\sqrt{10} \phi _2}\right) \left(\phi _1'\right){}^2 e^{2 A+4
   \sqrt{\frac{2}{5}} \chi +\frac{\phi _1}{\sqrt{2}}}\Bigg]\fa^{\f_1}\Big|_{\r=\r_i} +\nn \\
   && \Bigg[40 A' \phi _1' \phi _2' e^{2 A+\sqrt{\frac{2}{5}} \left(4 \chi +\phi _2\right)+\sqrt{2} \phi _1}\6_{\r}+e^{2 A} \phi _1' \Bigg(4 \sqrt{10} e^{2 \sqrt{\frac{2}{5}} \phi _2} A' \left(e^{2 \sqrt{2} \phi _1} \left({\cA_7^{(1)}}'\right)^2+\left({\cA_7^{(2)}}'\right)^2\right)
     -\nn\\
   &&
   8 e^{4 \sqrt{\frac{2}{5}} \chi +\sqrt{2} \phi _1} \left(4 \sqrt{10} A'-5 \phi _2'\right)+\nn \\
   &&4 \left(e^{\sqrt{2} \phi _1}+1\right)
   e^{\frac{8 \chi +5 \phi _2}{\sqrt{10}}+\frac{\phi _1}{\sqrt{2}}} \left(6 \sqrt{10} A'+5 \phi _2'\right)-e^{\sqrt{\frac{2}{5}} \left(4 \chi +5 \phi
   _2\right)+\sqrt{2} \phi _1} \left(16 \sqrt{10} A'+5 \phi _2'\right)\Bigg)\Bigg]\fa^{\f_2}\Big|_{\r=\r_i}+ \nn \\
   &&\Bigg[40 e^{2 A+2 \sqrt{2} \phi _1+2 \sqrt{\frac{2}{5}} \phi _2} A' \phi _1'{\cA_7^{(1)}}'\6_{\r}+5 e^{2 A+2 \sqrt{2} \phi _1+\frac{7 \phi _2}{\sqrt{10}}} \phi _1'{\cA_7^{(1)}}'\left(-9 \sinh \left(\sqrt{\frac{5}{2}} \phi _2\right)
     +\right.\nn\\
   &&
   \left.
   8 \cosh
   \left(\frac{\phi _1}{\sqrt{2}}\right)+7 \cosh \left(\sqrt{\frac{5}{2}} \phi _2\right)\right)\Bigg]\fa^{\cA_7^{(1)}}\Big|_{\r=\r_i}+ \nn \\
   &&\Bigg[40 e^{2 A+2 \sqrt{\frac{2}{5}} \phi _2} A' \phi _1' {\cA_7^{(2)}}'(\rho )\6_{\r}-5 e^{2 A+\sqrt{\frac{2}{5}} \phi _2} \phi _1' \left(e^{\sqrt{10} \phi _2}-8 e^{\sqrt{\frac{5}{2}} \phi _2} \cosh \left(\frac{\phi
   _1}{\sqrt{2}}\right)-8\right){\cA_7^{(2)}}'(\rho )\Bigg]\fa^{\cA_7^{(2)}}\Big|_{\r=\r_i}+\nn \\
   &&\Bigg[160 A' \chi ' \phi _1' e^{2 A+\sqrt{\frac{2}{5}} \left(4 \chi +\phi _2\right)+\sqrt{2} \phi _1}\6_{\r}+4 e^{2 A} \phi _1' \left(\left(-4 e^{\sqrt{\frac{5}{2}} \phi _2} \left(e^{\sqrt{2} \phi _1}+1\right)
     -\right.\right.\nn\\
   &&
   \left.\left.
   8 e^{\frac{\phi _1}{\sqrt{2}}}+e^{\frac{\phi
   _1}{\sqrt{2}}+\sqrt{10} \phi _2}\right) e^{4 \sqrt{\frac{2}{5}} \chi +\frac{\phi _1}{\sqrt{2}}} \left(\sqrt{10} A'-5 \chi '\right)
   -
   4 \sqrt{10}
   e^{2 \sqrt{\frac{2}{5}} \phi _2} A' \left(e^{2 \sqrt{2} \phi _1} \left({\cA_7^{(1)}}'\right)^2+\left({\cA_7^{(2)}}'\right)^2\right)\right)\Bigg]\fa^{\c}\Big|_{\r=\r_i}\nn \,  ,
\eeqs

\beqs
\label{eq: BCphi2}
0&=&\Bigg[40 A' \phi _1' \phi _2' e^{2 A+\sqrt{\frac{2}{5}} \left(4 \chi +\phi _2\right)+\sqrt{2} \phi _1}\6_{\r}+5 e^{2 A} \phi _2' \Bigg(4 \sqrt{2} e^{2 \sqrt{\frac{2}{5}} \phi _2} A' \left(e^{2 \sqrt{2} \phi _1} \left({\cA_7^{(1)}}'\right)^2-
  \right.\nn\\
   &&
   \left.
\left({\cA_7^{(2)}}'\right)^2+2\left(e^{\sqrt{2} \phi _1}-1\right) e^{\frac{8 \chi +\phi _2}{\sqrt{10}}+\frac{\phi _1}{\sqrt{2}}}\right)+ \\
&&\left(4 e^{\sqrt{\frac{5}{2}} \phi _2}
\left(e^{\sqrt{2} \phi _1}+1\right)+8 e^{\frac{\phi _1}{\sqrt{2}}}-e^{\frac{\phi _1}{\sqrt{2}}+\sqrt{10} \phi _2}\right) e^{4 \sqrt{\frac{2}{5}}\chi +\frac{\phi _1}{\sqrt{2}}} \phi _1'\Bigg)\Bigg]\fa^{\f_1}\Big|_{\r=\r_i}+ \nn \\
&&\Bigg[40 A' \left(\phi _2'\right){}^2 e^{2 A+\sqrt{\frac{2}{5}} \left(4 \chi +\phi _2\right)+\sqrt{2} \phi _1}\6_{\r}+4 \sqrt{10} e^{2 A} A' \phi _2' \left(e^{2 \sqrt{\frac{2}{5}} \phi _2} \left(e^{2 \sqrt{2} \phi _1} \left({\cA_7^{(1)}}'\right)^2+\left({\cA_7^{(2)}}'\right)^2\right)+ \right.  \nn \\
   && \left.2 \left(3 e^{\sqrt{\frac{5}{2}} \phi _2} \left(e^{\sqrt{2} \phi _1}+1\right)-4 e^{\frac{\phi _1}{\sqrt{2}}}-2 e^{\frac{\phi
   _1}{\sqrt{2}}+\sqrt{10} \phi _2}\right) e^{4 \sqrt{\frac{2}{5}} \chi +\frac{\phi _1}{\sqrt{2}}}\right)+320 M^2 \left(A'\right)^2
   e^{\sqrt{\frac{2}{5}} \left(5 \chi +\phi _2\right)+\sqrt{2} \phi _1}+\nn \\
   && 5 \left(4 e^{\sqrt{\frac{5}{2}} \phi _2} \left(e^{\sqrt{2} \phi
   _1}+1\right)+8 e^{\frac{\phi _1}{\sqrt{2}}}-e^{\frac{\phi _1}{\sqrt{2}}+\sqrt{10} \phi _2}\right) \left(\phi _2'\right){}^2 e^{2 A+4
   \sqrt{\frac{2}{5}} \chi +\frac{\phi _1}{\sqrt{2}}}\Bigg]\fa^{\f_2}\Big|_{\r=\r_i}+\nn \\
&&\Bigg[40 e^{2 A+2 \sqrt{2} \phi _1+2 \sqrt{\frac{2}{5}} \phi _2} A' \phi _2' {\cA_7^{(1)}}'\6_{\r}+5 e^{2 A+2 \sqrt{2} \phi _1+\frac{7 \phi _2}{\sqrt{10}}} \phi _2' {\cA_7^{(1)}}' \left(-9 \sinh \left(\sqrt{\frac{5}{2}} \phi _2\right)
  +\right.\nn\\
   &&
   \left.
8 \cosh
   \left(\frac{\phi _1}{\sqrt{2}}\right)+7 \cosh \left(\sqrt{\frac{5}{2}} \phi _2\right)\right) \Bigg]\fa^{\cA_7^{(1)}}\Big|_{\r=\r_i}+\nn \\
&&\Bigg[40 e^{2 A+2 \sqrt{\frac{2}{5}} \phi _2} A' \phi _2'  {\cA_7^{(2)}}'\6_{\r}-5 e^{2 A+\sqrt{\frac{2}{5}} \phi _2} \phi _2' \left(e^{\sqrt{10} \phi _2}-8 e^{\sqrt{\frac{5}{2}} \phi _2} \cosh \left(\frac{\phi
   _1}{\sqrt{2}}\right)-8\right) {\cA_7^{(2)}}'\Bigg]\fa^{\cA_7^{(2)}}\Big|_{\r=\r_i}+\nn \\
&&\Bigg[160 A' \chi ' \phi _2' e^{2 A+\sqrt{\frac{2}{5}} \left(4 \chi +\phi _2\right)+\sqrt{2} \phi _1}\6_{\r}+4 e^{2 A} \phi _2' \left(\left(-4 e^{\sqrt{\frac{5}{2}} \phi _2} \left(e^{\sqrt{2} \phi _1}+1\right)-
  \right.\right.\nn\\
   &&
   \left.\left.
8 e^{\frac{\phi _1}{\sqrt{2}}}+e^{\frac{\phi
   _1}{\sqrt{2}}+\sqrt{10} \phi _2}\right) e^{4 \sqrt{\frac{2}{5}} \chi +\frac{\phi _1}{\sqrt{2}}} \left(\sqrt{10} A'-5 \chi '\right)-\right. \nn \\
   && \left. 4 \sqrt{10}
   e^{2 \sqrt{\frac{2}{5}} \phi _2} A' \left(e^{2 \sqrt{2} \phi _1}\left({\cA_7^{(1)}}'\right)^2+\left({\cA_7^{(2)}}'\right)^2\right)\right)\Bigg]\fa^{\c}\Big|_{\r=\r_i}\nn \,,
\eeqs

\beqs
\label{eq: BCA71}
0&=&\Bigg[40 A' \chi ' \phi _1' e^{2 A+\sqrt{\frac{2}{5}} \left(4 \chi +\phi _2\right)+\sqrt{2} \phi _1}\6_{\r}+\\
&& 5 e^{2 A} \chi ' \Bigg(4 \sqrt{2} e^{2 \sqrt{\frac{2}{5}} \phi _2} A' \left(e^{2 \sqrt{2} \phi _1} \left({\cA_7^{(1)}}'\right)^2-\left({\cA_7^{(2)}}'\right)^2+2 \left(e^{\sqrt{2} \phi _1}-1\right) e^{\frac{8 \chi +\phi _2}{\sqrt{10}}+\frac{\phi_1}{\sqrt{2}}}\right)+ \nn \\ 
&&\left(4 e^{\sqrt{\frac{5}{2}} \phi _2} \left(e^{\sqrt{2} \phi _1}+1\right)+8 e^{\frac{\phi _1}{\sqrt{2}}}-e^{\frac{\phi _1}{\sqrt{2}}+\sqrt{10} \phi _2}\right) e^{4 \sqrt{\frac{2}{5}} \chi +\frac{\phi _1}{\sqrt{2}}}
   \phi _1'\Bigg)\Bigg]\fa^{\f_1}\Big|_{\r=\r_i}+ \nn \\
&&\Bigg[40 A' \chi ' \phi _2' e^{2 A+\sqrt{\frac{2}{5}} \left(4 \chi +\phi _2\right)+\sqrt{2} \phi _1}\6_{\r}+ e^{2 A} \chi ' \Bigg(4 \sqrt{10} e^{2 \sqrt{\frac{2}{5}} \phi _2} A' \left(e^{2 \sqrt{2} \phi _1} \left({\cA_7^{(1)}}'\right)^2+\left({\cA_7^{(2)}}'\right)^2\right)-\nn \\
&& 8 e^{4 \sqrt{\frac{2}{5}} \chi +\sqrt{2} \phi _1} \left(4 \sqrt{10} A'-5 \phi _2'\right)+4
   \left(e^{\sqrt{2} \phi _1}+1\right) e^{\frac{8 \chi +5 \phi _2}{\sqrt{10}}+\frac{\phi _1}{\sqrt{2}}} \left(6 \sqrt{10} A'+5 \phi _2'\right)-
 \nn\\
   &&
   e^{\sqrt{\frac{2}{5}} \left(4 \chi +5 \phi _2\right)+\sqrt{2} \phi _1} \left(16 \sqrt{10} A'+5
   \phi _2'\right)\Bigg)\Bigg]\fa^{\f_2}\Big|_{\r=\r_i}+\nn \\
&&\Bigg[40 e^{2 A+2 \sqrt{2} \phi _1+2 \sqrt{\frac{2}{5}} \phi _2} A' \chi ' {\cA_7^{(1)}}'\6_{\r}+5 e^{2 A+2 \sqrt{2} \phi _1+\frac{7 \phi _2}{\sqrt{10}}} \chi ' {\cA_7^{(1)}}'\left(-9 \sinh \left(\sqrt{\frac{5}{2}} \phi _2\right)
  +\right.\nn\\
   &&
   \left.
8 \cosh \left(\frac{\phi _1}{\sqrt{2}}\right)+7 \cosh \left(\sqrt{\frac{5}{2}} \phi _2\right)\right)\Bigg]\fa^{\cA_7^{(1)}}\Big|_{\r=\r_i}+\nn \\
&&\Bigg[40 e^{2 A+2 \sqrt{\frac{2}{5}} \phi _2} A' \chi ' {\cA_7^{(2)}}'\6_{\r}-5 e^{2 A+\sqrt{\frac{2}{5}} \phi _2} \chi ' \left(e^{\sqrt{10} \phi _2}-8 e^{\sqrt{\frac{5}{2}} \phi _2} \cosh \left(\frac{\phi _1}{\sqrt{2}}\right)-8\right) {\cA_7^{(2)}}'\Bigg]\fa^{\cA_7^{(2)}}\Big|_{\r=\r_i}+\nn \\
&&\Bigg[160 A' \left(\chi '\right)^2 e^{2 A+\sqrt{\frac{2}{5}} \left(4 \chi +\phi _2\right)+\sqrt{2} \phi _1}\6_{\r}+4 \Bigg(-\sqrt{10} e^{2 A} A' \chi ' \left(4 e^{2 \sqrt{\frac{2}{5}} \phi _2} \left(e^{2 \sqrt{2} \phi _1}\left({\cA_7^{(1)}}'\right)^2+\left({\cA_7^{(2)}}'\right)^2\right)+\right. \nn \\
&& \left.  \left(4 e^{\sqrt{\frac{5}{2}} \phi _2} \left(e^{\sqrt{2} \phi _1}+1\right)+8
   e^{\frac{\phi _1}{\sqrt{2}}}-e^{\frac{\phi _1}{\sqrt{2}}+\sqrt{10} \phi _2}\right) e^{4 \sqrt{\frac{2}{5}} \chi +\frac{\phi _1}{\sqrt{2}}}\right)+80 M^2 \left(A'\right)^2 e^{\sqrt{\frac{2}{5}} \left(5 \chi +\phi _2\right)+\sqrt{2} \phi
   _1}+\nn \\
   && 5 \left(4 e^{\sqrt{\frac{5}{2}} \phi _2} \left(e^{\sqrt{2} \phi _1}+1\right)+8 e^{\frac{\phi _1}{\sqrt{2}}}-e^{\frac{\phi _1}{\sqrt{2}}+\sqrt{10} \phi _2}\right) \left(\chi '\right)^2 e^{2 A+4 \sqrt{\frac{2}{5}} \chi +\frac{\phi
   _1}{\sqrt{2}}}\Bigg)\Bigg]\fa^{\c}\Big|_{\r=\r_i}\nn \, ,
\eeqs

\beqs
\label{eq: BCA72}
0&=&\Bigg[40 A' \phi _1' e^{2 A+\sqrt{\frac{2}{5}} \left(4 \chi +\phi _2\right)+\frac{3 \phi _1}{\sqrt{2}}} {\cA_7^{(1)}}'\6_{\r}+5 e^{2 A+\frac{\phi _1}{\sqrt{2}}} {\cA_7^{(1)}}' \Bigg(4 \sqrt{2} e^{2 \sqrt{\frac{2}{5}} \phi _2} A' \left(e^{2 \sqrt{2} \phi _1} \left({\cA_7^{(1)}}\right)'^2-\left({\cA_7^{(2)}}'\right)^2+\right. \\
&& \left. 2 \left(e^{\sqrt{2} \phi _1}-1\right) e^{\frac{8 \chi +\phi _2}{\sqrt{10}}+\frac{\phi
   _1}{\sqrt{2}}}\right)+\left(4 e^{\sqrt{\frac{5}{2}} \phi _2} \left(e^{\sqrt{2} \phi _1}+1\right)+8 e^{\frac{\phi _1}{\sqrt{2}}}-e^{\frac{\phi _1}{\sqrt{2}}+\sqrt{10} \phi _2}\right) e^{4 \sqrt{\frac{2}{5}} \chi +\frac{\phi _1}{\sqrt{2}}} \phi _1'\Bigg)\Bigg]\fa^{\f_1}\Big|_{\r=\r_i}+\nn \\
&&\Bigg[40 A' \phi _2' e^{2 A+\sqrt{\frac{2}{5}} \left(4 \chi +\phi _2\right)+\frac{3 \phi _1}{\sqrt{2}}} {\cA_7^{(1)}}'\6_{\r}+e^{2 A+\frac{\phi _1}{\sqrt{2}}} {\cA_7^{(1)}}'\Bigg(4 \sqrt{10} e^{2 \sqrt{\frac{2}{5}} \phi _2} A' \left(e^{2 \sqrt{2} \phi _1} \left({\cA_7^{(1)}}'\right)^2+\left({\cA_7^{(2)}}'\right)^2\right)-\nn \\
&& 8 e^{4 \sqrt{\frac{2}{5}} \chi +\sqrt{2} \phi _1} \left(4 \sqrt{10} A'-5 \phi _2'\right)+4
   \left(e^{\sqrt{2} \phi _1}+1\right) e^{\frac{8 \chi +5 \phi _2}{\sqrt{10}}+\frac{\phi _1}{\sqrt{2}}} \left(6 \sqrt{10} A'+5 \phi _2'\right)-
     \nn\\
   &&
   e^{\sqrt{\frac{2}{5}} \left(4 \chi +5 \phi _2\right)+\sqrt{2} \phi _1} \left(16 \sqrt{10} A'+5 \phi _2'\right)\Bigg)\Bigg]\fa^{\f_2}\Big|_{\r=\r_i}+\nn \\
&&\Bigg[40 e^{2 A+\frac{5 \phi _1}{\sqrt{2}}+2 \sqrt{\frac{2}{5}} \phi _2} A' \left({\cA_7^{(1)}}'\right)^2\6_{\r}+5 e^{\frac{3 \phi _1}{\sqrt{2}}+\sqrt{\frac{2}{5}} \phi _2} \Bigg(64 M^2 e^{\sqrt{10} \chi } \left(A'\right)^2+\nn \\
&& e^{2 A} \left(8 e^{\sqrt{2} \phi _1}+4 e^{\frac{\phi _1+\sqrt{5} \phi _2}{\sqrt{2}}}-e^{\sqrt{2} \left(\phi _1+\sqrt{5} \phi _2\right)}+4 e^{\frac{3 \phi _1+\sqrt{5} \phi
   _2}{\sqrt{2}}}\right) \left({\cA_7^{(1)}}'\right)^2\Bigg)\Bigg]\fa^{\cA_7^{(1)}}\Big|_{\r=\r_i}+\nn \\
&&\Bigg[40 e^{2 A+\frac{\phi _1}{\sqrt{2}}+2 \sqrt{\frac{2}{5}} \phi _2} A' {\cA_7^{(1)}}' {\cA_7^{(2)}}'\6_{\r}-5 \left(-4 e^{\sqrt{\frac{5}{2}} \phi _2} \left(e^{\sqrt{2} \phi _1}+1\right)-8 e^{\frac{\phi _1}{\sqrt{2}}}+e^{\frac{\phi _1}{\sqrt{2}}+\sqrt{10} \phi _2}\right) e^{2 A+\sqrt{\frac{2}{5}} \phi _2}{\cA_7^{(1)}}'{\cA_7^{(2)}}'\Bigg]\fa^{\cA_7^{(2)}}\Big|_{\r=\r_i}+\nn \\
&&\Bigg[160 A' \chi ' e^{2 A+\sqrt{\frac{2}{5}} \left(4 \chi +\phi _2\right)+\frac{3 \phi _1}{\sqrt{2}}} {\cA_7^{(1)}}'\6_{\r}+\nn \\
&& 4 e^{2 A+\frac{\phi _1}{\sqrt{2}}} {\cA_7^{(1)}}' \Bigg(\left(-4 e^{\sqrt{\frac{5}{2}} \phi _2} \left(e^{\sqrt{2} \phi _1}+1\right)-8 e^{\frac{\phi _1}{\sqrt{2}}}+e^{\frac{\phi _1}{\sqrt{2}}+\sqrt{10} \phi _2}\right) e^{4 \sqrt{\frac{2}{5}} \chi +\frac{\phi _1}{\sqrt{2}}}
   \left(\sqrt{10} A'-5 \chi '\right)-\nn \\
&& 4 \sqrt{10} e^{2 \sqrt{\frac{2}{5}} \phi _2} A' \left(e^{2 \sqrt{2} \phi _1} \left({\cA_7^{(1)}}'\right)^2+\left({\cA_7^{(2)}}'\right)^2\right)\Bigg)\Bigg]\fa^{\c}\Big|_{\r=\r_i}\nn \, ,
\eeqs

\beqs
\label{eq: BCchi2}
0&=&\Bigg[40 A' \phi _1' e^{2 A+\sqrt{\frac{2}{5}} \left(4 \chi +\phi _2\right)+\frac{3 \phi _1}{\sqrt{2}}} {\cA_7^{(2)}}'\6_{\r}+5 e^{2 A+\frac{\phi _1}{\sqrt{2}}} {\cA_7^{(2)}}' \Bigg(4 \sqrt{2} e^{2 \sqrt{\frac{2}{5}} \phi _2} A' \left(e^{2 \sqrt{2} \phi _1} \left({\cA_7^{(1)}}'\right)^2-\left({\cA_7^{(2)}}'\right)^2+\right. \\
&& \left. 2 \left(e^{\sqrt{2} \phi _1}-1\right) e^{\frac{8 \chi +\phi _2}{\sqrt{10}}+\frac{\phi
   _1}{\sqrt{2}}}\right)+\left(4 e^{\sqrt{\frac{5}{2}} \phi _2} \left(e^{\sqrt{2} \phi _1}+1\right)+8 e^{\frac{\phi _1}{\sqrt{2}}}-e^{\frac{\phi _1}{\sqrt{2}}+\sqrt{10} \phi _2}\right) e^{4 \sqrt{\frac{2}{5}} \chi +\frac{\phi _1}{\sqrt{2}}} \phi _1'\Bigg)\Bigg]\fa^{\f_1}\Big|_{\r=\r_i}+ \nn \\
&&\Bigg[40 A' \phi _2' e^{2 A+\sqrt{\frac{2}{5}} \left(4 \chi +\phi _2\right)+\frac{3 \phi _1}{\sqrt{2}}} {\cA_7^{(2)}}'\6_{\r}+e^{2 A+\frac{\phi _1}{\sqrt{2}}} {\cA_7^{(2)}}'\Bigg(4 \sqrt{10} e^{2 \sqrt{\frac{2}{5}} \phi _2} A' \left(e^{2 \sqrt{2} \phi _1}\left({\cA_7^{(1)}}'\right)^2+\left({\cA_7^{(2)}}'\right)^2\right)-\nn \\
&& 8 e^{4 \sqrt{\frac{2}{5}} \chi +\sqrt{2} \phi _1} \left(4 \sqrt{10} A'-5 \phi _2'\right)+4
   \left(e^{\sqrt{2} \phi _1}+1\right) e^{\frac{8 \chi +5 \phi _2}{\sqrt{10}}+\frac{\phi _1}{\sqrt{2}}} \left(6 \sqrt{10} A'+5 \phi _2'\right)-
   \nn\\
   &&
   e^{\sqrt{\frac{2}{5}} \left(4 \chi +5 \phi _2\right)+\sqrt{2} \phi _1} \left(16 \sqrt{10} A'+5 \phi _2'\right)\Bigg)\Bigg]\fa^{\f_2}\Big|_{\r=\r_i}+\nn \\
&&\Bigg[40 e^{2 A+\frac{5 \phi _1}{\sqrt{2}}+2 \sqrt{\frac{2}{5}} \phi _2} A' {\cA_7^{(1)}}'{\cA_7^{(2)}}'\6_{\r}+\nn \\
&& 5 e^{2 A+\frac{5 \phi _1}{\sqrt{2}}+\frac{7 \phi _2}{\sqrt{10}}} {\cA_7^{(1)}}' {\cA_7^{(2)}}' \left(-9 \sinh \left(\sqrt{\frac{5}{2}} \phi _2\right)+8 \cosh \left(\frac{\phi _1}{\sqrt{2}}\right)+7 \cosh \left(\sqrt{\frac{5}{2}} \phi _2\right)\right)\Bigg]\fa^{\cA_7^{(1)}}\Big|_{\r=\r_i}+\nn \\
&&\Bigg[40 e^{2 A+\frac{\phi _1}{\sqrt{2}}+2 \sqrt{\frac{2}{5}} \phi _2} A' \left({\cA_7^{(2)}}'\right)^2\6_{\r}+\nn \\
&& 5 e^{\sqrt{\frac{2}{5}} \phi _2} \left(64 M^2 \left(A'\right)^2 e^{\sqrt{10} \chi +\frac{3 \phi _1}{\sqrt{2}}}+e^{2 A} \left(4 e^{\sqrt{\frac{5}{2}} \phi _2} \left(e^{\sqrt{2} \phi _1}+1\right)+8 e^{\frac{\phi _1}{\sqrt{2}}}-e^{\frac{\phi _1}{\sqrt{2}}+\sqrt{10} \phi _2}\right)
   \left({\cA_7^{(2)}}'\right)^2\right)\Bigg]\fa^{\cA_7^{(2)}}\Big|_{\r=\r_i}+\nn \\
&&\Bigg[160 A' \chi ' e^{2 A+\sqrt{\frac{2}{5}} \left(4 \chi +\phi _2\right)+\frac{3 \phi _1}{\sqrt{2}}} {\cA_7^{(2)}}'\6_{\r}+\nn \\
&& 4 e^{2 A+\frac{\phi _1}{\sqrt{2}}} {\cA_7^{(2)}}'\Bigg(\left(-4 e^{\sqrt{\frac{5}{2}} \phi _2} \left(e^{\sqrt{2} \phi _1}+1\right)-8 e^{\frac{\phi _1}{\sqrt{2}}}+e^{\frac{\phi _1}{\sqrt{2}}+\sqrt{10} \phi _2}\right) e^{4 \sqrt{\frac{2}{5}} \chi +\frac{\phi _1}{\sqrt{2}}}
   \left(\sqrt{10} A'-5 \chi '\right)-\nn \\ 
&& 4 \sqrt{10} e^{2 \sqrt{\frac{2}{5}} \phi _2} A' \left(e^{2 \sqrt{2} \phi _1} \left({\cA_7^{(1)}}'\right)^2+\left({\cA_7^{(2)}}'\right)^2\right)\Bigg)\Bigg]\fa^{\c}\Big|_{\r=\r_i}\nn \, .
\eeqs
In the numerical routines written for the purposes of this paper, these equations are rewritten after the further change of variable to the $\varrho$ parametrisation of the holographic direction. We decided not to report such expressions, in view of the uninspiring complexity of their explicit form.

%%%%%%%%%%%%%%%%%%%%%%%
%%%%%%%%%%%%%%%%%%%%%%%
\subsection{UV Expansions}
\label{Sec:UVexpansions}

We provide the leading terms in the UV expansions of the fluctuations used to improve the numerical calculation of the spectrum, in terms of variable $\fz=\frac{1}{\vr}$. In the numerical calculations of the spectra reported in the main body, we retained terms  up to tenth order in these expansions. The expansions depend on ten free parameters: the dominant $\left\{\mathfrak{a}_2^{\phi_1},\,\mathfrak{a}_2^{\phi_2},\,\mathfrak{a}_0^{\chi},\,\mathfrak{a}_0^{{\cal A}_7^{(1)}},\,\mathfrak{a}_0^{{\cal A}_7^{(2)}} \right\}$ and the subdominant 
$\left\{\mathfrak{a}_4^{\phi_1},\,\mathfrak{a}_4^{\phi_2},\,\mathfrak{a}_6^{\chi},\,\mathfrak{a}_4^{{\cal A}_7^{(1)}},\,\mathfrak{a}_4^{{\cal A}_7^{(2)}} \right\}$.

\beqs
\label{eq:FlucUVexp A71}
\fa^{\mathcal{A}_7^{(1)}}_{UV}(\fz)&=&\fa^{\mathcal{A}_7^{(1)}}_{0} + 4\fa^{\mathcal{A}_7^{(1)}}_{0}M^2\fz^2 + \fz^4 \left(\fa^{\mathcal{A}_7^{(1)}}_{ 4}-16\fa^{\mathcal{A}_7^{(1)}}_{0}M^4 \log[\fz]\right)+ \\ 
 & &\fz^6\left(\frac{64}{3}\fa^{\mathcal{A}_7^{(1)}}_{0}M^6\log[\fz]-\frac{2}{45}\sqrt{\mu}Q_1\left(30\fa^{\mathcal{A}_7^{(1)}}_{4}M^2-3\sqrt{2}(5\fa^{\f_1}_{UV, 2}+\sqrt{5}(\fa^{\f_1}_{2}-8\fa^{\c}_{0}M^2))\right)+10\fa^{\mathcal{A}_7^{(1)}}_0M^2(32 M^4-9Q_1^2+3Q_2^2)\right)+ \nn  \\
 && \fz^8\left(\frac{}{} Q_1^2 (-16 \fa^{\mathcal{A}_7^{(1)}}_{0} M^4 \log (\fz)-2 \fa^{\mathcal{A}_7^{(1)}}_{0} M^4+\fa^{\mathcal{A}_7^{(1)}}_{4})-\right.  \nn \\
 && \frac{2}{9} M^2 \left(\fa^{\mathcal{A}_7^{(1)}}_{0} \left(6 Q_2^2 M^2+9 \mu -50 M^6\right)+48 \fa^{\mathcal{A}_7^{(1)}}_{0} M^6 \log (\fz)-3 \fa^{\mathcal{A}_7^{(1)}}_{4} M^2\right)+  \nn \\
 && \left. \frac{Q_1 \sqrt{\mu }}{15 \sqrt{2}} \left(M^2 \left(5  \fa^{\f_1}_2+\sqrt{5} \left( \fa^{\f_2}_2-32 \fa^{\c}_0 M^2\right)\right)-24 M^2 \left(5  \fa^{\f_1}_2+\sqrt{5}  \fa^{\f_2}_2\right) \log (\fz)+15 \fa^{\f_1}_4+3 \sqrt{5}  \fa^{\f_2}_4\right)\frac{}{}\right)+ \nn \\
 && \frac{\fz^{10}}{3375}\left(\frac{}{}-24 M^6 \left(5 \left(-120  \fa^{\mathcal{A}_7^{(1)}}_{0} Q_2^2 \log (z)+22  \fa^{\mathcal{A}_7^{(1)}}_{0} Q_2^2+5  \fa^{\mathcal{A}_7^{(1)}}_{4}\right)+ \nn \right.\right. \\
 && \left. 15  \fa^{\mathcal{A}_7^{(1)}}_{0}
  Q_1^2 (139-240 \log (\fz))+8 \sqrt{10} \fa^{\c}_0 Q_1 \sqrt{\mu } (1-60 \log (\fz))\right)-\nn  \\
  && \left. 90 M^2 \left(10
    \fa^{\mathcal{A}_7^{(1)}}_{0} \left(-15 Q_1^4+5 Q_1^2 Q_2^2+Q_2^4\right)+10  \fa^{\mathcal{A}_7^{(1)}}_{4} \left(6
   Q_1^2+Q_2^2\right)+\sqrt{2} Q_1 \sqrt{\mu } \left(35 \fa^{\f_1}_4+\sqrt{5} \left(48  \fa^{\c}_0
  Q_1^2+7 \fa^{\f_2}_4\right)\right)\right)+\nn \right. \\
  && \left. 18 Q_1 \sqrt{\mu } \left(-240 \sqrt{\mu } ( \fa^{\mathcal{A}_7^{(1)}}_{0}
   Q_1+ \fa^{\mathcal{A}_7^{(2)}}_{0} Q_2)-144 \sqrt{10} \fa^{\c}_0 \mu +\sqrt{2} \left(-60 \sqrt{5}  \fa^{\c}_6+5
   \fa^{\f_1}_2 \left(38 Q_1^2+7 Q_2^2\right)+\sqrt{5}  \fa^{\f_2}_2 \left(38 Q_1^2-37
   Q_2^2\right)\right)\right)+\nn \right. \\
   && \left. 80  \fa^{\mathcal{A}_7^{(1)}}_{0} M^{10} (120 \log (\fz)-157)-12600  \fa^{\mathcal{A}_7^{(1)}}_{0} \mu  M^4+6 \sqrt{2} Q_1
   \sqrt{\mu } M^4 \left(5  \fa^{\f_1}_2+\sqrt{5} \fa^{\f_2}_2\right) (840 \log (\fz)-649)\frac{}{}\right)+\cO(\fz^{12})\nn \, , \\
\label{eq:FlucUVexp A72}
\fa^{\mathcal{A}_7^{(2)}}_{UV}(\fz)&=&\fa^{\mathcal{A}_7^{(2)}}_{0} + 4\fa^{\mathcal{A}_7^{(2)}}_{0}M^2\fz^2 + \fz^4 \left(\fa^{\mathcal{A}_7^{(2)}}_{ 4}-16\fa^{\mathcal{A}_7^{(2)}}_{0}M^4 \log[\fz]\right)+ \\ 
 & &\fz^6\left(\frac{64}{3}\fa^{\mathcal{A}_7^{(2)}}_{0}M^6\log[\fz]-\frac{2}{45}\sqrt{\mu}Q_1\left(30\fa^{\mathcal{A}_7^{(2)}}_{4}M^2-3\sqrt{2}(5\fa^{\f_1}_{UV, 2}+\sqrt{5}(\fa^{\f_1}_{2}-8\fa^{\c}_{0}M^2))\right)+ 10\fa^{\mathcal{A}_7^{(2)}}_0M^2(32 M^4-9Q_1^2+3Q_2^2)\right) + \nn\\
  && \fz^8\left(\frac{}{} Q_2^2 \left(-16 \fa^{\mathcal{A}_7^{(2)}}_{0} M^4 \log (\fz)-2 \fa^{\mathcal{A}_7^{(2)}}_{0} M^4+\fa^{\mathcal{A}_7^{(2)}}_{4}\right)- \nn \right. \\
  && \frac{2}{9} M^2 \left(\fa^{\mathcal{A}_7^{(2)}}_{0} \left(6 Q_1^2 M^2+9 \mu -50 M^6\right)+48 \fa^{\mathcal{A}_7^{(2)}}_{0} M^6 \log (\fz)-3 \fa^{\mathcal{A}_7^{(2)}}_{4} M^2\right)+  \nn \\
 && \left. \frac{Q_2 \sqrt{\mu }}{15 \sqrt{2}} \left(M^2 \left(-5  \fa^{\f_1}_2+\sqrt{5} \left( \fa^{\f_2}_2-32 \fa^{\c}_0 M^2\right)\right)+24 M^2 \left(5  \fa^{\f_1}_2-\sqrt{5}  \fa^{\f_2}_2\right) \log (\fz)-15 \fa^{\f_2}_4+3 \sqrt{5}  \fa^{\f_2}_4\right)\frac{}{}\right)+ \nn \\
 && \frac{\fz^{10}}{3375}\left(\frac{}{}-24 M^6 \left(5 \left(-120  \fa^{\mathcal{A}_7^{(2)}}_{0} Q_1^2 \log (\fz)+22  \fa^{\mathcal{A}_7^{(2)}}_{0} Q_1^2+5  \fa^{\mathcal{A}_7^{(2)}}_{4}\right)+\nn \right. \right. \\
 && \left. 15  \fa^{\mathcal{A}_7^{(2)}}_{0}
  Q_2^2 (139-240 \log (\fz))+8 \sqrt{10} \fa^{\c}_0 Q_2 \sqrt{\mu } (1-60 \log (\fz))\right)-\nn  \\
  && \left. 90 M^2 \left(10
    \fa^{\mathcal{A}_7^{(2)}}_{0} \left(-15 Q_2^4+5 Q_1^2 Q_2^2+Q_1^4\right)+10  \fa^{\mathcal{A}_7^{(2)}}_{4} \left(6
   Q_2^2+Q_1^2\right)+\sqrt{2} Q_2 \sqrt{\mu } \left(-35 \fa^{\f_1}_4+\sqrt{5} \left(48  \fa^{\c}_0
  Q_2^2+7 \fa^{\f_2}_4\right)\right)\right)+\nn \right. \\
  && \left. 18 Q_2 \sqrt{\mu } \left(-240 \sqrt{\mu } ( \fa^{\mathcal{A}_7^{(1)}}_{0}
   Q_1+ \fa^{\mathcal{A}_7^{(2)}}_{0} Q_2)-144 \sqrt{10} \fa^{\c}_0 \mu -\sqrt{2} \left(60 \sqrt{5}  \fa^{\c}_6+5
   \fa^{\f_1}_2 \left(38 Q_2^2+7 Q_1^2\right)+\sqrt{5}  \fa^{\f_2}_2 \left(-38 Q_2^2+37
   Q_1^2\right)\right)\right)+\nn \right. \\
   && \left. 80  \fa^{\mathcal{A}_7^{(2)}}_{0} M^{10} (120 \log (\fz)-157)-12600  \fa^{\mathcal{A}_7^{(2)}}_{0} \mu  M^4+6 \sqrt{2} Q_2
   \sqrt{\mu } M^4 \left(5  \fa^{\f_1}_2-\sqrt{5} \fa^{\f_2}_2\right) (840 \log (\fz)-649)\frac{}{}\right)+\cO(\fz^{12})\nn \, , \\
 \label{eq:FlucUVexp phi1}
 \fa^{\f_1}_{UV}(\fz)&=& \fa^{\f_1}_2 \fz^2 +\fz^4(\fa^{\f_1}_4-8\fa^{\f_1}_2M^2\log[\fz])+  \\
 && \fz^6\left(16\fa^{\f_1}_2 M^4\log[\fz]-2\fa^{\f_1}_4 M^2+\frac{3}{2\sqrt{5}}\fa^{\f_2}_2(Q_1-Q_2)(Q_1+Q_2)+\frac{1}{2}\fa^{\f_1}_2(Q_1^2+Q_2^2-24M^4)\right)+\nn \\
 &&  \frac{\fz^8}{90} \left(-12 M^2 \left(40 \sqrt{2} \sqrt{\mu } ( \fa^{\mathcal{A}_7^{(1)}}_{0} Q_1- \fa^{\mathcal{A}_7^{(2)}}_{0}Q_2)+5 \fa^{\f_1}_2
   \left(Q_1^2+Q_2^2\right)+\sqrt{5} \fa^{\f_2}_2 \left(Q_2^2-Q_1^2\right)\right)- \right.   \nn \\ 
   && \left. 24 M^2 \log (z)
   \left(5 \fa^{\f_1}_2 \left(3 \left(Q_1^2+Q_2^2\right)+8 M^4\right)+3 \sqrt{5} \fa^{\f_2}_2
   (Q_1-Q_2) (Q_1+Q_2)\right)+60 \fa^{\f_1}_2 \mu +1120 \fa^{\f_1}_2 M^6+ \nn \right. \\
   && \left. 15 \fa^{\f_1}_4
 \left (3 \left(Q_1^2+Q_2^2\right)+8 M^4\right)+9 \sqrt{5} \fa^{\f_2}_4 (Q_1-Q_2)
   (Q_1+Q_2) \frac{}{}\right)+\nn \\
   &&\frac{\fz^{10}}{1350}\left(\frac{}{}3\sqrt{5}\fa^{\f_2}_2(Q_1-Q_2)(Q_1+Q_2)(57(Q_1^2+Q_2^2)-430M^4) +\nn \right. \\
   &&12\left(15\sqrt{5}\fa^{\f_2}_4(Q_2^2-Q_1^2)M^2-50\fa^{\f_1}_4(3(Q_1^2+Q_2^2)M^2+M^6-3\m)+192\sqrt{5}\fa^{\c}_0(Q_2^2-Q_1^2)\m- \nn \right. \\
   && \left. 10\sqrt{2}\left(15\fa^{\cA_7^{(1)}}_4Q_1-15\fa^{\cA_7^{(2)}}_4Q_2+6(Q_1-Q_2)(Q_1+Q_2)(\fa^{\cA_7^{(1)}}_0Q_1+\fa^{\cA_7^{(2)}}_0Q_2)+10(-\fa^{\cA_7^{(2)}}_0Q_1+\fa^{\cA_7^{(2)}}_0Q_2)M^4\right)\right)\frac{}{}\nn \\
   && 5\fa^{\f_1}_2(81Q_1^4+81Q_2^4-1770Q_2^2M^4+6Q_1^2(18Q_2^2-295M^4)-20(70M^8+33M^2\m))+\nn \\
   && \left.480M^2\log[\fz]\left(3M^2(\sqrt{5}\fa^{\f_2}_2(Q_1-Q_2)(Q1+Q_2)+20\sqrt{2}(\fa^{\cA_7^{(1)}}_0Q_1-\fa^{\cA_7^{(2)}}_0Q_2)\sqrt{\m})+10\fa^{\f_1}_2(2(Q_1^2+Q_2^2)M^2+M^6-3\m)\right)\right) + \nn \\
  && + \co(\fz^{12})\nn\, ,
   \eeqs
   \beqs
 \label{eq:FlucUVexp phi2}
\fa^{\f_2}_{UV}(\fz)&=& \fa^{\f_2}_2 \fz^2 +\fz^4(\fa^{\f_2}_4-8\fa^{\f_2}_2M^2\log[\fz])+  \\
 && \fz^6\left(-12\fa^{\f_2}_2 M^4-2\fa^{\f_2}_4 M^2+16\fa^{\f_2}_2M^4\log[\fz]+\frac{3}{2\sqrt{5}}\fa^{\f_2}_2(Q_1-Q_2)(Q_1+Q_2)-\frac{7}{10}\fa^{\f_2}_2(Q_1^2+Q_2^2-24M^4)\right)+ \nn \\
  &&  \fz^8\left(\frac{}{}\frac{1}{90}\left(\frac{}{}9\sqrt{5}\fa^{\f_1}_4(Q_1-Q_2)(Q_1+Q_2)+9\fa^{\f_2}_4(Q_1^2+Q_2^2)+120\fa^{\f_2}_4M^4+1120\fa^{\f_2}_2M^6 \right. \right. - \nn \\
  && \left. 12M^2\left(\sqrt{5}\fa^{\f_1}_2(Q_2^2-Q_1^2)+9\fa^{\f_2}_2(Q_1^2+Q_2^2)+8\sqrt{10}(\fa^{\cA_7^{(1)}}_0Q_1+\fa^{\cA_7^{(2)}}_0Q_2)\sqrt{\m}\right)+60\fa^{\f_2}_2\m\frac{}{}\right) -\nn \\
  && \left.\frac{4}{15}M^2\log[\fz](3\sqrt{5}\fa^{\f_1}_2(Q_1-Q_2)(Q_1+Q_2)+3\fa^{\f_2}_2(Q_1^2+Q_2^2)+40\fa^{\f_2}_2M^4) \frac{}{}\right) \nn + \\
  && \frac{\fz^{10}}{1350}\left(\frac{}{}180\sqrt{5}\fa^{\f_1}_4(Q_2^2-Q_1^2)M^2+3\sqrt{5}\fa^{\f_1}_2(Q_1-Q_2)(Q_1+Q_2)(57(Q_1^2+Q_2^2)-430M^4)\right. + \nn \\
  &&-120\fa^{\f_2}_4M^2(9(Q_1^2+Q_2^2)+5M^4)+24(75\fa^{\f_2}_4\m-96\fa^{\c}_0(Q_1^2+Q_2^2)\m)- \nn  \\
  &&  24\sqrt{10}\sqrt{\m}\left(15\fa^{\cA_7^{(1)}}_4Q_1+15\fa^{\cA_7^{(2)}}_4Q_2+2(\fa^{\cA_7^{(1)}}_0Q_1+\fa^{\cA_7^{(2)}}_0Q_2)(3(Q_1^2+Q_2^2)-5M^4)\right)\nn -\\
 && \fa^{\f_2}_2(279Q_1^4+279Q_2^4+3690Q_2^2M^4+700M^8+18Q_1^2(46Q_2^2+205M^4)+3300M^2\m)+ \nn \\
 && \left.480M^2\log[\fz]\left(10\fa^{\f_2}_2M^6+3M^2(\sqrt{5}\fa^{\f_1}_2(Q_1-Q_2)(Q_1+Q_2)+6\fa^{\f_2}_2(Q_1^2+Q_2^2)+4\sqrt{10}(\fa^{\cA_7^{(1)}}_0Q_1+\fa^{\cA_7^{(2)}}_0Q_2)\sqrt{\m})-30\fa^{\f_2}_2\right)  \frac{}{}\right) \nn \\
 && + \cO(\fz^{12})\nn \, , \\
 \label{eq:FlucUVexp chi}
\fa^{\c}_{UV}(\fz)&=& \fa^{\c}_0+2\fa^{\c}_0M^2\fz^2+4\fa^{\c}_0M^4\fz^4+\fz^6\left(\fa^{\c}_6-\frac{32}{3}\fa^{\c}_0M^6\log[\fz]\right)+  \\
&& \fz^8 \bigg( 2\fa^{\c}_0(Q_1^2+Q_2^2)M^4-\frac{20}{3}\fa^{\c}_0\M^8+\frac{M^2}{5}(-5\fa^{\c}_6+8\sqrt{10}(\fa^{\cA_7^{(1)}}_0Q_1+\fa^{\cA_7^{(2)}}Q_2)\sqrt{\m}10\fa^{\c}_0\m)+\frac{32}{3}\fa^{\c}_0M^8\log[\fz]\bigg) \nn \\
&& \frac{\fz^{10}}{375}\left(\frac{}{} 15(15\fa^{\c}_6(Q_1^2+Q_2^2)+8(\sqrt{10}\fa^{\cA_7^{(1)}}_4Q_1\sqrt{\m}+\sqrt{10}\fa^{\cA_7^{(2)}}_4Q_2\sqrt{\m}+4\fa^{\c}_0(Q_1^2+Q_2^2)\m))  - \right. \nn \\
&&150\fa^{\c}_0Q_1^2Q_2^2M^2+40\fa^{\c}_0M^{10}(39-40\log[\fz])-20\fa^{\c}_0(Q_1^2+Q_2^2)M^6(23+120\log[\fz])+ \nn \\
&& \left. 6M^4(25\fa^{\c}_6+350\fa^{\c}_0\m-8\sqrt{10}(\fa^{\cA_7^{(1)}}_0Q_1+\fa^{\cA_7^{(2)}}_0Q_2)\sqrt{\m})\frac{}{}\right) + \cO(\fz^{12}) \nn\,.
\eeqs

For the tensor fluctuations, the expressions are simpler, and depend on two free parameters that we denote as $T_0$ and $T_6$. 
\beqs
T_{UV}(\fz)&=&T_0+2 M^2 T_0 \fz^2+4 M^4 T_0 \fz^4+ \fz^6 \left(T_6-\frac{32}{3} M^6 T_0 \log (\fz)\right)+ \\
&& \fz^8 \left(\frac{32}{3} M^8 T_0 \log (\fz)+\frac{1}{3} M^2 \left(-20 M^6 T_0+6 M^2 \left(Q_1^2+Q_2^2\right) T_0+6 \mu  T_0-3T_6\right)\right)+ \nn \\ 
&& \fz^{10} \Bigg(\frac{1}{75} \left(Q_2^2 \left(45 T_6-92 M^6 T_0\right)+6 M^4 \left(52 M^6 T_0+70 \mu  T_0+5
   T_6\right)+Q_1^2 \left(-92 M^6 T_0-30 M^2 Q_2^2 T_0+45T_6\right)\right)-\nn \\
   && \frac{32}{15} M^6 T_0 \left(2 M^4+3 \left(Q_1^2+Q_2^2\right)\right) \log (\fz)\Bigg)+\cO(\fz^{12})\,.\nn
\eeqs

%%%%%%%%%%%%%%%%%%%%%%%
%%%%%%%%%%%%%%%%%%%%%%%
\subsection{Probe Approximation}
\label{Sec:probeeqs}

We report here the equations obeyed by the scalar fluctuations treated in the probe approximation defined in the main body of the text. 
In this Appendix, we denote the fluctuations as $\left\{\mathfrak{p}^{\phi_1},\,\mathfrak{p}^{\phi_2},\,\mathfrak{p}^{\chi},\,
\mathfrak{p}^{{\cal A}_7^{(1)}},\,\mathfrak{p}^{{\cal A}_7^{(2)}} \right\}$, to distinguish them from the gauge invariant variables.
These fluctuations all obey Dirichlet boundary conditions, $\left.\frac{}{}\mathfrak{p}\right|_{\rho=\rho_i} =0$.

\beqs
0 &=& \Bigg[e^{2 A+\sqrt{\frac{2}{5}} \left(4 \chi +\phi _2\right)+\sqrt{2} \phi _1}\left( 10\6_{\r}^2+\left(50 A'-\sqrt{10} \chi '\right) \6_{\r}\right)+ \\
&& 10 \left(-e^{2 A+2 \sqrt{\frac{2}{5}} \phi _2} \left(e^{2 \sqrt{2} \phi _1}\left({\cA_7^{(1)}}'\right)^2+\left({\cA_7^{(2)}}'\right)^2\right)+\left(e^{\sqrt{2} \phi _1}+1\right) e^{2 A+\frac{8 \chi +5
   \phi _2}{\sqrt{10}}+\frac{\phi _1}{\sqrt{2}}}+M^2 e^{\sqrt{\frac{2}{5}} \left(5 \chi +\phi _2\right)+\sqrt{2} \phi _1}\right)\Bigg]\mathfrak{p}^{\f_1} +\nn \\
&&  \Bigg[2 \sqrt{5} e^{2 A+2 \sqrt{\frac{2}{5}} \phi _2} \left(-e^{2 \sqrt{2} \phi _1} \left({\cA_7^{(1)}}'\right)^2+\left({\cA_7^{(2)}}'\right)^2+3 \left(e^{\sqrt{2} \phi _1}-1\right) e^{\frac{8 \chi +\phi
   _2}{\sqrt{10}}+\frac{\phi _1}{\sqrt{2}}}\right)\Bigg]\mathfrak{p}^{\f_2} +\nn \\
&&  \Bigg[\sqrt{2} e^{2 A+2 \sqrt{2} \phi _1+2 \sqrt{\frac{2}{5}} \phi _2} \left(-10{\cA_7^{(1)}}'\6_{\r}- \left({\cA_7^{(1)}}' \left(25 \sqrt{2} A'+\sqrt{5} \left(2 \phi _2'-9 \chi '\right)+10 \phi _1'\right)+5 \sqrt{2}
   {\cA_7^{(1)}}''\right)\right)\Bigg]\mathfrak{p}^{\cA_7^{(1)}} + \nn \\
&&  \Bigg[ e^{2 A+2 \sqrt{\frac{2}{5}} \phi _2} \left(10 \sqrt{2}{\cA_7^{(2)}}'\6_{\r} +{\cA_7^{(2)}}' \left(25 \sqrt{2} A'+\sqrt{5} \left(2 \phi _2'-9 \chi '\right)-10 \phi _1'\right)+5 \sqrt{2} {\cA_7^{(2)}}''\right)\Bigg]\mathfrak{p}^{\cA_7^{(2)}} +\nn  \\
&&  \Bigg[4 \sqrt{5} e^{2 A+2 \sqrt{\frac{2}{5}} \phi _2} \left(-2 e^{2 \sqrt{2} \phi _1}\left({\cA_7^{(1)}}'\right)^2+2 \left({\cA_7^{(2)}}'\right)^2+\left(e^{\sqrt{2} \phi _1}-1\right) e^{\frac{8 \chi +\phi
   _2}{\sqrt{10}}+\frac{\phi _1}{\sqrt{2}}}\right)\Bigg]\mathfrak{p}^{\c} \nn \, ,
\eeqs

\beqs
0 &=& \Bigg[2 \sqrt{5} e^{2 A+2 \sqrt{\frac{2}{5}} \phi _2} \left(-e^{2 \sqrt{2} \phi _1} \left({\cA_7^{(1)}}'\right)^2+\left({\cA_7^{(2)}}'\right)^2+3 \left(e^{\sqrt{2} \phi _1}-1\right) e^{\frac{8 \chi +\phi
   _2}{\sqrt{10}}+\frac{\phi _1}{\sqrt{2}}}\right)\Bigg]\mathfrak{p}^{\f_1}+ \\
&&  \Bigg[ e^{2 A+\sqrt{\frac{2}{5}} \left(4 \chi +\phi _2\right)+\sqrt{2} \phi _1}\left(10\6_{\r}^2+50 A'-\sqrt{10} \chi '\6_{\r}\right)+\nn \\
&& 2 e^{4 \sqrt{\frac{2}{5}} \chi +\frac{\phi _1}{\sqrt{2}}} \left(e^{2 A} \left(9 e^{\sqrt{\frac{5}{2}} \phi _2} \left(e^{\sqrt{2} \phi _1}+1\right)+8 e^{\frac{\phi _1}{\sqrt{2}}}-16
   e^{\frac{\phi _1}{\sqrt{2}}+\sqrt{10} \phi _2}\right)+5 M^2 e^{\sqrt{\frac{2}{5}} \left(\chi +\phi _2\right)+\frac{\phi _1}{\sqrt{2}}}\right)-\nn \\ 
   && 2 e^{2 A+2 \sqrt{\frac{2}{5}} \phi _2}
   \left(e^{2 \sqrt{2} \phi _1} \left({\cA_7^{(1)}}'\right)^2+\left({\cA_7^{(2)}}'\right)^2\right)\Bigg]\mathfrak{p}^{\f_2} +\nn \\
&&  \Bigg[ e^{2 A+2 \sqrt{2} \phi _1+2 \sqrt{\frac{2}{5}} \phi _2}\left( -2 \sqrt{10}{\cA_7^{(1)}}'\6_{\r}-{\cA_7^{(1)}}' \left(5 \sqrt{10} A'-9 \chi '+2 \sqrt{5} \phi _1'+2 \phi _2'\right)+\sqrt{10} {\cA_7^{(1)}}''\right)\Bigg]\mathfrak{p}^{\cA_7^{(1)}} +\nn \\
&&  \Bigg[ e^{2 A+2 \sqrt{\frac{2}{5}} \phi _2}\left(-2 \sqrt{10} {\cA_7^{(2)}}'\6_{\r}-{\cA_7^{(2)}}'\left(5 \sqrt{10} A'-9 \chi '-2 \sqrt{5} \phi _1'+2 \phi _2'\right)+\sqrt{10} {\cA_7^{(2)}}''\right)\Bigg]\mathfrak{p}^{\cA_7^{(2)}} +\nn  \\
&&  \Bigg[4 e^{2 A} \left(2 e^{2 \sqrt{\frac{2}{5}} \phi _2} \left(e^{2 \sqrt{2} \phi _1} \left({\cA_7^{(1)}}'\right)^2+\left({\cA_7^{(2)}}'\right)^2\right)-3 \left(e^{\sqrt{2} \phi _1}+1\right) e^{\frac{8 \chi
   +5 \phi _2}{\sqrt{10}}+\frac{\phi _1}{\sqrt{2}}}+2 \left(e^{\sqrt{10} \phi _2}+2\right) e^{4 \sqrt{\frac{2}{5}} \chi +\sqrt{2} \phi _1}\right)\Bigg]\mathfrak{p}^{\c} \nn \, ,
\eeqs

\beqs
0 &=& \Bigg[-2 \sqrt{5} e^{2 A+2 \sqrt{\frac{2}{5}} \phi _2} \left(-2 e^{2 \sqrt{2} \phi _1}\left({\cA_7^{(1)}}'\right)^2+2 \left({\cA_7^{(2)}}'\right)^2+\left(e^{\sqrt{2} \phi _1}-1\right) e^{\frac{8 \chi +\phi
   _2}{\sqrt{10}}+\frac{\phi _1}{\sqrt{2}}}\right)\Bigg]\mathfrak{p}^{\f_1}+ \\
&&  \Bigg[2 e^{2 A} \left(2 e^{2 \sqrt{\frac{2}{5}} \phi _2} \left(e^{2 \sqrt{2} \phi _1} \left({\cA_7^{(1)}}'\right)^2+\left({\cA_7^{(2)}}'\right)^2\right)-3 \left(e^{\sqrt{2} \phi _1}+1\right) e^{\frac{8 \chi
   +5 \phi _2}{\sqrt{10}}+\frac{\phi _1}{\sqrt{2}}}+2 \left(e^{\sqrt{10} \phi _2}+2\right) e^{4 \sqrt{\frac{2}{5}} \chi +\sqrt{2} \phi _1}\right)\Bigg]\mathfrak{p}^{\f_2}+ \nn \\
&&  \Bigg[e^{2 A+2 \sqrt{2} \phi _1+2 \sqrt{\frac{2}{5}} \phi _2}\left(4 \sqrt{10}  {\cA_7^{(1)}}'\6_{\r}+2 \left({\cA_7^{(1)}}'\left(5 \sqrt{10} A'-9 \chi '+2 \sqrt{5} \phi _1'+2 \phi _2'\right)+\sqrt{10} {\cA_7^{(1)}}''\right)\right)\Bigg]\mathfrak{p}^{\cA_7^{(1)}} +\nn \\
&&  \Bigg[e^{2 A+2 \sqrt{\frac{2}{5}} \phi _2} \left(4 \sqrt{10} {\cA_7^{(2)}}'\6_{\r} +2 \left({\cA_7^{(2)}}'\left(5 \sqrt{10} A'-9 \chi '-2 \sqrt{5} \phi _1'+2 \phi _2'\right)+\sqrt{10} {\cA_7^{(2)}}''\right)\right)\Bigg]\mathfrak{p}^{\cA_7^{(2)}}+ \nn  \\
&&  \Bigg[-16 e^{2 A+2 \sqrt{2} \phi _1+2 \sqrt{\frac{2}{5}} \phi _2} \left({\cA_7^{(1)}}'\right)^2-16 e^{2 A+2 \sqrt{\frac{2}{5}} \phi _2} \left({\cA_7^{(2)}}'\right)^2+e^{4 \sqrt{\frac{2}{5}} \chi +\sqrt{2}
   \phi _1} \left(20 M^2 e^{\sqrt{\frac{2}{5}} \left(\chi +\phi _2\right)}-e^{2 A} \left(e^{\sqrt{10} \phi _2}-8\right)\right)+\nn \\ 
   && 4 \left(e^{\sqrt{2} \phi _1}+1\right) e^{2 A+\frac{8
   \chi +5 \phi _2}{\sqrt{10}}+\frac{\phi _1}{\sqrt{2}}}\Bigg]\mathfrak{p}^{\c} \nn \, ,
\eeqs

\beqs
0 &=& \Bigg[e^{2 A+\sqrt{\frac{2}{5}} \left(4 \chi +\phi _2\right)}\left(10 \sqrt{2}  {\cA_7^{(1)}}'\6_{\r} +{\cA_7^{(1)}}' \left(25 \sqrt{2} A'+\sqrt{5} \left(2 \phi _2'-9 \chi '\right)+10 \phi _1'\right)+5 \sqrt{2} {\cA_7^{(1)}}'')\right)\Bigg]\mathfrak{p}^{\f_1}+ \\
&&  \Bigg[ e^{2 A+\sqrt{\frac{2}{5}} \left(4 \chi +\phi _2\right)} \left(2 \sqrt{10}{\cA_7^{(1)}}'\6_{\r}+{\cA_7^{(1)}}' \left(5 \sqrt{10} A'-9 \chi '+2 \sqrt{5} \phi _1'+2 \phi _2'\right)+\sqrt{10} {\cA_7^{(1)}}''\right)\Bigg]\mathfrak{p}^{\f_2} +\nn \\
&&  \Bigg[e^{2 A+\sqrt{\frac{2}{5}} \left(4 \chi +\phi _2\right)}\Bigg(10\6_{\r}^2 +\left(50 A'+\sqrt{10} \left(2 \phi _2'-9 \chi '\right)+10 \sqrt{2} \phi _1'\right)\6_{\r} +\nn \\
&& e^{2 A} \left(\phi _1' \left(25 \sqrt{2} A'-\sqrt{5} \chi '\right)-\left(\chi '-5 \sqrt{10} A'\right) \left(\phi _2'-4 \chi
   '\right)-4 \sqrt{10} \chi ''(\rho )+5 \sqrt{2} \phi_1''(\rho )+\sqrt{10} \phi_2''(\rho )-\nn \right. \\
   && \left. 5 e^{4 \sqrt{\frac{2}{5}} \phi _2}+20 e^{\frac{\phi _1}{\sqrt{2}}+\frac{3
   \phi _2}{\sqrt{10}}}\right)+10 M^2 e^{\sqrt{\frac{2}{5}} \chi }\Bigg)-10 e^{2 A+\sqrt{2} \phi _1+2 \sqrt{\frac{2}{5}} \phi _2} \left({\cA_7^{(1)}}'\right)^2\Bigg]\mathfrak{p}^{\cA_7^{(1)}} +\nn \\
&&  \Bigg[e^{2 A+\sqrt{\frac{2}{5}} \left(4 \chi +\phi _2\right)} \left(-8 \sqrt{10} {\cA_7^{(1)}}'\6_{\r}-4 \left({\cal A}_7^{(1)\prime}(\rho ) \left(5 \sqrt{10} A'-9 \chi '+2 \sqrt{5} \phi _1'+2 \phi _2'\right)+\sqrt{10} {\cA_7^{(1)}}''\right)\right)\Bigg]\mathfrak{p}^{\c} \nn \, ,
\eeqs

\beqs
0 &=& \Bigg[e^{2 A+\sqrt{\frac{2}{5}} \left(4 \chi +\phi _2\right)+\sqrt{2} \phi _1}\left(-10 \sqrt{2} {{\cal A}_7^{(2)}}'(\rho )\6_{\r}+{\cA_7^{(2)}}' \left(-25 \sqrt{2} A'+\sqrt{5} \left(9 \chi '-2 \phi _2'\right)+10 \phi _1'\right)-5 \sqrt{2} {\cA_7^{(2)}}''\right)\Bigg]\mathfrak{p}^{\f_1} +\\
&&  \Bigg[e^{2 A+\sqrt{\frac{2}{5}} \left(4 \chi +\phi _2\right)+\sqrt{2} \phi _1}\left(2 \sqrt{10} {\cA_7^{(1)}}'\6_{\r}+{\cA_7^{(2)}}'\left(5 \sqrt{10} A'-9 \chi '-2 \sqrt{5} \phi _1'+2 \phi _2'\right)+\sqrt{10} {\cA_7^{(2)}}''\right)\Bigg]\mathfrak{p}^{\f_2} +\nn \\
&&  \Bigg[e^{2 A+\sqrt{\frac{2}{5}} \left(4 \chi +\phi _2\right)+\sqrt{2} \phi _1}\left(10\6_{\r}^2+ \left(50 A'+\sqrt{10} \left(2 \phi _2'-9 \chi '\right)-10 \sqrt{2} \phi _1'\right)\6_{\r}\right)+\nn \\
&& e^{\sqrt{\frac{2}{5}} \left(4 \chi +\phi _2\right)+\frac{\phi _1}{\sqrt{2}}} \Bigg(e^{\frac{\phi _1}{\sqrt{2}}} \left(e^{2 A} \left(5 \sqrt{2} A' \left(\sqrt{5} \left(\phi _2'-4 \chi
   '\right)-5 \phi _1'\right)-5 \sqrt{2} \phi_1''(\rho )+\sqrt{10} \left(\phi_2''(\rho )-4 \chi ''(\rho )\right)+\chi ' \left(4 \chi '+\sqrt{5} \phi _1'-\phi
   _2'\right)\right)+\right. \nn \\
   && \left. 10 M^2 e^{\sqrt{\frac{2}{5}} \chi }\right)-5 e^{2 A+\frac{\phi _1}{\sqrt{2}}+4 \sqrt{\frac{2}{5}} \phi _2}+20 e^{2 A+\frac{3 \phi _2}{\sqrt{10}}}\Bigg)-10 e^{2
   A+2 \sqrt{\frac{2}{5}} \phi _2} \left({\cA_7^{(2)}}'\right)^2\Bigg]\mathfrak{p}^{\cA_7^{(2)}} +\nn  \\
&&  \Bigg[e^{2 A+\sqrt{\frac{2}{5}} \left(4 \chi +\phi _2\right)+\sqrt{2} \phi _1}\left(-8 \sqrt{10}  {\cA_7^{(2)}}'\6_{\r}-4 \left({\cA_7^{(2)}}' \left(5 \sqrt{10} A'-9 \chi '-2 \sqrt{5} \phi _1'+2 \phi _2'\right)+\sqrt{10} {\cA_7^{(2)}}''\right)\right)\Bigg]\mathfrak{p}^{\c} \nn \, ,
\eeqs

%%%%%%%%%%%%%%%%%%%%%%%
%%%%%%%%%%%%%%%%%%%%%%%
\section{Infrared cut-off artefacts}
\label{Sec:ircutoffdependence}

\begin{figure*}
    \centering
    \subfloat[$\th=0$,$\vr_0=0.067 \label{fig:IRcutoff pi0 10-6}$]{\includegraphics[width=0.5\linewidth]{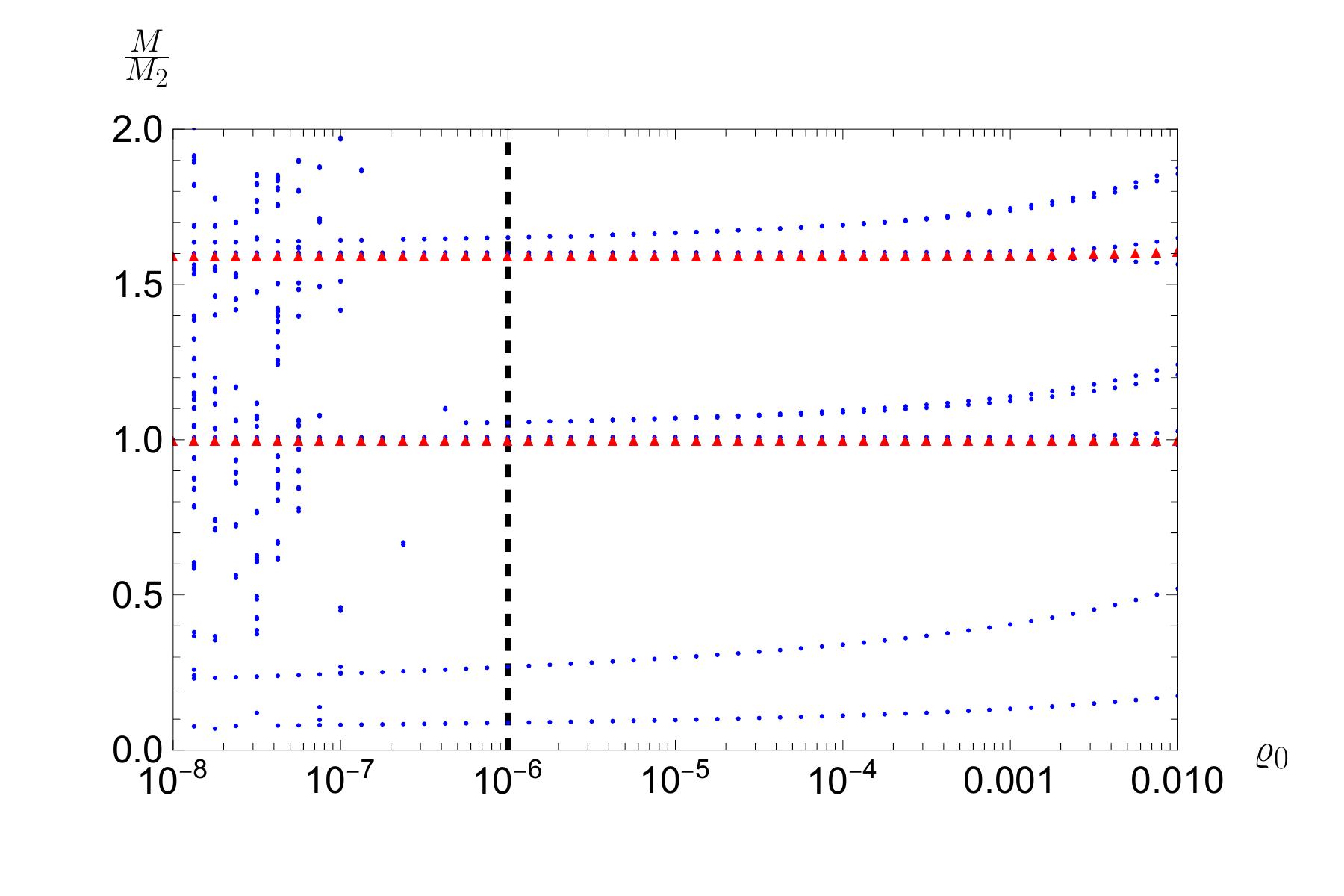}}
    \subfloat[$\th=\frac{\pi}{9}$,$\vr_0=0.912 \label{fig:IRcutoff tensor pi4 10-9}$]{\includegraphics[width=0.5\linewidth]{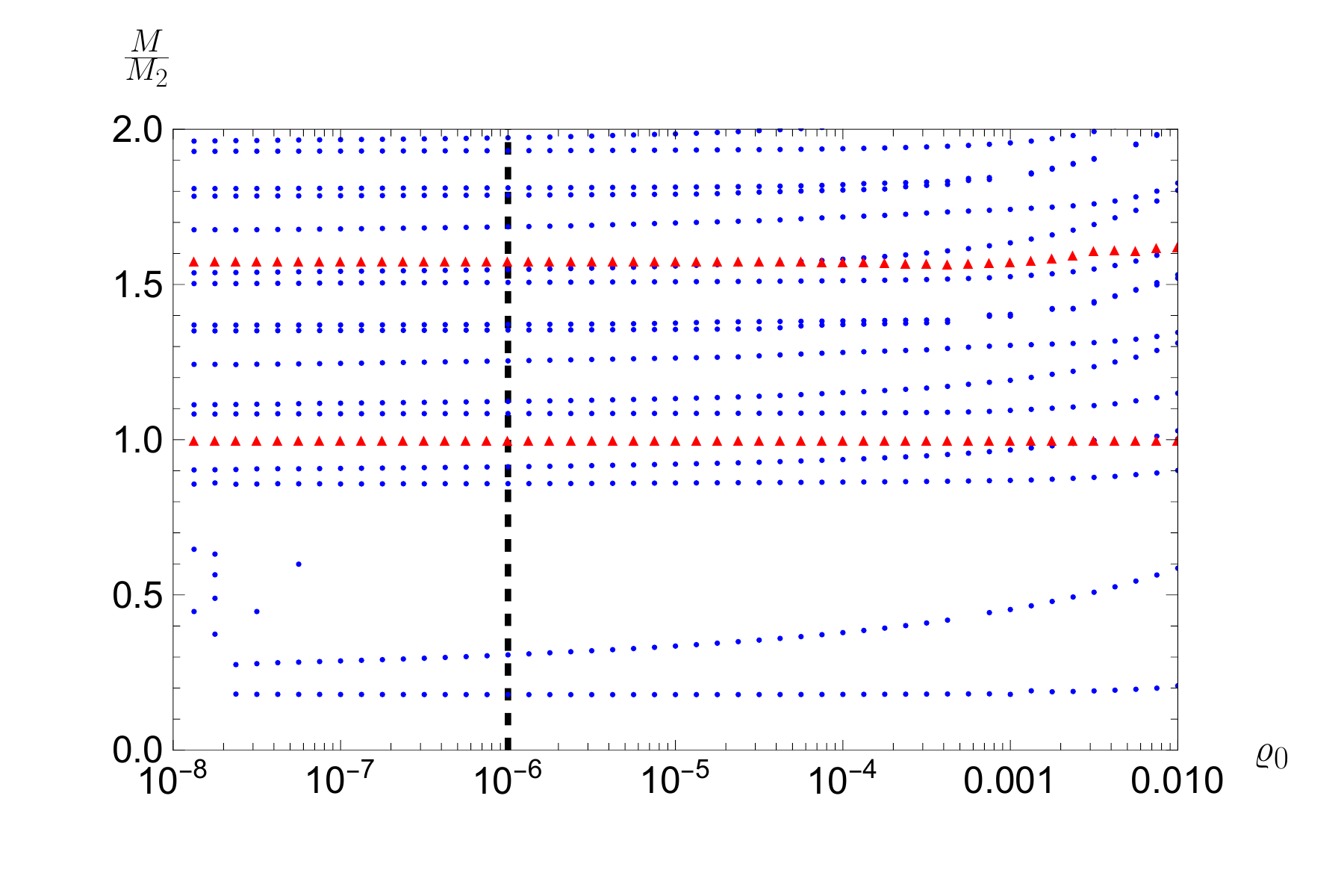}}\\
    \subfloat[$\th=\frac{\pi}{6}$,$\vr_0=0.982\label{fig:IRcutoff tensor pi4 10-6}$]{\includegraphics[width=0.5\linewidth]{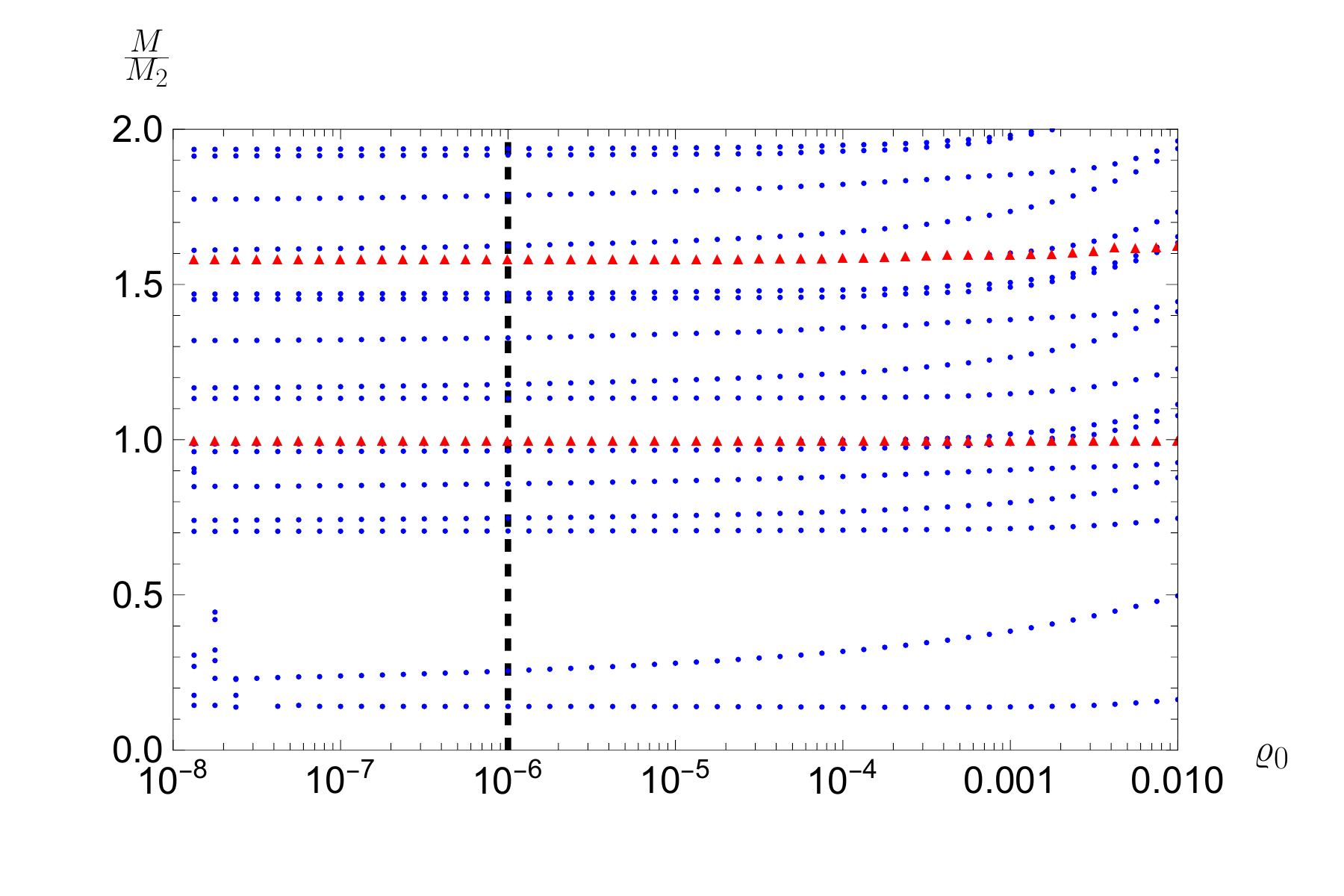}}
    \subfloat[$\th=\frac{\pi}{4},\vr_0=1.019$ \label{fig:IRcutoff tensor pi4 10-4}]{\includegraphics[width=0.5\linewidth]{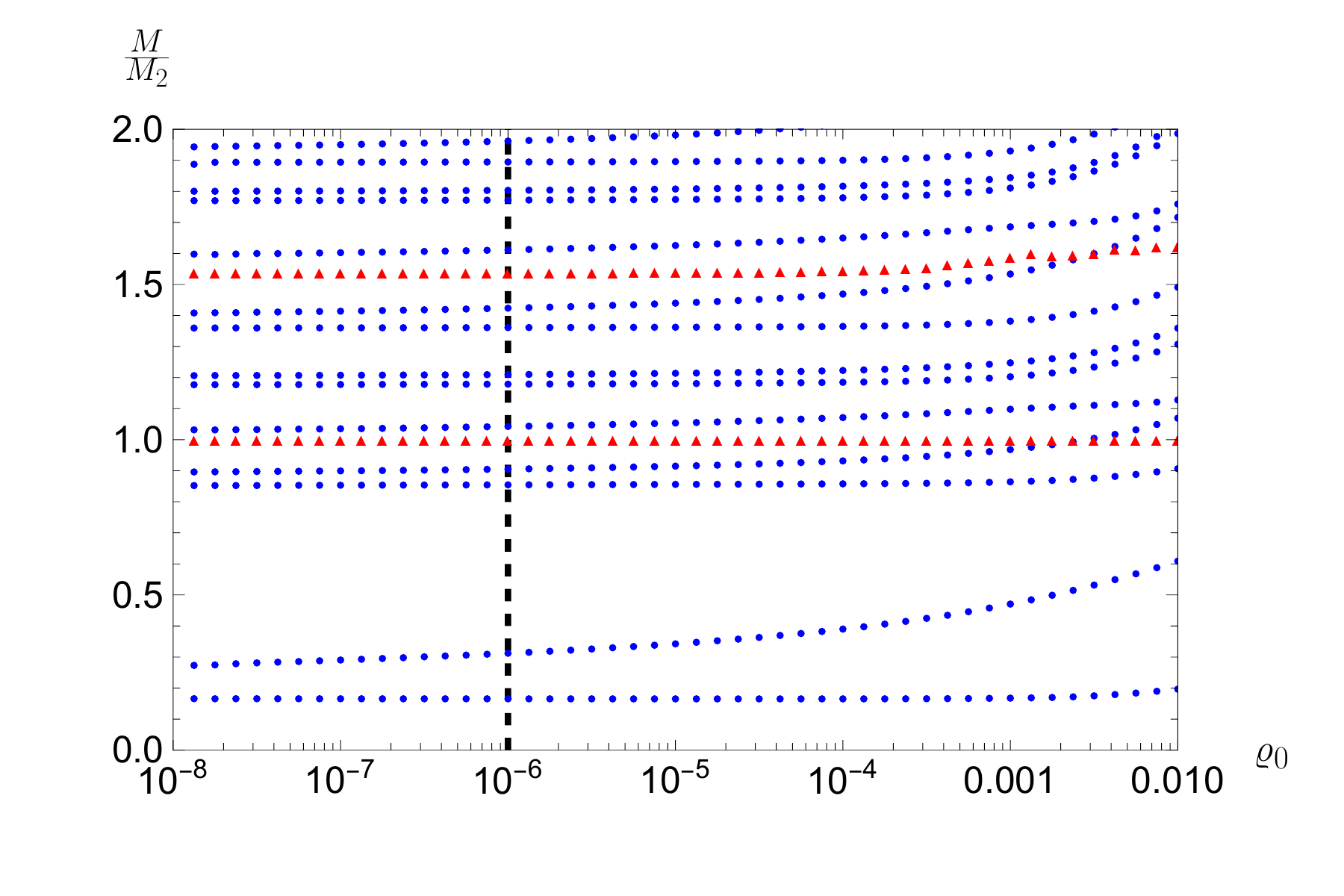}}   
   \caption{Examples of mass spectra, normalised to the mass of the lightest spin-2 fluctuation, $M_2$, as a function of $\varrho_0$, for four  examples of one-parameter subclasses of soliton (confining) solutions, obtained by fixing the ratio of the two magnetic fluxes---
see the free energy of the same solutions  in Fig. \ref{Fig:PhaseDiagram}---and choosing a representative value of $\varrho_0$.
    Spin-2 states shown as (red) triangles and Spin-0 as (blue) disks. 
   To demonstrate the suppressed IR cutoff dependence of the spectra, we evaluated them at values of $\vr_0$ close to, but not at, the phase transition. The black dashed vertical line shows the value of the cutoff used in the calculations presented in the main body of the paper, with $\vr_{IR}=10^{-6}\vr_0$}.
\label{fig:IRcutoffdependence}
\end{figure*}

In Fig.~\ref{fig:IRcutoffdependence}, we show representative examples  of the IR-cutoff dependence of our calculations of the spectrum of bound states. For four representative backgrounds, we repeated the calculation of the spectrum by changing the IR cutoff, denoted here as $\varrho_{IR}$.
The results show that  $\varrho_{IR}=10^{-6} \varrho_0$ is sufficiently small as to yield a good approximation of the physical  spectrum. At smaller values of the IR cutoff, we start to see the effect of the finite  working precision of our numerical routines. Although we notice the persistence of a small yet visible IR dependence in the second to lightest scalar state, these results are sufficiently accurate to the purposes of this study.

%%%%%%%%%%%%%%%%%%%%%%%
%%%%%%%%%%%%%%%%%%%%%%%
\newpage
\bibliographystyle{JHEP} 
\bibliography{7DSupergravity.bib}

\end{document}